\pdfoutput=1 
\documentclass[showpacs,prb,aps,superscriptaddress,twocolumn,floatfix]{revtex4}
\usepackage{graphicx,float,amsmath,amssymb}

\usepackage{color}

\usepackage{mathrsfs} 
\usepackage[utf8]{inputenc}  
\usepackage[francais,english]{babel}

\usepackage{multirow} 

\usepackage{url,hyperref}  

\usepackage{wasysym}

\newfont{\ensmathquatorze}{msbm10 scaled 1400}
\newfont{\ensmathonze}{msbm10 scaled 1100}
\newfont{\ensmathdix}{msbm10}
\newfont{\ensmathneuf}{msbm10 scaled 833}
\newfont{\ensmathhuit}{msbm10 scaled 694}
\newfam\ensmathfam                        
\textfont\ensmathfam=\ensmathonze        
\scriptfont\ensmathfam=\ensmathdix       
\scriptscriptfont\ensmathfam=\ensmathhuit

\newcommand{\ket}[1]{|\kern.3ex#1\kern.3ex\rangle}
\newcommand{\bra}[1]{\langle\kern.3ex #1 \kern.3ex|}
\newcommand{\scalar}[2]{\langle\kern.3ex #1 \kern.3ex|\kern.3ex#2\kern.3ex\rangle}
\newcommand{\mean}[1]{\left\langle #1 \right\rangle} 
\newcommand{\smean}[1]{\langle #1 \rangle} 

 \newcommand{\EXP}[1]{{\mathrm{e}}^{#1}}         

\newcommand{\re}{\mathop{\mathrm{Re}}\nolimits}      
\newcommand{\tr}[1]{\mathop{\mathrm{tr}}\nolimits\left\{ #1 \right\}}  
\renewcommand{\min}[2]{\mathop{\mathrm{min}}\nolimits\left( #1 , #2\right)}  
\renewcommand{\max}[2]{\mathop{\mathrm{max}}\nolimits\left( #1 , #2\right)}

 \newcommand{\heaviside}{\mathop{\theta_\mathrm{H}}\nolimits}  

\def\I{\mathrm{i}}                  
\def\D{\mathrm{d}}                  

\newcommand{\deriv}[2]{\frac{\mathrm{d}#1}{\mathrm{d}#2}}
\newcommand{\derivp}[2]{\frac{\partial #1}{\partial #2}}
\newcommand{\derivf}[2]{\frac{\delta #1}{\delta #2}}

\newcommand{\diagram}[3]{\raisebox{#3}{\includegraphics[scale=#2]{#1}}}

\newcommand{\diagramw}[3]{\raisebox{#3}{\includegraphics[width=#2]{#1}}}

\graphicspath{{FigPaperNL/}}

\def\Sm{\mathcal{S}}

\def\DoS{\nu}
\def\Nc{N_c}
\def\sw{w}
\def\xiloc{\xi_\mathrm{loc}}
\def\EThouless{E_\mathrm{Th}}
\def\Bcorr{\mathcal{B}_c}

\def\nonint{0}
\def\Gint{ \mathcal{G}^\mathrm{int} }
\def\Gnonint{ \mathcal{G}_0 }

\def\DifG{ g_d} 

\def\PotEnerg{U}

\begin{document}

\title{Non-linear conductance in mesoscopic weakly disordered wires \\
       Interaction and magnetic field asymmetry }

\author{Christophe Texier}
\affiliation{LPTMS, CNRS, Univ. Paris-Sud, Universit\'e Paris-Saclay, F-91405 Orsay, France}
\affiliation{Laboratoire de Physique des Solides, CNRS, Univ. Paris-Sud, Universit\'e Paris-Saclay, F-91405 Orsay, France}

\author{Johannes Mitscherling}
\affiliation{LPTMS, CNRS, Univ. Paris-Sud, Universit\'e Paris-Saclay, F-91405 Orsay, France}
\affiliation{Max Planck Institute for Solid State Research, Heisenbergstrasse 1, D-70569 Stuttgart, Germany}

\date{January 25, 2018}

\begin{abstract}
We study the non-linear conductance $\mathcal{G}\sim\partial^2I/\partial V^2|_{V=0}$ in coherent quasi-one-dimensional weakly disordered metallic wires.
Our analysis is based on the scattering approach and includes the effect of Coulomb interaction. 
The non-linear conductance correlations can be related to integrals of two fundamental correlation functions~:
the correlator of functional derivatives of the conductance and the correlator of injectivities (the injectivity is the contribution to the local density of states of eigenstates incoming from one contact).
These correlators are obtained explicitly by using diagrammatic techniques for weakly disordered metals.
In a coherent wire of length $L$, we obtain $\mathrm{rms}\big(\mathcal{G}\big)\simeq 0.006\,E_\mathrm{Th}^{-1}$ (and $\mean{\mathcal{G}}=0$), where $E_\mathrm{Th}=\hbar D/L^2$ is the Thouless energy of the wire and $D$ the diffusion constant~; the small dimensionless factor results from screening, i.e. cannot be obtained within a simple theory for non-interacting electrons.
Electronic interactions are also responsible for an asymmetry under magnetic field reversal~:
the antisymmetric part of the non-linear conductance (at high magnetic field) being much smaller than the symmetric one, $\mathrm{rms}\big(\mathcal{G}_a\big)\simeq  0.001\,(gE_\mathrm{Th})^{-1}$, where $g\gg1$ is the dimensionless (linear) conductance of the wire. 
In a weakly coherent wire (i.e. $L_\varphi\ll L$, where $L_\varphi$ is the phase coherence length), the non-linear conductance is of the same order than the result $\mathcal{G}_0$ of a free electron calculation (although screening again strongly reduces the dimensionless prefactor)~: 
we get $\mathcal{G}\sim\mathcal{G}_0\sim(L_\varphi/L)^{7/2}E_\mathrm{Th}^{-1}$, while the antisymmetric part (at high magnetic field) now behaves as 
$\mathcal{G}_a\sim(L_\varphi/L)^{11/2}(gE_\mathrm{Th})^{-1}\ll\mathcal{G}$.
The effect of thermal fluctuations is studied~: 
when the thermal length $L_T=\sqrt{\hbar D/k_BT}$ is the smallest length scale, $L_T\ll L_\varphi\ll L$, 
the free electron result $\mathcal{G}_0\sim(L_T/L)^3(L_\varphi/L)^{1/2}E_\mathrm{Th}^{-1}$ is negligible and the dominant contribution is provided by screening, 
$\mathcal{G}\sim(L_T/L)(L_\varphi/L)^{7/2}E_\mathrm{Th}^{-1}$~;
in this regime, the antisymmetric part is 
$\mathcal{G}_a\sim(L_T/L)^2(L_\varphi/L)^{7/2}(gE_\mathrm{Th})^{-1}$.
All the precise dimensionless prefactors are obtained.
Crossovers from zero to strong magnetic field regimes are also analysed.
\end{abstract}


\maketitle





\section{Introduction }

The analysis of non-linear electronic transport in mesoscopic devices is a powerful tool which can provide remarkable informations.
Among the most striking examples are the experimental techniques using the high non-linearity of the transport through a normal/superconducting interface~: this allows measuring the set of transmission probabilities characterising atomic contacts \cite{SchJoyEstUrbDev97}, or can give access to the local distribution function for electrons in metallic wires~\cite{PotGueBirEstDev97}.

Non-linear transport in normal metals has also been a subject of investigation and in particular the question of symmetries of electronic transport in the non-linear regime (symmetry under the reversal of the current flow or under the magnetic field reversal).
In the simple configuration of a two-terminal conductor, the current-voltage characteristic is expected to be an antisymmetric function, $I(-V)=-I(V)$, what in particular ensures that Joule heating does not depend on the direction of the current flow~\cite{VegTimManCunBehHow88}.
In small conductors of mesoscopic dimensions, it was however shown that the lack of inversion center symmetry, due to the geometry of the sample or to the presence of impurities, can lead to deviations from the perfect antisymmetry~: for a coherent and weakly disordered conductor at low temperature and low voltage $k_BT,\: eV\ll\EThouless$, Altshuler and Khmelnitskii obtained~\cite{AltKhm85}
$\smean{\big[I(V)+I(-V)\big]^2}\sim(e^2V/h)^2(eV/\EThouless)^2$, where $\EThouless$ is the Thouless energy (or correlation energy).
This can be reformulated by expanding the $I$-$V$ characteristic for small voltage as 
$I(V)=(2_se^2/h)\,g\,V+(2_se^3/h)\,\mathcal{G}\,V^2+\cdots$, 
where $g$ is the dimensionless (linear) conductance and $\mathcal{G}$ the rescaled non-linear conductance ($2_s$ denotes the spin degeneracy).
Despite the non-linear conductance vanishes on average, $\smean{\mathcal{G}}=0$,  in a coherent device it presents mesososcopic (sample to sample) fluctuations of order $\mathcal{G}\sim\EThouless^{-1}$.
Mesoscopic fluctuations of the current-voltage characteristic were further studied by Khmelnitskii and Larkin~\cite{LarKhm86,KhmLar86} who analysed the role of inelastic processes (decoherence) and thermal fluctuations on the correlations of the $I(V)$ curve (some of these results are recalled in Section~\ref{subsec:KLmainRes}). 
The problem was later reconsidered in Ref.~\onlinecite{LudBlaMir04} where the crossover between linear and non-linear regimes was analysed more precisely.
This question was studied experimentally in various types of samples in Refs.~\onlinecite{WebWasUmb88,VegTimManCunBehHow88}.

A second important symmetry of electric transport is the symmetry with respect to magnetic field reversal. In the linear regime, the Onsager-Casimir reciprocity relations~\cite{Ons31a,Ons31b,Cas45} for the local conductivity tensor were extended to the non-local four terminal resistances by B\"uttiker~\cite{But86a,But88a}, what leads in particular to the symmetry of the linear conductance of a two-terminal conductor, $g(\mathcal{B})=g(-\mathcal{B})$.
Such a symmetry has however no fundamental reason to hold at the level of non-linear transport, $I(V,-\mathcal{B})\neq I(V,\mathcal{B})$.
Several symmetry relations have been proposed and verified experimentally in Ref.~\onlinecite{LofMarShoTayOmlSamLin04} for mesoscopic samples with spatial symmetries.
This however leaves open the question of the origin of the asymmetry $\mathcal{G}(\mathcal{B})\neq\mathcal{G}(-\mathcal{B})$.
For example a theory for non-interacting electrons (Landauer-B\"uttiker scattering formalism) predicts that non-linear transport in a two-terminal conductor has the same symmetry as linear transport.
S\'anchez and B\"uttiker \cite{SanBut04} and Spivak and Zyuzin \cite{SpiZyu04} have proposed that the asymmetry of the non-linear transport has its origin in the electronic interactions.
The study of the non-linear asymmetry under magnetic field reversal was thus proposed as a new way to probe electronic interaction in coherent conductors~\cite{SanBut04,SpiZyu04,ZumMarHanGos06,AngZakDebGueBouCavGenPol07}.

\begin{figure}[!ht]
\centering
\includegraphics[scale=0.9]{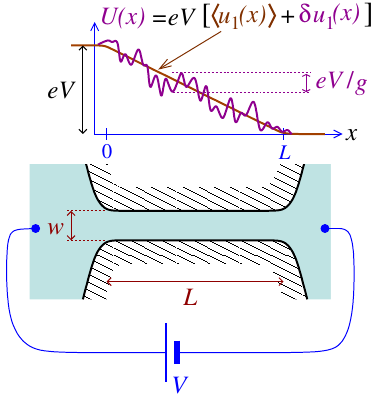}
\caption{(color online) \it A weakly disordered metallic wire of length $L$ and width $w$ between two large contacts.
 Above, we show a sketch of the electrostatic potential $U(x)$ in the disordered wire out of equilibrium.
 At linear order in the external voltage $V$, the potential is controlled by the injectivity $u_1(x)$.
 On the top of the expected linear behaviour, the disorder is responsible for mesoscopic fluctuations of order $1/g$, where $g$ is the dimensionless conductance.}
\label{fig:PotentialInWire}
\end{figure}

The study of electronic interactions in weakly disordered and coherent metals has a long history (see Refs.~\onlinecite{AltAro85,LeeRam85,AkkMon07} for reviews).
Electronic interactions were shown to be responsible for a small correction to the averaged current on the top of the classical (Drude) response, controlled by the the thermal length $L_T=\sqrt{\hbar D/k_BT}$, where  $D$ is the diffusion constant in the weakly disordered metal.
To be specific, we consider a weakly disordered quasi-1D metallic wire of length $L$ in the out-of-equilibrium situation. 
At small voltage $V$, the quantum interaction correction to the classical result $\mean{I}_\mathrm{class}=(2_se^2V/h)\,g$ is~\cite{LeaRaiSchCas00,SchRai01}~:
$\smean{\Delta I(V)}\simeq(2_se^2V/h)\,(L_T/L)\,\big[-1.57+0.067\,(eV)^2\EThouless/(k_BT)^3\big]$, where  $\EThouless=\hbar D/L^2$ is the Thouless energy of the wire.
The first term is the well-known Altshuler-Aronov correction~\cite{AltAro79,AltAro85,LeeRam85,AkkMon07,TexMon07c}, dominated by exchange (Fock contribution) for weak screening.~\cite{footnote1} 
The effect studied in the present article, initially predicted by S\'anchez and B\"uttiker \cite{SanBut04} and Spivak and Zyuzin \cite{SpiZyu04}, is rather due to Hartree contributions and can be understood as follows~:
whereas the electrostatic potential at equilibrium is a symmetric function of the magnetic field $U_\mathrm{eq}(\vec{r},\mathcal{B})=U_\mathrm{eq}(\vec{r},-\mathcal{B})$ (as it is a scalar field), in the out-of-equilibrium case the current and charge densities, and thus, due to interaction, the electrostatic potential, are non symmetric with respect to magnetic field reversal.
The potential in the wire can be written as 
$U(\vec{r},\mathcal{B})=U_\mathrm{class}(\vec{r})+\delta U(\vec{r},\mathcal{B})$, 
where the first classical term is $U_\mathrm{class}(\vec{r})=eV\,(1-x/L)$ ($x$ being the coordinate along the wire, cf.~Fig.~\ref{fig:PotentialInWire}) and the second contribution $\delta U(\vec{r},\mathcal{B})\neq\delta U(\vec{r},-\mathcal{B})$ describes the asymmetric mesoscopic (sample to sample) fluctuations, of order $\delta U\sim eV/g$~;  
these fluctuations arise from quantum interferences and are equivalent to Friedel oscillations. 
This asymmetry of the potential is of order $\mathcal{O}(V^1)$ and therefore affects contributions of order $\mathcal{O}(V^2)$ to the current, hence the asymmetry of the non-linear conductance.
We end this paragraph with two remarks concerning the specificity of the mesoscopic effect studied in Refs.~\onlinecite{SanBut04,SpiZyu04,PolBut06,DeySpiZyu06,PolBut07a} and the present article~:
first, as pointed out by Deyo, Spivak and Zyuzin~\cite{DeySpiZyu06}, this contribution to the non-linear conductance is of the same nature than the interaction corrections studied in Refs.~\onlinecite{AltAro79,AltAro85,LeeRam85,LeeStoFuk87}, as they both result from the renormalisation of the electrostatic potential due to electronic interactions, however the latter is related to the mesoscopic fluctuations of the \textit{equilibrium} potential, while the former is due to the fluctuations of the potential \textit{out-of-equilibrium} (Fig.~\ref{fig:PotentialInWire}). 
The second remark concerns the existence of classical effects~:
asymmetry with magnetic field in the non-linear transport also occurs in the classical regime in normal metal due to the bending of electronic trajectories in the presence of Coulomb interaction~\cite{AndGla06} and in chiral materials~\cite{RikFolWyd01,IvcSpi02,DeySpiZyu06}, however such contributions are proportional to the inelastic electron relaxation rate which vanishes at low temperature, whereas the effect discussed here is a mesoscopic effect which remains finite in this limit.

Interaction and the magnetic field asymmetry were also studied in another regime recently by considering Aharonov-Bohm interferometer with a quantum dot embedded in one arm in the Coulomb blockade regime, a problem motivated by several experiments (see Refs.~\onlinecite{PulMeiSigEnsIhn09,LimSanLop10} and references therein).

Magnetic field symmetry of the electric conduction in the non-linear regime was addressed in several experiments~: 
on quantum dots \cite{LofMarShoTayOmlSamLin04,MarTayFaiShoLin06,ZumMarHanGos06},
carbon nanotubes \cite{WeiShiWanRadDorCob05}, mesoscopic 2D metallic rings~\cite{LetSanGotIhnEnsDriGos06,AngZakDebGueBouCavGenPol07,Ang07,GusKvoOlsPlo09,HerGusKvoPor11} and monolayer graphene sheets~\cite{OjeMonWeiFerGueBou10}.
Motivated by the early experiments, the first theoretical works on ballistic~\cite{SanBut04} and diffusive~\cite{SpiZyu04} quantum dots were completed by investigating the role of dephasing, thermal smearing, etc~\cite{DeySpiZyu06,PolBut06,PolBut07,PolBut07a}.
However, all these theoretical studies describe zero-dimensional (0D) devices in the ergodic regime.
On the other hand, this regime is not always clearly reached in the experiments~:
in particular the rings analysed in \cite{AngZakDebGueBouCavGenPol07,Ang07} are weakly disordered and similar samples were successfully analysed by assuming diffusive regime (see for example the analysis of decoherence in Ref.~\onlinecite{CapTexMonMaiWieSam13}).
This raises several questions~: 
what is the role of the specific geometry of the system, in particular if one does not consider the ergodic regime but the \textit{diffusive} regime when the 1D character of the device is probed~?
What is the effect of dephasing in this case and what is the dependence of the non-linear conductance in the phase coherence length~? 
Are the effect of thermal fluctuations similar in diffusive and ergodic regimes~?
This is the aim of this article to provide answers to these questions.

The outline of the article is the following~:
in Section~\ref{sec:MainResults} we will introduce some of the notations and sketch the main results of the article.
Section~\ref{sec:ScatteringFormalism} presents the scattering formalism that we have adopted.
In Section~\ref{sec:Symmetries} we will introduce the two fundamental correlators on which relies the analysis, the correlator of conductance's functional derivatives and the correlator of injectivities, whose symmetries are discussed.
They will be analysed in the two rather technical Sections~\ref{sec:CondCorr} and \ref{sec:InjCorr}, which can be skipped for a first reading.
Section~\ref{sec:NLC} will combine these issues in order to derive the first part of the main results of the paper (correlations of the non-linear conductance in the weakly coherent regime).
The coherent regime will be discussed in Section~\ref{sec:Coherent}, as it requires to discuss the effect of contacts and boundary conditions, which was ignored for simplication in Section~\ref{sec:NLC}.
In Section~\ref{sec:Conclusion}, we summarise all results and close the paper by some concluding remarks.

\section{Main results}
\label{sec:MainResults}

\subsection{Non-linear conductance in disordered wires}

In a two-terminal device (Figs.~\ref{fig:PotentialInWire} and \ref{fig:qd}), the current-voltage relation can generally be expanded in powers of the applied voltage~:~\cite{footnote2} 
\begin{equation}
  I(V) = 
  \frac{2_se^2}{h} \, g\,V 
  + \frac{2_se^3}{h}\,\mathcal{G}\,V^2
  +\mathcal{O}(V^3)
\end{equation}
where $2_s$ is the spin degeneracy,
$g$ the dimensionless (linear) conductance and $\mathcal{G}$ the rescaled non-linear conductance which will be the subject of investigation of the present article.
It has thus dimension $[\mathcal{G}]=[\mathrm{Energy}]^{-1}$.

A possible starting point for the study of coherent electronic transport is the Landauer formula 
\begin{equation}
  \label{eq:Landauer}
  I(V) = \frac{2_se}{h}\int\D\varepsilon\,\left[f(\varepsilon-eV)-f(\varepsilon)\right]
  \,g(\varepsilon)
  \:,
\end{equation}
where $f(\varepsilon)$ is the Fermi-Dirac distribution and $V$ the voltage drop.
$g(\varepsilon_F)$ is the zero temperature linear dimensionless conductance at Fermi energy $\varepsilon_F$.
The expansion of the Landauer formula \eqref{eq:Landauer} gives the well-known expression of the dimensionless conductance
\begin{equation}
  g = \int\D\varepsilon\left(-\derivp{f}{\varepsilon} \right)\,g(\varepsilon)
\end{equation}
and the non-linear conductance
\begin{equation}
  \label{eq:DefG0}
  \Gnonint 
  =  \frac{1}{2}\int\D\varepsilon\left(-\derivp{f}{\varepsilon} \right)\,g'(\varepsilon)
\end{equation}
(the subscript ``$\:_\nonint\:$'' refers to the case where interaction effects are ignored).

We emphasize an important aspect of the present study~:
our analysis of non-linear transport concerns the properties of the $V\to0$ expansion of the $I$-$V$ characteristic, precisely its second derivative, $\mathcal{G}\sim\partial^2I(V)/\partial V^2\big|_{V=0}$.
In particular, we see that $\Gnonint $ provides an information on the sensitivity of the zero temperature linear conductance to the change of Fermi energy $g'(\varepsilon_F)$.
Non-linear transport was studied in Refs.~\onlinecite{LarKhm86,KhmLar86,LudBlaMir04} from the viewpoint of the correlations of current $\smean{\delta I(V_1)\delta I(V_2)}$ and the correlation of the differential conductance $\smean{\delta\DifG(V_1)\delta\DifG(V_2)}$~:  
these well established results are compared to ours in Appendix~\ref{subsec:KLDiffCond} (moreover, note that the effect of screening, which will be shown to provide the dominant contribution here, was not considered in Refs.~\onlinecite{LarKhm86,KhmLar86}).

When electronic interactions are included in the description, the non-linear conductance receives other contributions~:
\begin{equation}
  \label{eq:NLCsplittingIntNonint}
  \mathcal{G} = \Gnonint + \Gint
  \:.
\end{equation}
The distinction between the two contributions $\Gnonint$ and $\Gint$ is however purely a theoretical matter. 
In practice it is more useful to split the non-linear conductance with respect to the symmetry under magnetic field reversal~: $\mathcal{G} = \mathcal{G}_s + \mathcal{G}_a$,
where $\mathcal{G}_{s}(-\mathcal{B})=\mathcal{G}_{s}(\mathcal{B})$ and $\mathcal{G}_{a}(-\mathcal{B})=-\mathcal{G}_{a}(\mathcal{B})$.
As already mentioned, contrary to $\Gnonint$ which is symmetric under magnetic field reversal, the contribution from interactions has the remarkable property that it does not present a specific symmetry~:
it can be splitted into a symmetric and an antisymmetric part $\Gint=\Gint_s+\Gint_a$, so that the antisymmetric part is entirely due to the electronic interactions~:
\begin{equation}
  \label{eq:NLCsplittingSA}
  \mathcal{G} = \mathcal{G}_s + \mathcal{G}_a
  \hspace{0.25cm}\mbox{with }
  \begin{cases}
    \mathcal{G}_s = \Gnonint + \Gint_s
    \\
    \mathcal{G}_a = \Gint_a
  \end{cases}
 \:.
\end{equation}
The analysis of the symmetry under magnetic field reversal thus provides a practical way to identify the contribution from interactions in experiments.

In this article we consider the case of metallic (weakly disordered) wires (Fig.~\ref{fig:PotentialInWire}).
We now give the main dependences for the three contributions in terms of the characteristic length scales of the problem.
The detailed calculation of these quantities is the main purpose of the paper.
In paragraph~\ref{subsubsec:OurNewResults} we sketch our new results (a more precise summary will be provided in the concluding Section~\ref{sec:Conclusion}).


\subsubsection{Important parameters and length scales}

The study of quantum transport in a weakly disordered wires involves several length scales~:
the length of the wire $L$, the phase coherent length $L_\varphi$, which sets the scale below which quantum interferences take place. 
Thermal effects involve the thermal length $L_T=\sqrt{\hbar D/k_BT}$ (in the following, we will set $\hbar=1$ and $k_B=1$ for simplicity).
Finally we introduce the localisation length of the infinitely long weakly disordered wire~\cite{Bee97}
\begin{equation}
  \label{eq:LocalisationLength}
   \xiloc = \alpha_d\Nc\ell_e = 2\pi \rho_0 D \sw^{d-1} 
\end{equation} 
where $\Nc$ is the number of conducting channels and $\ell_e$ the elastic mean free path. $\alpha_d=V_d/V_{d-1}$ is a dimensionless constant involving the volume of the $d$-dimensional sphere of unit radius (hence $\alpha_3=4/3$ for a metallic wire deposited on a substrate and $\alpha_2=\pi/2$ for a wire etched in a 2D electron gas at the interface between two semiconductors).
We denote $\rho_0$ the DoS per unit volume and per spin channel and $\DoS_0=2_s\rho_0$ the DoS including spin degeneracy. $\sw^{d-1}$ is the cross-section of the wire and $d$ the dimension. 
The dimensionless (Drude) conductance of the wire may be expressed as 
\begin{equation}
  g = \frac{\xiloc}{L} = 2\pi\,\frac{\EThouless}{\Delta}
  \:,
\end{equation}
where $\Delta=1/(\rho_0 L\sw^{d-1})$ is the mean level spacing and $\EThouless=D/L^2$ the Thouless energy. 
$g$ is a large parameter of the problem, $g\gg1$, what ensures the validity of the diagrammatic approach~\cite{AltAro85,AkkMon07}.

\subsubsection{Non-linear conductance for free electrons}

For reference, we start by recalling the well-known behaviour of the linear conductance fluctuations.
Universal conductance fluctuations denotes (sample to sample) fluctuations of order unity $\delta g\sim1$ in a coherent device (of size $L\lesssim L_\varphi$)~\cite{Sto85,Alt85,LeeSto85,AltShk86,LeeStoFuk87,Bee97,AkkMon07}.
In a long wire ($L\gg L_\varphi$) we can use simple arguments as quantum interferences break the classical law of addition of resistances only below the scale $L_\varphi$. Thus if we slice the wire in $\mathcal{N}\sim L/L_\varphi$ pieces, different pieces can be considered as uncorrelated and we can add their resistances, leading to 
$\delta g\sim\mathcal{N}^{-2}\sum_i\delta g_i$, where $\delta g_i$ is a mesoscopic fluctuation arising from quantum interferences inside the piece~$i$.
Using $\smean{\delta g_i\delta g_j}\sim\delta_{ij}$, we end with 
$\smean{\delta g^2}\sim\mathcal{N}^{-3}$ i.e. $\delta g\sim(L_\varphi/L)^{3/2}$.
When thermal fluctuations become important ($L_T\ll L_\varphi\ll L$) the conductance fluctuations behave as $\smean{\delta g^2}\sim(L_T/L)^2(L_\varphi/L)$ (see Refs.~\onlinecite{AkkMon07,TexMon16} and references therein).

Let us now describe the main behaviours obtained within a theory of non-interacting electrons.
In the coherent limit at zero temperature, the correlator 
$\smean{\delta g(\varepsilon_F)\delta g(\varepsilon_F-\omega)}$ decays over a scale given by the Thouless energy (see Eq.~\eqref{eq:CorrelCondOmega} of Appendix~\ref{Appendix:KnownResultsNLC}), hence  $\smean{g'(\varepsilon_F)^2}\sim\smean{\delta g^2}\EThouless^{-2}$ and we recover the result of Ref.~\onlinecite{AltKhm85}, $\Gnonint=(1/2)g'(\varepsilon_F)\sim\EThouless^{-1}$ (this also follows from dimensionless analysis, as the Thouless energy is then the only relevant energy scale).
The mesoscopic fluctuation $\delta I_2\sim(e^3/h)\,\Gnonint\, V^2$ from the non-linear term remains small compared to the fluctuation $\delta I_1\sim(e^2/h)\,\delta g\,V$ from the linear term, as long as the voltage is smaller than the Thouless energy $eV\lesssim\EThouless$.

When dephasing becomes important, in the limit $L_\varphi\ll L$, 
the characteristic energy scale controlling the conductance correlator 
$\smean{\delta g(\varepsilon_F)\delta g(\varepsilon_F-\omega)}$ is the dephasing rate $1/\tau_\varphi=D/L_\varphi^2$ (see Eq.~\eqref{eq:CorrelCondOmegaIncoh} of Appendix~\ref{Appendix:KnownResultsNLC}), hence we have 
$
\smean{\Gnonint^2}
=
(1/4)\smean{g'(\varepsilon_F)^2}
\sim\smean{\delta g^2}\tau_\varphi^2\sim(L_\varphi/L)^{3+4}\EThouless^{-2}
$. 
The regime dominated by fluctuations linear in the voltage, $\delta I_1\gtrsim\delta I_2$, is therefore extended to $eV\lesssim\EThouless\,(L/L_\varphi)^2$. 

The thermal fluctuations bring a different reduction factor to $\Gnonint$ if the temperature becomes larger than the dephasing rate, i.e. when $L_T\ll L_\varphi\ll L$~; in this case 
we can write
$
\smean{\Gnonint^2}
=(1/4)\int\D(\varepsilon-\varepsilon')\,\delta_T(\varepsilon-\varepsilon')\,\smean{g'(\varepsilon)g'(\varepsilon')}
$
where $\delta_T(\omega)$ is a normalised function of width $T$.
Integration by parts gives
$
\smean{\Gnonint^2}
=-(1/4)\int\D(\varepsilon-\varepsilon')\,\delta_T''(\varepsilon-\varepsilon')\,\smean{\delta g(\varepsilon)\delta g(\varepsilon')}
\sim T^{-2}\mean{\delta g^2}
$. 
Using the expression of $\smean{\delta g^2}$ recalled above we end with
$\smean{\Gnonint^2}\sim(L_T/L)^{4+2}(L_\varphi/L)\EThouless^{-2}$.

\subsubsection{Contributions of interactions to the non-linear conductance: sketch of our main results}
\label{subsubsec:OurNewResults}

In this subsection we sketch our new results (a more precise summary will be provided in the concluding Section~\ref{sec:Conclusion} where all precise dimensionless factors are given).

The dominant effect of Coulomb interaction is the screening of the electrostatic potential. 
If the potential at contact $1$ is raised by $V$, one can write the change of electrostatic potential in the wire at lowest order in the voltage as 
$\delta U(\vec{r})=U(\vec{r})-U_\mathrm{eq}(\vec{r})\simeq u_1(\vec{r})\,eV$ where $u_1(\vec{r})$, known as the \textit{characteristic potential}, controls the response of the potential to the increase of the voltage at contact $1$.
In the diffusive wire, one has $\smean{u_1(\vec{r})}=1-x/L\sim1$ where $x\in[0,L]$ is the coordinate along the wire.
On the top of this behaviour, the characteristic potential presents mesoscopic fluctuations of the order of the DoS fluctuations~\cite{AltShk86}~: in the coherent regime,
$\delta u_1=u_1-\mean{u_1}\sim\delta\DoS/\DoS_0\sim1/g\ll1$ (Fig.~\ref{fig:PotentialInWire}), and with other reduction factors due to decoherence and/or thermal fluctuations in long wires. 
We can estimate the fluctuations of the contribution due to electronic interaction (screening) by writing $\Gint\sim\Gnonint\,u_1$. 
Making use of the fact that $\smean{\Gnonint}=0$ and $\smean{\Gnonint\,u_1}=0$ (Appendix~\ref{Appendix:CorrInjFdc}), we get two different contributions
$\smean{\big(\Gint\big)^2}\sim\smean{\Gnonint^2}\smean{u_1}^2+\smean{\Gnonint^2}\smean{\delta u_1^2}$. 
We will see that the symmetric part is dominated by the first contribution 
$\smean{\big(\Gint_s\big)^2}\sim\smean{\Gnonint^2}\smean{u_1}^2$ 
whereas the antisymmetric part is given by the subdominant contribution 
$\smean{\big(\Gint_a\big)^2}\sim\smean{\Gnonint^2}\smean{\delta u_1^2}$.
Thus $\mathcal{G}_a=\Gint_a\ll\Gint_s$. 
These simple arguments lead to the estimate $\smean{\big(\Gint\big)^2}\sim\smean{\Gnonint^2}$, which will be shown to be correct as long as thermal fluctuations can be ignored, i.e. in the regime $L_T\gg\min{L}{L_\varphi}$.
When thermal fluctuations are important, we obtain instead that $\smean{\big(\Gint\big)^2}\gg\smean{\Gnonint^2}$ (this is due to some subtle properties related to the spatial structure of the correlators, which go beyond this simple presentation and will be explained later). 
Note that $\smean{\mathcal{G}^2}=\smean{\big(\Gnonint+\Gint\big)^2}$ also receives the contribution of the anticorrelations $\smean{\Gint\Gnonint}<0$, however this does not affect the rough discussion given here.

\begin{widetext}

\begin{table}[!ht]
\centering
\begin{tabular}{|c|c|c|c|}
\hline
  & $L \ll L_\varphi,\: L_T$
  & $L_\varphi\ll L,\: L_T$             
  & $L_T\ll L_\varphi\ll L$ 
\\[0.15cm] \hline\hline
$\mathrm{rms}(\delta g)$ 
  & $\sqrt{1/15}$ (Ref.~\onlinecite{AkkMon07})
  & $\sqrt{3/2}\left(\frac{L_\varphi}{L}\right)^{3/2}$  (Ref.~\onlinecite{AkkMon07})
  & $\sqrt{\pi/3}\left(\frac{L_T}{L}\right)\left(\frac{L_\varphi}{L}\right)^{1/2}$  (Ref.~\onlinecite{AkkMon07})
\\[0.15cm] \hline\hline
$\mathrm{rms}\big(\Gnonint\big)$
  & $\simeq0.0178\,\EThouless^{-1}$ 
  $^\mathrm{(*)}$
  & $\simeq
     \frac{\sqrt{15}}{4}
     \left(\frac{L_\varphi}{L}\right)^{7/2}\EThouless^{-1}$ 
  $^\mathrm{(**)}$
  & $\simeq\frac{\sqrt{\pi}}{\sqrt{60}}\left(\frac{L_T}{L}\right)^{3}\left(\frac{L_\varphi}{L}\right)^{1/2}\EThouless^{-1}$  
\\[0.15cm]
$\mathrm{rms}\big(\Gint_s\big)$  
  & $\simeq0.0183\,\EThouless^{-1}$ 
  & $\simeq
     \frac{\sqrt{5}}{2}
     \left(\frac{L_\varphi}{L}\right)^{7/2}\EThouless^{-1}$ 
  & $\simeq0.202\left(\frac{L_T}{L}\right)\left(\frac{L_\varphi}{L}\right)^{7/2}\EThouless^{-1}$  
\\[0.15cm]
$\mathrm{rms}\big(\mathcal{G}^\mathrm{int,\,fluc}_s\big)$  
  & $\simeq0.0029\,\left(\frac{L}{\xiloc}\right)\EThouless^{-1}$ 
  & $\simeq
     \frac{\sqrt{15}}{4}
     \left(\frac{L}{\xiloc}\right)\left(\frac{L_\varphi}{L}\right)^{5}\EThouless^{-1}$ 
  & $\simeq0.203\left(\frac{L}{\xiloc}\right)\left(\frac{L_T}{L}\right)^{2}\left(\frac{L_\varphi}{L}\right)^{7/2}\EThouless^{-1}$  
\\[0.15cm]
$\mathrm{rms}\big(\mathcal{G}_a\big)$  
  & $\simeq0.0012\,\left(\frac{L}{\xiloc}\right)\EThouless^{-1}$ 
  & $\simeq
     \frac{\sqrt{27}}{8\sqrt{2}}
     \left(\frac{L}{\xiloc}\right)\left(\frac{L_\varphi}{L}\right)^{11/2}\EThouless^{-1}$ 
  & $\simeq0.055\left(\frac{L}{\xiloc}\right)\left(\frac{L_T}{L}\right)^{2}\left(\frac{L_\varphi}{L}\right)^{7/2}\EThouless^{-1}$  
\\[0.15cm] \hline
$\mathrm{rms}\big(\mathcal{G}\big)$  
  & $\simeq0.0041\,\EThouless^{-1}$ 
  & $\simeq
     \frac{\sqrt{5}}{4}
     \left(\frac{L_\varphi}{L}\right)^{7/2}\EThouless^{-1}$ 
  & $\simeq0.202\left(\frac{L_T}{L}\right)\left(\frac{L_\varphi}{L}\right)^{7/2}\EThouless^{-1}$   
\\[0.15cm]
\hline
\end{tabular}
\caption{\it 
Some of the main results obtained in the article~: the root mean square (rms) of the several contributions to the non-linear conductance $\mathcal{G}=\Gnonint+\Gint_s+\Gint_a$ of diffusive wires (numerical constants correspond to the strong magnetic field regime). 
$\Gnonint$ is the result for free electrons.
$\Gint_s$ and $\Gint_a\equiv\mathcal{G}_a$ are the contributions due to the electronic interaction, 
symmetric and antisymmetric under magnetic field reversal, respectively. 
$\mathcal{G}^\mathrm{int,\,fluc}_s$ is the subdominant contribution to $\Gint_s$ originating from the mesoscopic fluctuations of the screened electrostatic potential, with the same physical origin than $\mathcal{G}_a\equiv\Gint_a$.
The various regimes are controlled by the length of the wire $L$, the phase coherence length $L_\varphi$ and the thermal length $L_T=\sqrt{D/T}$.
$\xiloc/L=g\gg1$ is the dimensionless conductance and $\EThouless=D/L^2$ is the Thouless energy.
The behaviours for the mesoscopic fluctuations of the linear conductance $\delta g$ are also recalled for reference.
$^\mathrm{(*)}$~: our result agrees with the estimate $\Gnonint\sim\EThouless^{-1}$ of Ref.~\onlinecite{AltKhm85}.
$^\mathrm{(**)}$~: LK have shown in Ref.~\onlinecite{LarKhm86} that the differential conductance is controlled by the same power $\DifG(V)\sim\sqrt{eV/\EThouless}\,(L_\varphi/L)^{7/2}$ for high voltage $eV\gg1/\tau_\varphi$~; the behaviour of the table, $\DifG(V)-\DifG(0)\simeq2eV\mathcal{G}\sim (eV/\EThouless)\,(L_\varphi/L)^{7/2}$ describes the low voltage regime $eV\ll1/\tau_\varphi$ (cf. Appendix~\ref{subsec:KLDiffCond}).
}
\label{tab:MainRes}
\end{table}

\end{widetext}

\vspace{0.25cm}

\paragraph{Zero field.---}



In the coherent regime, we have found
\begin{equation}
  \label{eq:Result0}
  \smean{ \mathcal{G}^2 } 
  \sim \mean{\Gnonint^2}
  \sim\smean{ \left(\Gint_s\right)^2 } 
  \sim \EThouless^{-2}
 \quad\mbox{for }
 L\ll L_\varphi  ,\, L_T
\end{equation}
in agreement with Altshuler and Khmelnitskii~\cite{AltKhm85}. 
In the present article, we have also derived the precise dimensionless factors in all regimes (cf. Sections~\ref{sec:Coherent} and~\ref{sec:Conclusion}).
It will be also useful for the following to characterise the subdominant contribution to $\smean{ (\Gint_s)^2 }\sim\smean{\Gnonint^2}\smean{u_1}^2+\smean{\Gnonint^2}\smean{\delta u_1^2}$ related to the mesoscopic fluctuations of the electrostatic potential, which we denote
\begin{equation}
  \smean{ \left(\mathcal{G}^\mathrm{int,\,fluc}_s\right)^2 } 
  \sim \mean{\Gnonint^2}\mean{\delta u_1^2}
  \sim (g\EThouless)^{-2}
\end{equation}
(it will also be denoted $\smean{ \left(\mathcal{G}_s^\mathrm{int,\,fluc}\right)^2 }\equiv\smean{\big(\Gint_s\big)^2}_\mathrm{corr}$ in Subsection~\ref{subsec:Section7D}, for reasons that will be clear there).

In a weakly coherent wire we have found 
\begin{align}
  \label{eq:MainRes1}
  \smean{ \mathcal{G}^2 } 
  \sim \mean{\Gnonint^2}
  \sim\smean{ \left(\Gint_s\right)^2 } 
  &\sim (\EThouless)^{-2} (L_\varphi/L)^{7}
  \nonumber\\
 & \quad\mbox{for }
  L_\varphi\ll L  ,\, L_T
  \:.
\end{align}
the fluctuations of the characteristic potential are reduced by the same factor as the DoS or conductance fluctuations, therefore $\delta u_1\sim(1/g)(L_\varphi/L)^{3/2}$.
As a result
\begin{equation}
  \smean{ \left(\mathcal{G}^\mathrm{int,\,fluc}_s\right)^2 } 
  \sim \mean{\Gnonint^2}\mean{\delta u_1^2}
  \sim (g\EThouless)^{-2} (L_\varphi/L)^{7+3}
  \:.
\end{equation}

The regime where thermal fluctuations become important ($L_T\ll L_\varphi$) cannot be analysed in such simple terms~: the $L_T$ dependence is not simply related to the one of $\mean{\Gnonint^2}\propto L_T^6$ and the specific structure of the spatial correlations controlling $\smean{ \left(\Gint_s\right)^2 }$ plays a non-trivial role. We obtain
\begin{align}
  \label{eq:MainRes2}
  \smean{ \mathcal{G}^2 } 
  \sim\smean{ \left(\Gint_s\right)^2 } 
  \sim \EThouless^{-2}(L_T/L)^{2} & (L_\varphi/L)^{7}
  \gg \mean{\Gnonint^2}
  \\\nonumber
 &\mbox{for }
 L_T\ll L_\varphi  \ll L
\end{align}
which thus decays with temperature as $T^{-1}$, that is \textit{slower} than $\mean{\Gnonint^2}\propto T^{-3}$ (note that $L_\varphi$ may also be responsible for additional temperature dependence, cf. Section~\ref{sec:Conclusion}).
The two results \eqref{eq:MainRes1} and \eqref{eq:MainRes2} do not match at first sight when $L_T\sim L_\varphi$, however we will discuss how the crossover at $L_T\sim L_\varphi$ is realised. 
The fluctuation part receives additional reduction factors coming from the DoS fluctuations~:
\begin{align}
  \smean{ \left(\mathcal{G}^\mathrm{int,\,fluc}_s\right)^2 } 
  \sim (g\EThouless)^{-2}
  (L_T/L)^{2+2}(L_\varphi/L)^{4+2+1}
  \:.
\end{align}
The exponents are splitted in order to identify the contributions of the two correlators in $\smean{ \left(\mathcal{G}^\mathrm{int,\,fluc}_s\right)^2 }\sim\smean{\Gnonint^2}\mean{\delta u_1^2}$~; the last contribution to the second exponent comes from the spatial integration of the correlators.

\vspace{0.25cm}

\paragraph{Antisymmetric part at high field.---}

A remarkable property of the non-linear conductance, when interaction effects are taken into account, is the existence of an antisymmetric part $\mathcal{G}_{a}$ under magnetic field reversal.
This antisymmetric contribution arises from the absence of symmetry of the fluctuating part of the characteristic potential $\delta u_1$.
Thus $\mathcal{G}_{a}\equiv\Gint_a$ has the same origin as the subdominant contribution to the symmetric part $\mathcal{G}^\mathrm{int,\,fluc}_s$.
In the coherent regime we obtain 
\begin{equation}
  \label{eq:MainResAsym0}
  \mean{\mathcal{G}_a^2} 
  \sim \smean{ \left(\mathcal{G}^\mathrm{int,\,fluc}_s\right)^2 }\sim
  \left(g\EThouless\right)^{-2}
\quad\mbox{for }
 L\ll  \ L_\varphi\, ,  L_T
  \:.
\end{equation}
In the weakly coherent wire, $\mathcal{G}_a$ is smaller than $\mathcal{G}^\mathrm{int,\,fluc}_s$ as the antisymmetric part of the characteristic potential correlations is \textit{short range} (i.e. decays exponentially on the scale $L_\varphi$) whereas its symmetric part is \textit{long range}. 
This produces a reduction factor
\begin{align}
  \label{eq:MainRes3}
  \mean{\mathcal{G}_a^2} 
  &\sim \smean{ \left(\mathcal{G}^\mathrm{int,\,fluc}_s\right)^2 } \frac{L_\varphi}{L}
 \hspace{1cm}\mbox{for }
 L_\varphi\ll L_T\, , \ L  
 \nonumber\\
 &\sim
  \left(g\EThouless\right)^{-2}(L_\varphi/L)^{11}
\:.
\end{align}
In the ``high temperature'' regime, the analysis is more subtle and we have obtained 
\begin{align}
  \label{eq:MainRes4}
  \mean{\mathcal{G}_a^2} 
  &\sim \smean{ \left(\mathcal{G}^\mathrm{int,\,fluc}_s\right)^2 } 
 \hspace{2cm}\mbox{for }
 L_T\ll L_\varphi  \ll L
 \nonumber\\
 &\sim
  \left(g\EThouless\right)^{-2}(L_T/L)^{4}(L_\varphi/L)^{7}
\:.
\end{align}
All these behaviours are summarised in Table~\ref{tab:MainRes}.
They will be derived below by a careful analysis of conductance's functional derivative correlations and characteristic potential's correlations.
All the precise numerical factors involved in the correlators will be determined neatly.

\vspace{0.25cm}

\paragraph{Magnetic field dependence.---}

The antisymmetric contribution $\mathcal{G}_a$ obviously vanishes as $\mathcal{B}\to0$.
We will analyse the expressions for the correlators 
$\smean{\mathcal{G}_{s,a}(\mathcal{B})\mathcal{G}_{s,a}(\mathcal{B}')}$ 
in order to describe the full crossover between the high field regime discussed so far and the low field regime.
The linear behaviour
\begin{equation}
  \label{eq:MainRes5}
  \mathcal{G}_a(\mathcal{B}) \sim \mathcal{G}_a(\infty)\,\frac{\mathcal{B}}{\Bcorr}
  \hspace{0.5cm}\mbox{for }
  \mathcal{B}\ll\Bcorr
\end{equation}
is generically expected, where the crossover field which separates the two regimes is~\cite{AltAro81} $\Bcorr\sim\phi_0/(L_\varphi\sw)$ for $L_\varphi\ll L$, where $\sw$ is the width of the wire, and  $\Bcorr\sim\phi_0/(L\sw)$ (rather denoted $\mathcal{B}_{c0}$ later) for $L_\varphi\gg L$.
The linear behaviour \eqref{eq:MainRes5} is obtained in the coherent regime $L\ll L_\varphi$ and in the regime dominated by thermal fluctuations $L_T\ll L_\varphi\ll L$. 
Quite surprisingly, in the regime $L_\varphi\ll L_T,\,L$ where thermal fluctuations are negligible, the linear term unexpectedly vanishes and we obtain a quadratic behaviour 
$\mathcal{G}_a(\mathcal{B})\sim\mathrm{sign}(\mathcal{B})\,\mathcal{B}^2$.

\subsection{Comparison with known results for quantum dots}

\begin{figure}[!ht]
\centering
\includegraphics[scale=1]{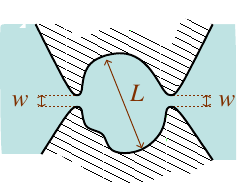}
\caption{(color online) \it A quantum dot (QD) of size $L$ closed by two constrictions of width $w$.}
\label{fig:qd}
\end{figure}

We compare our results with the ones previously obtained for quantum dots.
As quantum dots (Fig.\ref{fig:qd}) have more complex geometrical properties than a simple wire (Fig.\ref{fig:PotentialInWire}), the set of characteristic parameters is richer (for a review, see Ref.~\onlinecite{Bee97}).

The Thouless energy, or the Thouless time $\tau_D=1/\EThouless$, and the dimensionless conductance $g=2\pi\EThouless/\Delta$, where $\Delta$ is the mean level spacing, do not carry information about the nature of the contacts and thus characterise a closed quantum dot (QD).
In open QDs, such as the one of Fig.~\ref{fig:qd}, two other important parameters are the two-terminal (dimensionless) conductance, denoted here $g_\mathrm{Drude}$, and the dwell time $\tau_\mathrm{dwell}$. This leads to introduce another Thouless energy~\cite{Bee97} $\EThouless^\mathrm{open}=1/\tau_\mathrm{dwell}$. 
In ballistic QDs, one has $1/\tau_\mathrm{dwell}\sim\Nc\Delta\sim\EThouless\,(w/L)$, where  $\Nc$ is the number of channels at the contact (see Refs.~\onlinecite{Bee97,Tex16} for example), i.e. the escape rate is the inverse of the Thouless time multiplied by the probability to find the contact of width $w$ once on the boundary of the QD~; we expect that the expression $1/\tau_\mathrm{dwell}\sim\EThouless\,(w/L)$ also applies to diffusive QDs. In ballistic QDs the resistance is dominated by the resistances of the constrictions, hence $g_\mathrm{Drude}=\Nc/2$ (for symmetric contacts). 
In the diffusive regime, the 2D dimensionless conductance $g=\pi\DoS_0D=k_F\ell_e/2$ (the conductivity in unit $2_se^2/h$) characterises the transport through a conductor of length equal to the width of the contacts~; for narrow constrictions the resistance between the two narrow contacts receives an additional factor $\ln(L/w)$ (see Refs.~\onlinecite{TexDelMon09,Tex10hdr} where the transport through a planar device with narrow contacts was studied).
These scales are summarised in Tab.~\ref{tab:QD}.

\begin{table}[!ht]
\begin{center}
\begin{tabular}{|rc|cc|c|}
\hline
 \multicolumn{2}{|c|}{ } & \multicolumn{2}{|c|}{QDs} & wires \\ 
  & & ballistic & diffusive & \\
\hline 
\rotatebox[origin=c]{90}{closed}     
           & $\EThouless$ & $v_F/L$ & $D/L^2$     & $D/L^2$ \\
           & $g\sim\frac{\EThouless}{\Delta}$          
                          & $k_FL$  & $k_F\ell_e$ & $\frac{\xiloc}{L}\sim k_F\ell_e\,\frac{w}{L}$ \\
 \hline
\rotatebox[origin=c]{90}{open}     
           & $\EThouless^\mathrm{open}=\frac{1}{\tau_\mathrm{dwell}}$
                          & $\Nc\Delta\sim\EThouless\,\frac{w}{L}$
                                    & $\EThouless\,\frac{w}{L}$ & $\EThouless$ \\
           & $g_\mathrm{Drude}$ 
                          & $\Nc\sim g\,\frac{w}{L}$
                                    & $\frac{g}{\ln(L/w)}$ & $g$ \\
\hline                                    
\end{tabular}
\end{center}
\caption{\it 
Comparison between characteristic parameters for ballistic QDs,  diffusive QDs and diffusive wires, in two dimension. The QD has a size $L$ and is closed by constrictions of width $w$ (i.e. with $\Nc=k_Fw/\pi$ open channels).
  $v_F$ and $k_F$ are the Fermi velocity and the Fermi wavevector, $\ell_e$ the elastic mean free path and $D=v_F\ell_e/2$ the diffusion constant.
  }
\label{tab:QD}
\end{table}

Non-linear transport in ballistic QDs was considered first by S\'anchez and B\"uttiker (SB)~\cite{SanBut04} and later by Polianski and B\"uttiker (PB)~\cite{PolBut06,PolBut07,PolBut07a}, within a random matrix approach.
If the QD has contacts with $N_1$ and $N_2$ channels.  
At $T=0$, using $1/\tau_\mathrm{dwell}=N\Delta/(2\pi)$, we can summarise the results for the symmetric and antisymmetric parts of the non-linear conductance as~\cite{PolBut07a}
\begin{align}
  \label{eq:Polianski2007s}
  \smean{ \mathcal{G}_s ^2} &= \: \tau_\mathrm{dwell}^2\frac{4}{\beta}
  \frac{N_1^2N_2^2}{N^4}
  \nonumber\\
  &\hspace{0.5cm}
  \times\left[ 
    \left(\frac{1}{2} - \gamma_\mathrm{int}\frac{N_1}{N}\right)^2
    + \frac{\gamma_\mathrm{int}^2}{\beta N^2} \frac{N_1N_2}{N^2}
  \right]
  \\
  \label{eq:Polianski2007a}
  \smean{ \mathcal{G}_a ^2}
  & = \frac{\tau_\mathrm{dwell}^2}{N^2}\frac{4}{\beta}\left(1-\frac{1}{\beta}\right)
  \gamma_\mathrm{int}^2\frac{N_1^3N_2^3}{N^6}
\end{align}
where $N=N_1+N_2$. 
Note that the second term in $\smean{ \mathcal{G}_s ^2}$ (subdominant for $N\gg1$) corresponds to $\smean{\big(\mathcal{G}^\mathrm{int,\,fluc}_s\big)^2}$ introduced above.
We have simplified the discussion of the magnetic field dependence by introducing the Dyson index~: $\beta=1$ describes the zero field case and $\beta=2$ the strong field regime (the full dependence in $\mathcal{B}$-field can be found in Ref.~\onlinecite{PolBut07a}).
The parameter $\gamma_\mathrm{int}=C_\mu/C$ is the ratio of the mesoscopic and the geometrical capacitance. It controls the efficiency of screening ($\gamma_\mathrm{int}=1$ for perfect screening and $\gamma_\mathrm{int}\ll1$ for weak screening).  
For $N_1=N_2\equiv\Nc$, we can rewrite SB's result as $\mathcal{G}_a\sim\big(g_\mathrm{Drude}\EThouless^\mathrm{open}\big)^{-1}$ in order to make connection with \eqref{eq:MainResAsym0} (see also Ref.~\onlinecite{AngZakDebGueBouCavGenPol07}). 
Note however that screening can suppress the dominant contribution $\mathcal{G}\sim \mathcal{G}_s\sim\big(\EThouless^\mathrm{open}\big)^{-1}$, if $\gamma_\mathrm{int}=N/(2N_1)$. A similar effect in diffusive wires would require some control on the connections between the wire and the reservoirs, what will not be considered here.

The case of QDs in the diffusive regime was considered by Spivak and Zyuzin (SZ)~\cite{SpiZyu04}, within a diagrammatic approach similar to our approach, although less detailed. 
They considered the coherent limit for which they got 
$\smean{\mathcal{G}_a(\mathcal{B})^2}\sim\big(\DoS_0\,\mathrm{Surf})^{-2}\EThouless^{-4}\,(\mathcal{B}\,\mathrm{Surf}/\phi_0)^2$. 
We can rewrite SZ's result
$\smean{\mathcal{G}_a(\mathcal{B})^2}\sim(g\EThouless)^{-2}(\mathcal{B}\,\mathrm{Surf}/\phi_0)^2$ (see also Ref.~\onlinecite{AngZakDebGueBouCavGenPol07}),
which is consistent with our result \eqref{eq:MainRes5} with~$\Bcorr\to\mathcal{B}_{c0}\sim\phi_0/(L\sw)$.
However, the SZ's estimation~\cite{SpiZyu04,DeySpiZyu06} has not taken into account the geometry of the QD (the presence of the narrow contacts)~: 
the analogy with SB's result suggests that $\mathcal{G}_a$ should rather involve  $\EThouless^\mathrm{open}=1/\tau_\mathrm{dwell}$. 
The precise dimensionless factor was not obtained either.

The temperature dependence was analysed by PB~\cite{PolBut06,PolBut07a} who obtained a suppression of $\smean{\mathcal{G}_{a}^2}$ by a factor $(\EThouless^\mathrm{open}/T)^{2}$ when $T\gg\EThouless^\mathrm{open}$ (the $\tau_\varphi$ dependence was only investigated by simulations in Ref.~\onlinecite{PolBut06}).
Rewriting our Eq.~\eqref{eq:MainRes4} as 
$\mean{\mathcal{G}_a^2} 
\sim(\tau_\varphi/\tau_D)^{7/2}(gT)^{-2}$,
and extrapolating to the coherent limit ($\tau_\varphi\sim\tau_D$), we conclude that our result agrees with the one of~PB in this limit.

We can already stress some important differences between the results obtained in 0D (quantum dots) and the results sketched above for 1D devices (wires). 
In the first case the symmetric and antisymmetric contributions arising from interaction were shown to be equal (in the limit of perfect screening)~\cite{PolBut06} whereas in the wire they are always different~:
in the coherent wire they are controlled by quite different numerical factors and, moreover, in the incoherent limit the $L_\varphi$ dependences are different.
With the existence of two regimes (\ref{eq:MainRes3},\ref{eq:MainRes4}), which arises from the importance of the spatial structure of the correlators, this makes the extension of the results obtained in 0D to higher dimensions quite non-trivial.

\subsection{Strategy of the analysis}

Having sketched the main ideas and results in Subsection~\ref{subsubsec:OurNewResults}, we can now describe more precisely the strategy of the analysis~:
\begin{enumerate}
\item
in Section~\ref{sec:ScatteringFormalism}, we will introduce the general expressions for the two contributions $\mathcal{G}_\nonint$ and $\Gint$.
\item
In Section~\ref{sec:Symmetries}, we will analyse the structure of the correlator $\mean{\mathcal{G}_{s,a}^2}$ for weakly disordered metals and will show the connection with two fundamental correlators $\chi_g$ and $\chi_\nu$ (these two correlators are in correspondence with $\smean{\mathcal{G}_\nonint^2}$ and $\smean{\delta u_1^2}$, respectively, introduced in the qualitative discussion, Subsection~\ref{subsubsec:OurNewResults}).
\item
The two correlators $\chi_g$ and $\chi_\nu$ will be computed in detail in Sections~\ref{sec:CondCorr} and~\ref{sec:InjCorr}
\item
They are combined in Section~\ref{sec:NLC} in order to obtain the final result for $\mean{\mathcal{G}_{s,a}^2}$.
\end{enumerate}
The same logic will be repeated in Section~\ref{sec:Coherent} for the coherent regime, which moreover requires a careful treatment of boundary conditions.

\section{Scattering formalism}
\label{sec:ScatteringFormalism}

B\"uttiker has developed a scattering formalism for non-linear transport in coherent conductors \cite{But93,ChrBut96a,ButChr97}, including a Hartree treatment of electronic interaction within a Thomas-Fermi approximation.
Our analysis will be based on this approach, which we briefly recall in this section (the formalism for ergodic systems has been reviewed in \cite{PolBut07a,Tex16}).
The relation with the non-equilibrium Green's function formalism has been discussed in Ref.~\onlinecite{HerLew13}.
On the top of the B\"uttiker scattering formalism, we will use diagrammatic techniques necessary in order to study disordered metallic devices (next sections), similarly as Spivak and Zyuzin~\cite{SpiZyu04,DeySpiZyu06}.

In this section we consider the general case of multiterminal conductors.
We denote by greek letters $\alpha,\,\beta,$ etc, the contacts through which currents are injected and collected.
In general, it is possible to present the relation between external applied voltages $V_\beta$'s and currents $I_\alpha$'s as an expansion
\begin{equation}
  \label{eq:ExpansionCurrent}
  I_\alpha = 
  \frac{2_se^2}{h}\sum_\beta g_{\alpha\beta}\,V_\beta
  + \frac{2_se^3}{h}\sum_{\beta,\,\gamma} g_{\alpha\beta\gamma}\,V_\beta\,V_\gamma
  +\cdots
  \:,
\end{equation}
where $g_{\alpha\beta}$ are the dimensionless linear conductances and $g_{\alpha\beta\gamma}$ the rescaled non-linear conductances (with dimension of the inverse of energy). 
If necessary, we impose the symmetry with respect to voltage indices
\begin{equation}
  \label{eq:SymmetryVoltageIndices}
  g_{\alpha\beta\gamma}=g_{\alpha\gamma\beta}
  \:,
\end{equation}
as any non-symmetric contribution would not contribute to the expansion~\eqref{eq:ExpansionCurrent}.
Our aim in the section is to recall some general formulae for the linear and non-linear conductances obtained within the scattering formalism.

\subsection{Non-interacting electrons (Landauer-B\"uttiker formula)}

For non-interacting electrons, the currents can be obtained from the Landauer-B\"uttiker formula for a multiterminal coherent conductor~\cite{But92}
\begin{equation}
  \label{eq:LandauerButtiker}
  I_\alpha = \frac{2_se}{h}
  \int\D\varepsilon \, \sum_\beta
  g_{\alpha\beta}(\varepsilon) \, f(\varepsilon-eV_\beta)
\end{equation}
where $f(\varepsilon)$ is the Fermi-Dirac distribution and $V_\beta$ the voltage at contact $\beta$.
The zero temperature dimensionless conductance at Fermi energy $\varepsilon$ is related to the scattering matrix $\Sm$ encoding the scattering properties of an electronic wave at energy~$\varepsilon$~: 
\begin{equation}
  \label{eq:LinearConductance}
  g_{\alpha\beta}(\varepsilon)
  = N_\alpha\,\delta_{\alpha\beta}
  - \tr{ \Sm_{\alpha\beta}^\dagger(\varepsilon)\Sm_{\alpha\beta}(\varepsilon) }
  \:,
\end{equation}
where the trace runs over conducting channels, $N_\alpha$ being the number of conducting channels in contact $\alpha$.
Expanding the Landauer-B\"uttiker formula \eqref{eq:LandauerButtiker}, one gets the non-interaction non-linear conductances
\begin{equation}
  \label{eq:NoninteractingNLC}
   g_{\alpha\beta\gamma}^\nonint
   = \frac12\delta_{\beta\gamma}
  \int\D\varepsilon\,\left(-\derivp{f}{\varepsilon}\right)\,
   g_{\alpha\beta}'(\varepsilon)
   \:.
\end{equation}

\subsection{Characteristic potentials}
\label{subsec:CharacPot}

B\"uttiker proposed a self-consistent theory describing the effect of screening (Coulomb interaction)~\cite{But93}.
This theory accounts for the fact that the modification of the external potentials redefines the electron density inside the conductor, and hence, due to Coulomb interaction, the electrostatic potential. 
Electronic interactions thus make the scattering matrix and the conductances $g_{\alpha\beta}(\varepsilon)$ voltage-dependent, what leads to another contribution to the non-linear conductance.
This can be formalised by introducing the characteristic potential $u_\alpha(\vec{r})$, which measures the response of the electrostatic potential $U(\vec{r})$ inside the conductor to a change of the external voltage at contact $\alpha$ to linear order~:
\begin{equation}
  \label{eq:DefCharacPot}
  \delta U(\vec{r}) = U(\vec{r})-U_\mathrm{eq}(\vec{r})
   \simeq \sum_\alpha eV_\alpha \, u_\alpha(\vec{r}) 
  \:,
\end{equation}
where $U_\mathrm{eq}(\vec{r})$ is the potential at equilibrium.
They obey the sum rule~\cite{But93}
\begin{equation}
  \label{eq:ConstraintCharacPot}
  \sum_\alpha u_\alpha(\vec{r}) =1 
\end{equation}
which ensures that the potential is simply shifted by a constant when all voltages are equal.

The characteristic potentials are determined as follows~:
in the out-of-equilibrium situation, we can write the charge in excess introduced from the leads in terms of the injectivities~:
\begin{equation}
  \delta n_\mathrm{ext}(\vec{r}) 
  \simeq   \sum_\alpha eV_\alpha
  \int\D\varepsilon\,\left(-\derivp{f}{\varepsilon}\right)\,    
  \DoS_\alpha(\vec{r};\varepsilon)
  \:.
\end{equation}
The injectivity $\DoS_\alpha(\vec{r};\varepsilon)$ measures the contribution to the local density of states (DoS) 
$\DoS(\vec{r};\varepsilon)=\bra{\vec{r}}\delta(\varepsilon-H)\ket{\vec{r}}$ of the scattering states describing electrons incoming from the contact $\alpha$.
Obviously, the injectivities satisfy the sum rule
\begin{equation}
  \label{eq:SumRuleInjectivities}
  \sum_\alpha \DoS_\alpha(\vec{r};\varepsilon) = \DoS(\vec{r};\varepsilon)
  \:.
\end{equation}
We explain below how the injectivities are determined, and in particular their representations in terms of Green's functions.

The electrostatic potential is related to the charge in excess through the (static) screened interaction 
\begin{equation}
  \delta U(\vec{r}) =  \int\D\vec{r}\,'\,
  U_\mathrm{RPA}(\vec{r},\vec{r}\,')\,
  \delta n_\mathrm{ext}(\vec{r}\,')
  \:, 
\end{equation}
which is obtained by solving the Coulomb equation
\begin{align}
  \label{eq:CoulombEquation}
  -&\frac{1}{4\pi e^2}\Delta \, U_\mathrm{RPA}(\vec{r},\vec{r}\,')
  \\\nonumber
  &
  +\int\D\vec{r}\,'\,
  \Pi(\vec{r},\vec{r}\,')\, U_\mathrm{RPA}(\vec{r},\vec{r}\,')
  =\delta(\vec{r}-\vec{r}\,')
  \:.
\end{align}
The right hand side corresponds to the external charge and the integral term to the induced charge, where $\Pi(\vec{r},\vec{r}\,')$ is the static compressibility --Lindhard function-- characterising the linear response (zero frequency density-density correlation)~\cite{BruFle04}~:
$\delta n_\mathrm{ind}(\vec{r})=-\int\D\vec{r}\,'\,\Pi(\vec{r},\vec{r}\,')\,\delta U(\vec{r}\,')$ (the presence of the potential in the right hand side, and not the potential related to $\delta n_\mathrm{ext}(\vec{r})$, makes the approach self-consistent).
We deduce the expression of the characteristic potential
\begin{align}
  \label{eq:CharactPot}
  u_\alpha(\vec{r}) = 
  \int\D\varepsilon\,\left(-\derivp{f}{\varepsilon}\right)
  \int\D\vec{r}\,'\, U_\mathrm{RPA}(\vec{r},\vec{r}\,')\,
  \DoS_\alpha(\vec{r}\,';\varepsilon)
\end{align}
(Refs.~\onlinecite{But93,ChrBut96a} gave an integro-differential equation of the form \eqref{eq:CoulombEquation} directly for $u_\alpha$, and hence with a source term given by the injectivity).
In a good metal with a high DoS, the response can be considered as local~: 
$\Pi(\vec{r},\vec{r}\,')\simeq\DoS_0\,\delta(\vec{r}-\vec{r}\,')$.
Since the Thomas-Fermi screening length $\ell_\mathrm{TF}=1/\sqrt{4\pi\DoS_0e^2}$ is usually very small~\cite{footnote3}, 
the first term $\Delta \, U_\mathrm{RPA}$ in Eq.~\eqref{eq:CoulombEquation} can be neglected and we deduce the local form 
$U_\mathrm{RPA}(\vec{r},\vec{r}\,')\simeq(1/\DoS_0)\,\delta(\vec{r}-\vec{r}\,')$, which describes perfect screening (see also Ref.~\onlinecite{TreTexYevDelLer11}). As a consequence~:
\begin{equation}
  \label{eq:CharacPotPerfectScreeningLimit}
  u_\alpha(\vec{r}) 
  =
  \int\D\varepsilon\,\left(-\derivp{f}{\varepsilon}\right)\,
  \frac{\DoS_\alpha(\vec{r};\varepsilon)}{\DoS_0(\varepsilon)}
  \:.
\end{equation}
We can check that the characteristic potentials obey the sum rule~\eqref{eq:ConstraintCharacPot}.

In the following we will restrict ourselves to the case of perfect screening.

\subsection{Non-linear conductances }

Expanding the Landauer-B\"uttiker formula \eqref{eq:LandauerButtiker} in powers of the external potentials $V_\alpha$'s, one can now account for the dependence of the conductances on the external potentials~:
\begin{align}
  I_\alpha &= \frac{2_se}{h}
  \int\D\varepsilon \, \sum_\beta
  \\ \nonumber
  &\left[
     g_{\alpha\beta}(\varepsilon)
     + 
     \int\D\vec{r}\,\derivf{g_{\alpha\beta}(\varepsilon)}{U(\vec{r})} \, 
     \sum_\gamma u_\gamma(\vec{r})\,eV_\gamma+\cdots
  \right]
  \\ \nonumber
  \times
  &\left[
     \left(-\derivp{f}{\varepsilon}\right) eV_\beta
     + \frac{1}{2}\left(\derivp{^2f}{\varepsilon^2}\right) (eV_\beta)^2
     +\cdots
  \right]
  \:.
\end{align}
Additionally to the non-interaction contribution \eqref{eq:NoninteractingNLC}, we obtain a second contribution from electronic interactions
\begin{equation}
  \label{eq:NLC}
  g_{\alpha\beta\gamma}
  =g_{\alpha\beta\gamma}^\nonint + g_{\alpha\beta\gamma}^\mathrm{int}
\end{equation}
with
\begin{align}
  \label{eq:NLCinteractingpart}
  g_{\alpha\beta\gamma}^\mathrm{int}
  &= \frac12
  \int\D\varepsilon\,\left(-\derivp{f}{\varepsilon}\right)\,
  \\ \nonumber
  &\times  
  \int\D\vec{r}\,
  \left[
    \derivf{g_{\alpha\beta}(\varepsilon)}{U(\vec{r})} \, u_\gamma(\vec{r})
    + \derivf{g_{\alpha\gamma}(\varepsilon)}{U(\vec{r})} \, u_\beta(\vec{r})
  \right] 
  \:
\end{align}
The complete expression may be written in a symmetric form by using 
$
g_{\alpha\beta}'(\varepsilon)
+
\int\D\vec{r}\,\delta g_{\alpha\beta}(\varepsilon)/\delta U(\vec{r})=0
$~: 
\begin{align}
  \label{eq:ButtikerChristen}
  &g_{\alpha\beta\gamma}
  = \frac12
  \int\D\varepsilon\,\left(-\derivp{f}{\varepsilon}\right)\,
  \\ \nonumber
  &\times\int\D\vec{r}\,\left[
    \derivf{g_{\alpha\beta}(\varepsilon)}{U(\vec{r})} \, u_\gamma(\vec{r})
    + \derivf{g_{\alpha\gamma}(\varepsilon)}{U(\vec{r})} \, u_\beta(\vec{r})
    -    \derivf{g_{\alpha\beta}(\varepsilon)}{U(\vec{r})} \, \delta_{\beta\gamma}
  \right]
\end{align}
which is the expression given by Christen and B\"uttiker~\cite{ChrBut96a}, further symmetrised with respect to the voltage indices.

\subsection{Few remarks and diagrammatic representation}

\begin{figure}[!ht]
\centering
\includegraphics[scale=0.75]{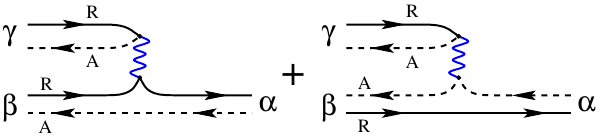}
\caption{\textit{Diagram for the interaction part $g_{\alpha\beta\gamma}^\mathrm{int}$ of the non-linear conductance, i.e. first term of Eq.~\eqref{eq:ButtikerChristen}.
  The upper part represents the injectivity $\DoS_\gamma(\vec{r}\,';\varepsilon)$.
  The wavy line is the screened interaction $U_\mathrm{RPA}(\vec{r},\vec{r}\,')$ (local in good metals). Injectivity and interaction represent the characteristic potential $u_\gamma(\vec{r})$.
  The lower part corresponds to the functional derivative of the conductance $\delta g_{\alpha\beta}/\delta U(\vec{r})$, which gives rise to the two diagrams as the functional derivative acts either on the retarded or the advanced Green's function line.}}
  \label{fig:DiagramNonLinear}
\end{figure}

Let us close the section with a few remarks.
\begin{itemize}
\item
  The most important remark concerns the symmetry under magnetic field reversal~:
  it is clear from \eqref{eq:NoninteractingNLC} that both the linear conductance $g_{\alpha\beta}$ and $g_{\alpha\beta\gamma}^\nonint$ present the same symmetry. 
  On the other hand, the contribution $g_{\alpha\beta\gamma}^\mathrm{int}$ depends on the injectances $u_\beta$ and $u_\gamma$ which do not have any specific symmetry under magnetic field reversal.
  
\item 
  Injectivities are important ingredients of the scattering approach.
  A convenient representation is given by the relation with the $\Sm$-matrix~\cite{GasChrBut96} (see \cite{Tex10hdr,Tex16} for recent reviews)
  \begin{equation}
    \DoS_\alpha(\vec{r};\varepsilon) = - \frac{1}{2\I\pi}
    \left(\Sm^\dagger \derivf{\Sm}{U(\vec{r})} \right)_{\alpha\alpha}
    \:.
  \end{equation}
  Therefore, all quantities involved in the non-linear conductance \eqref{eq:ButtikerChristen} can be expressed in terms of the scattering matrix.

\item  
  For a 1D contact, one may write simple representations for the injectivity.
  Being the contribution to the DoS of the electrons incoming from contact $\alpha$, it can be related to the stationary scattering state $\psi_\varepsilon^{(\alpha)}(\vec{r})$ describing electrons incoming from contact $\alpha$~:
$\DoS_\alpha(\vec{r};\varepsilon)=2_s|\psi_\varepsilon^{(\alpha)}(\vec{r})|^2$
This makes clear the relation with the Green's function~\cite{Tex10hdr}  
$\DoS_\alpha(\vec{r};\varepsilon)=\big[2_s/(2\pi)\big]\,v_\alpha\,\big|G^\mathrm{R}(\vec{r},\alpha;\varepsilon)\big|^2$, where the second argument of the Green's function is the position of the contact and $v_\alpha$ the group velocity in the contact wire.

\item 
  Using the Fisher and Lee relation~\cite{FisLee81}, 
  $
    \Sm_{\alpha\beta}(\varepsilon)=-\delta_{\alpha\beta}
  + \I\sqrt{v_\alpha v_\beta}\,
  G^\mathrm{R}(\alpha,\beta;\varepsilon)
  $
  (written here for 1D contact wires)  
  we can deduce the expression of the functional derivative of the  $\Sm$-matrix~:
  $\delta\Sm_{\alpha\beta}/\delta U(\vec{r})=\I\sqrt{v_\alpha v_\beta}\,
  G^\mathrm{R}(\alpha,\vec{r};\varepsilon)G^\mathrm{R}(\vec{r},\beta;\varepsilon)$.
  This will be useful later in order to express the functional derivative of the conductance \eqref{eq:LinearConductance} involved in the non-linear conductance in terms of Green's functions.
\end{itemize}
When the contact wires are characterised by many conducting channels, these relations can be easily generalised. 
These two remarks lead to the diagrammatic representation of Fig.~\ref{fig:DiagramNonLinear}, where the first term of \eqref{eq:NLCinteractingpart} is represented.
Continuous lines represent retarded Green's function and dashed lines advanced Green's function.

\begin{itemize}
\item
Finally, we come back to the question of screening evoked above.
In the article we will assume \textit{perfect screening}.
In Refs.~\onlinecite{PolBut06,PolBut07a} (see also \cite{Tex16}), it has been shown that, for quantum dots in the ergodic regime, the crossover between perfect and weak screening can be accounted for through an additional dimensionless factor 
$\gamma_\mathrm{int}=\big[1+1/(\DoS_0E_C)\big]^{-1}$ in the characteristic potential, where $E_C=e^2/C$ is the charging energy of the quantum dot, $C$ being its capacitance.
Note that when screening is not efficient enough, one must take into account the presence of the nearby metallic gates and in particular the response of the potential \eqref{eq:DefCharacPot} to a change of the gate voltage via an additional characteristic potential~\cite{But93,PolBut07a,Tex16} $u_\mathrm{gate}$.
\\
For the weakly disordered rings of perimeter $L=4.8\:\mu$m studied in Ref.~\onlinecite{AngZakDebGueBouCavGenPol07,Ang07}, the interaction parameter was found to be $\gamma_\mathrm{int}=0.90\pm0.05$.
This justifies to consider the perfect screening limit ($\gamma_\mathrm{int}=1$).~\cite{footnote4}
\end{itemize}


\section{Symmetries}
\label{sec:Symmetries}

The symmetry with respect to magnetic field reversal is of special interest here, as it is related to the renewal of the interest for non-linear transport in normal metals~\cite{LofMarShoTayOmlSamLin04,SpiZyu04,SanBut04,PolBut06,LofMarHumShoTayOmlNewLinLin06,
LetSanGotIhnEnsDriGos06,MarTayFaiShoLin06,ZumMarHanGos06,AngZakDebGueBouCavGenPol07}.
Before discussing this matter, we come back to the important question of current conservation and gauge invariance, which introduce two types of constraints on the non-linear conductance.

\subsection{Current conservation and gauge invariance}

Current conservation takes the form $\sum_\alpha I_\alpha=0$.
Using the expansion \eqref{eq:ExpansionCurrent}, we deduce the two constraints 
\begin{align}
  \label{eq:CurrentConservationLinear}
  \sum_\alpha g_{\alpha\beta} &=  0
  \\
  \label{eq:CurrentConservationNonLinear}
  \sum_\alpha g_{\alpha\beta\gamma} &=  0
  \:.
\end{align}

We feel useful to rediscuss gauge invariance, as the statement of Refs.~\onlinecite{ChrBut96a,BlaBut00} that Coulomb interaction would be necessary in order to satisfy gauge invariance seems at odds to us.
 From the second reference~:
``\textit{In general, for non-linear and non-stationary problems, current conservation and gauge invariance are not automatically fulfilled. Indeed, in ac-transport a direct calculation of average particle currents does not yield a current conserving theory. Only the introduction of displacement currents, determined by  the long-range Coulomb interaction, leads to a theory which satisfies these basic requirements}.''
This statement is not satisfactory as the Landauer formula is based on calculation of currents from the Schr\"odinger equation which, when properly written in terms of electromagnetic potentials, is well-known to be gauge invariant, as can be found in quantum mechanics textbooks (see for instance the books~\cite{Sak94,Tex15book}).
We will reconciliate below the two point of view and derive an equation expressing gauge invariance for the theory of free electrons.

\subsubsection{Gauge invariance for free electrons}

Let us first recall the usual formulation of gauge invariance in quantum mechanics.
Gauge invariance of the Schr\"odinger equation 
$\I\hbar\partial_t\psi(\vec{r},t)
=\big[-(\hbar^2/2m)\Delta+\PotEnerg(\vec{r},t)\big]\psi(\vec{r},t)$
is the invariance under transformation of the wave function
$\psi(\vec{r},t)\to\tilde{\psi}(\vec{r},t)=\psi(\vec{r},t)\,\EXP{-\I\chi(t)/\hbar}$,
provided that the potential is changed as 
$\PotEnerg(\vec{r},t)\to
\tilde{\PotEnerg}(\vec{r},t)=\PotEnerg(\vec{r},t)+\partial_t\chi(t)$ (we disregard the possible spatial dependence of the phase in order to simplify the discussion)~: 
i.e. the transformed wave function obeys
$\I\hbar\partial_t\tilde\psi(\vec{r},t)
=\big[-(\hbar^2/2m)\Delta+\tilde\PotEnerg(\vec{r},t)\big]\tilde\psi(\vec{r},t)$.
In a stationary problem, only the transformation with phase $\chi(t)=U_0t$ respects the invariance under time translation~; in this case the gauge transformation corresponds to the simple addition of a constant potential $\PotEnerg(\vec{r})\to\tilde{\PotEnerg}(\vec{r})=\PotEnerg(\vec{r})+U_0$. 
Accordingly the gauge transformation of the stationary state takes the form of a translation in energy $\psi_\varepsilon(\vec{r})\to\tilde\psi_\varepsilon(\vec{r})
=\psi_{\varepsilon-U_0}(\vec{r})$. 
In a scattering situation, the $\Sm$-matrix, which encodes the information on the scattering stationary states, is changed in the same way
$\Sm_{\alpha\beta}(\varepsilon)\to\Sm_{\alpha\beta}(\varepsilon-U_0)$.

In order to see clearly the implication of this discussion on the Landauer formula, it is convenient to express that the conductance is also a functional of the potential. We rewrite Eq.~\eqref{eq:LandauerButtiker}
\begin{align}
  \label{eq:LandauerButtiker2}
  I_\alpha = \frac{2_se}{h}
  \int\D\varepsilon \, \sum_\beta
  g_{\alpha\beta}(\varepsilon;U_\mathrm{eq}(\vec{r})) \, 
  f(\varepsilon-eV_\beta)
  \:.
\end{align}
This expression is invariant under the transformations
\begin{align}
  \begin{cases}
    U_\mathrm{eq}(\vec{r}) &\to U_\mathrm{eq}(\vec{r}) + U_0 \\
    V_\alpha & \to V_\alpha + U_0/e
  \end{cases}
\end{align}
The gauge transformation of the scattering matrix discussed above implies
\begin{equation}
g_{\alpha\beta}(\varepsilon;U_\mathrm{eq}(\vec{r})+ U_0)
=g_{\alpha\beta}(\varepsilon-U_0;U_\mathrm{eq}(\vec{r}))
\:.
\end{equation}
This makes the invariance of the Landauer-B\"uttiker formula under these transformations completely clear. 
After transformation, we can also expand \eqref{eq:LandauerButtiker2} in powers of $U_0$ in order to see the implications for the non-linear conductance~:
\begin{align}
  &I_\alpha
  =\frac{2_se^2}{h}\sum_\beta 
  \left(g_{\alpha\beta}+U_0\derivp{g_{\alpha\beta}}{U_0}+\cdots\right)\,
  \left( V_\beta+\frac{U_0}{e} \right)
  \nonumber\\
  &+ \frac{2_se^3}{h}\sum_{\beta,\,\gamma} 
  \left( g_{\alpha\beta\gamma}^\nonint + \cdots \right)\,  
  \left(V_\beta+\frac{U_0}{e} \right)\, 
  \left(V_\gamma+\frac{U_0}{e}\right)\, 
    \nonumber\\
  &+\cdots
  \:.
\end{align}
We now impose the vanishing of all terms depending on $U_0$.
At linear order we recover the well-known condition 
\begin{equation}
  \label{eq:GaugeInvarianceLC}
  \sum_\beta g_{\alpha\beta} = 0
  \:.
\end{equation}
The vanishing of quadratic terms gives
\begin{align}
  \label{eq:IntermediateGaugeInv}
  &\sum_\beta V_\beta\,\derivp{g_{\alpha\beta}}{U_0} 
  + \frac{U_0}{e}\sum_\beta \derivp{g_{\alpha\beta}}{U_0}
  \nonumber\\
  + &2 \sum_{\beta,\gamma} V_\beta\, g_{\alpha\beta\gamma}^\nonint 
  + \frac{U_0}{e}\sum_{\beta,\gamma} g_{\alpha\beta\gamma}^\nonint = 0 
\end{align}
The second term cancels by virtue of \eqref{eq:GaugeInvarianceLC}.
The three remaining contributions must vanish $\forall\ V_\beta$ and $\forall\ U_0$, therefore we deduce the condition
\begin{equation}
  \label{eq:GaugeInvarianceNonIntNLC}
  \derivp{g_{\alpha\beta}}{U_0} 
  + 2 \sum_{\gamma} g_{\alpha\beta\gamma}^\nonint = 0 
  \:,
\end{equation}
which is the expression of gauge invariance for the theory of free electrons.
Summation over $\beta$ leads to $\sum_{\beta,\gamma} g_{\alpha\beta\gamma}^\nonint = 0$ and ensures the vanishing of the last term in Eq.~\eqref{eq:IntermediateGaugeInv}.

Using $\partial g_{\alpha\beta}(\varepsilon)/\partial U_0=-\partial g_{\alpha\beta}(\varepsilon)/\partial\varepsilon$, we can now check that the non-linear conductances \eqref{eq:NoninteractingNLC} deduced from the Landauer-B\"uttiker formula fulfill gauge invariance \eqref{eq:GaugeInvarianceNonIntNLC}, as it should.

\subsubsection{Gauge invariance in the theory with Coulomb interaction}

B\"uttiker's theory~\cite{But93,ChrBut96a} includes the response of the potential to the shift of the external voltages, caused by the Coulomb interaction. In this case we should rewrite~\eqref{eq:LandauerButtiker} as 
\begin{align}
  \label{eq:LandauerButtiker3}
  &I_\alpha = \frac{2_se}{h}
  \int\D\varepsilon \, \sum_\beta
  \nonumber\\
  &g_{\alpha\beta}\Big(
    \varepsilon;
    U_\mathrm{eq}(\vec{r})
    +\sum_\gamma u_\gamma(\vec{r})eV_\gamma+\cdots
   \Big) \, 
  f(\varepsilon-eV_\beta)
  \:.
\end{align}
Due to the voltage dependence of the potential, the currents are now invariant under the shift of the voltages \textit{alone} 
\begin{equation}
  V_\alpha  \to V_\alpha + U_0/e
  \:.
\end{equation}
As a consequence, gauge invariance for the non-linear conductance now take the simpler form
\begin{equation}
  \label{eq:GaugeInvarianceIntNLC}
  \sum_{\gamma} g_{\alpha\beta\gamma} = 0 
\end{equation}
for the symmetrised non-linear conductance $g_{\alpha\beta\gamma}=g_{\alpha\gamma\beta}$,
which is the condition given by B\"uttiker in Refs.~\onlinecite{But93,ChrBut96a}.

We now check that \eqref{eq:NLC} satisfies the condition \eqref{eq:GaugeInvarianceIntNLC} (we set $T=0$ for simplicity).
Using \eqref{eq:ConstraintCharacPot} and \eqref{eq:GaugeInvarianceLC} we obtain that 
\begin{equation}
  \sum_\gamma g_{\alpha\beta\gamma}^\mathrm{int}
   = \frac{1}{2}\int\D\vec{r}\,
    \derivf{g_{\alpha\beta}(\varepsilon_F)}{U(\vec{r})} 
    = -\frac{1}{2} g'_{\alpha\beta}(\varepsilon_F)
    \:.
\end{equation}
Therefore this term exactly compensates $\sum_\gamma g_{\alpha\beta\gamma}^\nonint$ by virtue of \eqref{eq:GaugeInvarianceNonIntNLC}.
{\sc Qed}.

\vspace{0.25cm}

\paragraph*{Illustration~: two-terminal conductor.---}

For the two-terminal conductor we have $g_{11}=g_{22}=-g_{12}=-g_{21}\equiv g$.
As a consequence of \eqref{eq:SymmetryVoltageIndices}, \eqref{eq:CurrentConservationNonLinear} and~\eqref{eq:GaugeInvarianceIntNLC}, we deduce that all elements are equal, up to a sign~:
$g_{\alpha\beta\gamma}=(-1)^{1+\alpha+\beta+\gamma}g_{111}$, where $\alpha,\:\beta,\:\gamma\in\{1,\,2\}$.
In particular $g_{111}=-g_{222}$ implies that the non-linear conductance vanishes in a device which is symmetric under the exchange of the contacts $1\leftrightarrow2$ (left/right symmetry).
It is worth emphasizing that this is a property of the theory including the effect of interactions. 
In the theory for free electrons, as $g'(\varepsilon_F)$ has no specific reason to vanish, apart for some particular values of the Fermi energy, we deduce from \eqref{eq:GaugeInvarianceNonIntNLC} that $g_{111}^\nonint+g_{222}^\nonint=g'(\varepsilon_F)$, which is different from zero in general.

\subsection{Symmetry of the correlators under magnetic field reversal}
\label{subsec:SUMFR}

The aim of the article is to discuss the statistical properties of the non-linear conductance \eqref{eq:NLC} in weakly disordered metals. 
As we will show below, the disordered averaged non-linear conductance vanishes, what will lead us to concentrate ourselves on \textit{correlators}.
For simplicity we will consider a two-terminal conductor~: in this case, setting the two potentials as $V_1=V$ and $V_2=0$, the non-linearity of the current is controlled by $g_{111}=-g_{211}$~:
$I_2=(2_se^2/h)\,g_{21}V+(2_se^3/h)\,g_{211}V^2+\cdots$.
Below, we rather use the notations introduced in the introduction and Section~\ref{sec:MainResults}~:
$g\equiv-g_{21}$ and $\mathcal{G}\equiv-g_{211}$.
The non-linear conductance is splitted into two contributions
$  \mathcal{G} = \Gnonint + \Gint $,
where the first term 
\begin{align}
  \Gnonint \equiv -g_{211}^\nonint = -\frac{1}{2}
  \int\D\varepsilon\,\left(-\derivp{f}{\varepsilon}\right)\,
  \int\D\vec{r}\,\derivf{g(\varepsilon)}{U(\vec{r})}
\end{align}
is the non-linear conductance of the theory for free electrons and the second term 
\begin{align}
  \Gint \equiv -g_{211}^\mathrm{int} = 
  \int\D\varepsilon\,
  &\left(-\derivp{f}{\varepsilon}\right)\,
  \int\D\varepsilon'\,\left(-\derivp{f}{\varepsilon'}\right)\,
  \nonumber\\
  \times
  &\int\D\vec{r}\,\derivf{g(\varepsilon)}{U(\vec{r})}\,
  \frac{\DoS_1(\vec{r};\varepsilon')}{\DoS_0}
\end{align}
is due to electronic interactions.

Our task will be to obtain the correlations at different magnetic fields 
$\mean{\mathcal{G}(\mathcal{B})\mathcal{G}(\mathcal{B}')}$.
Correlations between conductance and injectivity can be ignored (see Appendix~\ref{Appendix:CorrInjFdc}), hence
\begin{align}
  \label{eq:CorrelatorGintGint}
  \mean{\Gint(\mathcal{B})\Gint(\mathcal{B}')} 
  = \int\D\vec{r}\D\vec{r}\,'\,
  \chi_g(\vec{r},\vec{r}\,')\,\chi_\DoS(\vec{r},\vec{r}\,')
\end{align}
with 
\begin{align}
  \label{eq:DefChiG}
  \chi_g(\vec{r},\vec{r}\,') &= \int\D\omega\,\delta_T(\omega)\,
  \mean{ \derivf{g(\varepsilon)}{U(\vec{r})}\,\derivf{g(\varepsilon-\omega)}{U(\vec{r}\,')} }  
  \\
  \label{eq:DefChiU}
  \chi_\DoS(\vec{r},\vec{r}\,') &= \int\D\omega\,\delta_T(\omega)\,
  \frac{ \mean{  \DoS_1(\vec{r};\varepsilon)\,\DoS_1(\vec{r}\,';\varepsilon-\omega)}_c }{ \DoS_0^2 }
  \:,
\end{align}
where $g(\varepsilon)$ is the two-terminal linear dimensionless conductance at zero temperature. 
$\delta_T(\omega)$ is a function of width $T$ such that $\int\D\omega\,\delta_T(\omega)=1$, arising from the property \eqref{eq:ThermFctProperty1}.
We introduce the notation $\mean{XY}_c=\mean{XY}-\mean{X}\mean{Y}$.
Note that $\smean{\delta g/\delta U(\vec{r})}=0$ (see below). 
The correlations for the non-interaction part can also be deduced from the correlator~$\chi_g$~:
\begin{equation}
  \label{eq:CorrelatorGnonintGnonint}
  \mean{\Gnonint(\mathcal{B})\Gnonint(\mathcal{B}')} 
  = \frac14\int\D\vec{r}\D\vec{r}\,'\,  \chi_g(\vec{r},\vec{r}\,')
  \:.
\end{equation}
The correlator $\mean{\Gnonint\Gint}$ has the same symmetry as $\mean{\Gnonint\Gnonint}$ and will be also considered.

The study of the two correlators $\chi_g(\vec{r},\vec{r}\,')$ and $\chi_\DoS(\vec{r},\vec{r}\,')$ will be the main issue of the two next sections.
For the moment we concentrate ourselves on the magnetic field dependence. For this reason, we ignore the spatial and energy dependences of the correlators until the end of this subsection. 
It will be straightforward to reintroduce them later.
As we will see in section~\ref{sec:CondCorr}, the correlations of conductances in weakly disordered metals are given by Diffuson and Cooperon contributions, what leads to the magnetic field dependence~:
\begin{equation}
  \chi_g(\mathcal{B},\mathcal{B}') 
  = 
    \chi_1(\mathcal{B}-\mathcal{B}') 
  + \chi_1(\mathcal{B}+\mathcal{B}')
  \:.
\end{equation}
The fact that the Diffuson and Cooperon contributions involve the \textit{same} function $\chi_1$ reflects the symmetry of the conductance $g(-\mathcal{B})=g(\mathcal{B})$, resulting into
$\chi_g(-\mathcal{B},\mathcal{B}')=\chi_g(\mathcal{B},-\mathcal{B}')=\chi_g(\mathcal{B},\mathcal{B}')$.
On the other hand the injectivity $\DoS_1(\vec{r};\varepsilon)$, i.e. the characteristic potential $u_1(\vec{r})$, is \textit{not} a symmetric function of the magnetic field.
As a result, the Diffuson and Cooperon contributions to the injectivity correlator involve two \textit{different} functions $\chi_{2}^d\neq\chi_{2}^c$~:
\begin{equation}
  \chi_\DoS(\mathcal{B},\mathcal{B}') 
  = \chi_{2}^d(\mathcal{B}-\mathcal{B}') + \chi_{2}^c(\mathcal{B}+\mathcal{B}')
  \:.
\end{equation}
This important difference between conductance correlations (Fig.~\ref{fig:ConductanceCorr12}) and injectivity correlations (Fig.~\ref{fig:InjectivityCorr12}) can be related to the structure of the diagrams.
For the wire, the external Diffusons of the conductance correlation diagrams of Fig.~\ref{fig:ConductanceCorr12} are just constant,~\cite{footnote5} 
so that the two diagrams only differ by exchanging Diffusons and Cooperons in the loop, i.e. are in exact correspondence provided $\mathcal{B}-\mathcal{B}'\leftrightarrow\mathcal{B}+\mathcal{B}'$.
On the other hand, the external Diffusons of the injectivity correlation diagrams of Fig.~\ref{fig:InjectivityCorr12} carry some non-trivial spatial dependence. 
As a consequence, the two diagrams of Fig.~\ref{fig:InjectivityCorr12} are not related by a simple substitution of Diffusons into Cooperons in the loop, which is the reason why $\chi_{2}^d\neq\chi_{2}^c$.

Splitting the interaction part of the non-linear conductance into symmetric and antisymmetric part as $\Gint=\Gint_s+\Gint_a$, we introduce the corresponding correlators
  \begin{equation}
    \label{eq:GaCorrelator}
    \mean{ \Gint_{s,a}(\mathcal{B})\,\Gint_{s,a}(\mathcal{B}') } = 
    \chi_g(\mathcal{B},\mathcal{B}')\, 
    \chi_\DoS^{s,a}(\mathcal{B},\mathcal{B}')  
    \:,
  \end{equation}
which presents the following structure
\begin{align}
  \chi_\DoS^{s,a}(\mathcal{B},\mathcal{B}')
   = \frac{1}{2}
   \Big[
     &\chi_{2}^d(\mathcal{B}-\mathcal{B}') \pm \chi_{2}^c(\mathcal{B}-\mathcal{B}')
     \nonumber\\
    +&\chi_{2}^c(\mathcal{B}+\mathcal{B}') \pm \chi_{2}^d(\mathcal{B}+\mathcal{B}')
   \Big]
   \:.
\end{align}  
Spatial and energy integrations will be reintroduced below.

\section{Correlations of conductance's functional derivatives ($\chi_g$)}
\label{sec:CondCorr}

We start by considering the correlator $\chi_g(\vec{r},\vec{r}\,')$ which is the most simple to obtain as it can be deduced from the known expression of the conductance correlator.

\subsection{Conductance correlator and diagrammatic rules for quasi-1D devices}
\label{subsec:DiagramRules}

\subsubsection{Conductance}

Conductivity correlations in weakly disordered metals have been studied in Refs.~\onlinecite{AltKhm85,LeeSto85,AltShk86,LeeStoFuk87,KanSerLee88} (for recent reviews, see Refs.~\onlinecite{AkkMon07,TexMon16}), where the contributions to the conductance correlations in multiterminal devices was studied.
Here we will give a simplified description, based on Ref.~\onlinecite{TexMon16}.

\begin{figure}[!ht]
\centering
\includegraphics[width=0.4\textwidth]{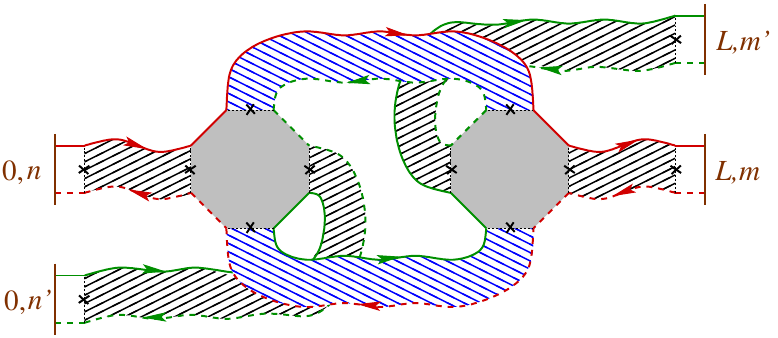}
\\[0.25cm]
\includegraphics[width=0.4\textwidth]{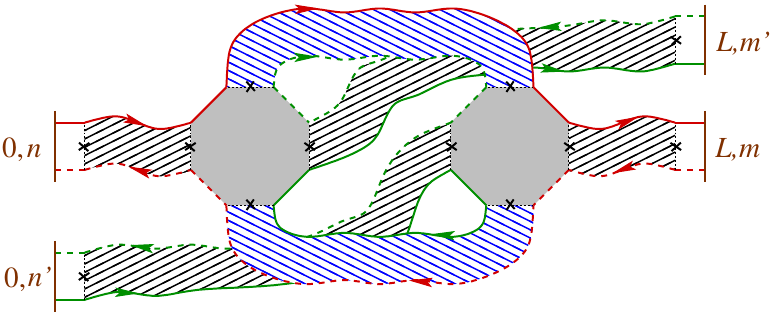}
\caption{(color online) \it Two contributions to the conductance correlations (``diffusion constant correlation diagrams'').
  Retarded Green's functions are represented by continuous lines and advanced ones by dashed lines.
  The two colors (red versus green) correspond to the two conductances.
  The dashed areas represent the ladder diagrams (in black, ``external'' Diffusons $P_d$ and in blue the Diffusons $P_\omega^{(d)}$ and the Cooperons $P_\omega^{(c)}$). The dotted line and the cross is the first (or the last) disorder interaction line of the ladder. Two Hikami boxes ensure the branching between Diffusons/Cooperons.}
\label{fig:ConductanceCorr12}
\end{figure}

We consider the simple case of a quasi-one-dimensional (1D) wire with $\Nc$ conducting channels.
Our starting point is the Fisher and Lee formula~\cite{FisLee81} for the dimensionless conductance (here at zero temperature, for simplicity)
\begin{equation}
  \label{eq:GfromFisherLee}
  g = \sum_{n,m=1}^{\Nc}
  v_nv_m\,G^\mathrm{R}_{nm}(L,0;\varepsilon_F)\,G^\mathrm{A}_{mn}(0,L;\varepsilon_F)
\end{equation}
where  
\begin{align}
  G^\mathrm{R,A}_{nm}(L,0;\varepsilon_F)
  =\int\D y\D y'\,\chi_n(y)\,G^\mathrm{R,A}(\vec{r},\vec{r}\,';\varepsilon_F)\,\chi_m(y')
\end{align}
with $x=L$ and $x'=0$.
$\chi_n(y)$ is a transverse mode wave function and $v_n$ the group velocity in channel $n$.
The calculation of conductance correlations thus requires to correlate four Green's function lines with disorder impurity lines.
In the weak disorder limit, dominant contributions arise from ladder diagrams (Diffuson and Cooperon). We obtain six diagrams~:
two represented in Fig.~\ref{fig:ConductanceCorr12} (one with Diffusons and one with Cooperons), interpreted as \textit{diffusion constant correlations}~\cite{AltShk86,AkkMon07}.
The other diagrams, represented in Fig.~\ref{fig:ConductanceCorr34}, are interpreted as \textit{DoS correlations}~\cite{AltShk86,AkkMon07} (although there are four distinct diagrams, two are simply obtained by exchanging retarded and advanced lines). 

\subsubsection{Ladders}
\label{subsec:Ladders}

Few simple rules allow one to determine straightforwardly the expressions of the correlator $\mean{g(\mathcal{B})\,g(\mathcal{B}')}_c$ from the diagrams.
For this purpose we will assume that the system has a quasi-one-dimensional geometry and will perform all traces over transverse modes in order to keep only the 1D structure of the propagators.

An important remark illustrated by the representation of Fig.~\ref{fig:ConductanceCorr12} is that diagrams involve two types of ladders~: the ``external'' Diffusons which start from the boundaries and correlate lines from the \textit{same} conductance, and the ``internal'' Diffusons and Cooperons which correlate Green's function lines from \textit{different} conductances and describe correlations.
These latters are the propagators
\begin{equation}
  P_\omega^{(d,c)}(x,x')
  =\bra{x} \frac{1}{\gamma_\omega-\big[\partial_x-2\I eA_\mp(x)\big]^2} \ket{x'}
  \:,
\end{equation}
where we assume quasi-1D limit so that the Diffusons/Cooperons depend only on the coordinate along the wire.
\begin{equation}
  \gamma_\omega=\gamma-\I\frac{\omega}{D}
\end{equation} 
involves the phase coherence length $L_\varphi$ in $\gamma=1/L_\varphi^2$ and the energy difference 
$\omega=\varepsilon-\varepsilon'$, where $\varepsilon$ and $\varepsilon'$ are the energies of the two Green's functions. 
The vector potential involves the two vector potentials associated with the two magnetic fields
$A_\pm(x)=\big[A(x)\pm A'(x)\big]/2$, where $-$ is chosen for a Diffuson and $+$ for a Cooperon.
In a narrow wire, we can account for the perpendicular magnetic field through an effective phase coherence length thanks to the substitution
\begin{equation}
  \label{eq:LdLc}
  \gamma\to \gamma_{d,c} =\frac{1}{L_{d,c}^2}   
  = \frac{1}{L_\varphi^2} + \frac{1}{L_{(\mathcal{B}\mp\mathcal{B})/2}^2}
\end{equation}
where the magnetic length is~\cite{AltAro81,footnote6} 
\begin{equation}
  L_\mathcal{B} = \frac{\sqrt{3}}{2\pi}\frac{\phi_0}{|\mathcal{B}|\sw}
  \:;
\end{equation}
$\phi_0=h/e$ is the flux quantum and $\sw$ the width of the wire.

The ladders starting from a boundary ($x=0$ or $L$ here) correlate two Green's function lines with same Fermi energy and same magnetic field, thus they correspond to the Diffuson
\begin{equation}
  P_d(x,x') =\bra{x} \frac{1}{-\partial_x^2} \ket{x'}
  \:.
\end{equation}

In order to model the connection to the reservoirs we impose that the Diffusons/Cooperons vanish at the boundary~: $P_\omega^{(d,c)}(0,x')=P_\omega^{(d,c)}(L,x')=P_d(0,x')=P_d(L,x')=0$ (see Refs.~\onlinecite{HasStoBar94,TexMon04} for a more precise discussion~; see also chapter~5 of Ref.~\onlinecite{AkkMon07}).
In the narrow wire, we get~:
\begin{align}
  \label{eq:PropagatorFiniteWire}
  &P_\omega^{(d,c)}(x,x') 
  =  \bra{x} \frac{1}{\gamma_\omega-\partial_x^2} \ket{x'}
  \nonumber\\
  &= 
  \frac{\sinh(\sqrt{\gamma_\omega}x_<)\sinh(\sqrt{\gamma_\omega}(L-x_>))}
       {\sqrt{\gamma_\omega}\sinh(\sqrt{\gamma_\omega}L)}
       \:,
\end{align}
where $x_<=\min{x}{x'}$ and $x_>=\max{x}{x'}$.
Setting $\gamma_\omega=0$ we get 
\begin{equation}
  \label{eq:PdWire}
  P_d(x,x') = \min{x}{x'}-\frac{xx'}{L}
  \:.
\end{equation}
In the following we consider a weakly coherent wire, $L_\varphi\ll L$ (the case of a fully coherent wire will be considered later in Section~\ref{sec:Coherent}). This simplifies many calculations as it allows to neglect the effect of boundaries on the $\gamma_\omega$-dependent Diffuson/Cooperon which then takes the simple form
\begin{equation}
  \label{eq:DiffusonInfiniteWire}
  P_\omega^{(d,c)}(x,x') 
  \simeq\frac{\EXP{-\sqrt{\gamma_\omega}|x-x'|}}{2\sqrt{\gamma_\omega}}
  \:. 
\end{equation}

\subsubsection{Simplified diagrammatic rules for quasi-1D systems}

We now give three simple diagrammatic rules for quasi-1D devices obtained after tracing over transverse modes.
The two first rules concern the ladders~:
  \begin{align}
    \label{eq:Rule1}
    &
    \begin{cases}
      \diagram{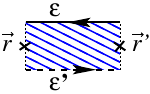}{0.75}{-0.4cm} \\[0.4cm]
      \diagram{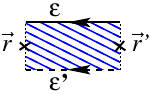}{0.75}{-0.4cm}
    \end{cases}
    \hspace{-0.5cm} = \frac{1}{\tau_e^2\,\xiloc}
    \begin{cases}
      P^{(d)}_{\varepsilon-\varepsilon'}(x,x') \\[0.5cm]
      P^{(c)}_{\varepsilon-\varepsilon'}(x,x') 
    \end{cases}
    \\
    \label{eq:Rule2}
    &\sum_{n}\diagram{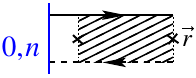}{0.75}{-0.4cm}
    = \frac{1}{\tau_e}\,\frac{P_d(x,\ell_e)}{\ell_e}
\end{align}    
where summation runs over channel index.
The localisation length is given by \eqref{eq:LocalisationLength}.
$\tau_e$ is the elastic mean free time and $\ell_e$ the elastic mean free path.
Diffusons and Cooperons describe arbitrary long sequences of scattering events on the disorder and therefore decay over scale $\gg\ell_e$.
In the diagrams, they are plugged with each others through ``Hikami boxes'' with four entries, which involves average Green's functions of extension $\lesssim\ell_e$ (see Ref.~\onlinecite{AkkMon07}). Because we are here interested in large scale properties, compared to $\ell_e$, the Hikami boxes can simply be considered as purely local.
After tracing over channels, we obtain the 1D structure~:
  \begin{align}
    \label{eq:Rule3}
    \diagram{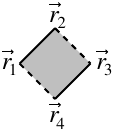}{0.75}{-0.8cm} =
    \tau_e^4\,\xiloc\, H(x_1,x_2,x_3,x_4)
  \:,
\end{align}    
with
\begin{align}
  \label{eq:H4}
  H(x_1,x_2,x_3,x_4) \simeq
  \int\D X\,\left(\prod_{i=1}^4\delta(X-x_i)\right)
  \times
  \begin{cases} 
    2\,  \partial_1\partial_3 \\[0.25cm]
    - \partial_1\partial_2
  \end{cases}
\end{align}
where the choice between the two expressions is made such that the gradients act on ``external'' Diffusons (first choice is made for diagrams of Fig.~\ref{fig:ConductanceCorr12} and second choice for diagrams of Fig.~\ref{fig:ConductanceCorr34}).
These expressions differ from the one obtained originally in Ref.~\onlinecite{Hik81} by a direct calculation of the diagram~\eqref{eq:Rule3}, which gives $\partial_1\partial_3+\partial_2\partial_4-(1/2)\sum_i\partial_i^2$ (see also \cite{AkkMon07}). This expression produces divergences and leads to expressions of the correlators which do not fulfill the requirement of current conservation.
Expressions \eqref{eq:H4} follow from a procedure proposed by Kane, Serota and Lee~\cite{KanSerLee88} allowing to relate short range conductivity correlation diagrams to long range contributions, which are related to conductance correlators (see Ref.~\onlinecite{TexMon04} where this point was discussed on the more simple case of the weak localisation).

\subsubsection{Application to conductance correlations}

As an application of the rules (\ref{eq:Rule1},\ref{eq:Rule2},\ref{eq:Rule3}), we deduce straightforwardly the expression of the contribution to the conductance correlator corresponding to the first diagram of Fig.~\ref{fig:ConductanceCorr12}~:
\begin{widetext} 
\begin{align}
  \label{eq:ConductanceCorrelator1}
  \mean{g(\mathcal{B})\,g(\mathcal{B}')}^{(1)}
  = 4 \int\D\omega\,\delta_T(\omega)\int_0^L\D x\D x'\,
  \left[\frac{\partial_xP_d(L-\ell_e,x)}{\ell_e}\right]^2\,
  P^{(d)}_\omega(x,x')P^{(d)}_{-\omega}(x,x')\,
  \left[\frac{\partial_{x'}P_d(x',\ell_e)}{\ell_e}\right]^2
\end{align}
(see also Refs.~\onlinecite{Tex10hdr,TexMon16}).
The correlator receives another contribution where one pair of Green's function lines are reversed. 
This leads to replace the two Diffusons inside the loop by Cooperons (second diagram of Fig.~\ref{fig:ConductanceCorr12}), and exchange two ``external'' Diffusons~:
\begin{align}
  \label{eq:ConductanceCorrelator2}
  \mean{g(\mathcal{B})\,g(\mathcal{B}')}^{(2)}
  = 4 \int\D\omega\,\delta_T(\omega)\int_0^L\D x\D x'\,
  \frac{\partial_xP_d(L-\ell_e,x)}{\ell_e}\,\frac{\partial_xP_d(\ell_e,x)}{\ell_e}\,
  &P^{(c)}_\omega(x,x')P^{(c)}_{-\omega}(x',x)\,
  \\\nonumber
  &\times
  \frac{\partial_{x'}P_d(x',L-\ell_e)}{\ell_e}\,\frac{\partial_{x'}P_d(x',\ell_e)}{\ell_e}
\end{align}
The two contributions (\ref{eq:ConductanceCorrelator1},\ref{eq:ConductanceCorrelator2}) are interpreted as diffusion constant correlations.
Finally, the correlator receives two other contributions corresponding to the diagrams of Fig.~\ref{fig:ConductanceCorr34}~:
\begin{align}
  \label{eq:ConductanceCorrelator3}
  \mean{g(\mathcal{B})\,g(\mathcal{B}')}^{(3)}
  = 2 \int\D\omega\,\delta_T(\omega)\int_0^L\D x\D x'\,
  \frac{\partial_xP_d(L-\ell_e,x)}{\ell_e}\,\frac{\partial_xP_d(\ell_e,x)}{\ell_e}\,
  &\re\left[ P^{(d)}_\omega(x,x')P^{(d)}_{\omega}(x',x)\right]\,
  \\\nonumber
  &\times
  \frac{\partial_{x'}P_d(x',L-\ell_e)}{\ell_e}\,\frac{\partial_{x'}P_d(x',\ell_e)}{\ell_e}
\end{align}
\begin{align}
  \label{eq:ConductanceCorrelator4}
  \mean{g(\mathcal{B})\,g(\mathcal{B}')}^{(4)}
  = 2 \int\D\omega\,\delta_T(\omega)\int_0^L\D x\D x'\,
  \frac{\partial_xP_d(L-\ell_e,x)}{\ell_e}\,\frac{\partial_xP_d(\ell_e,x)}{\ell_e}\,
  &\re\left[ P^{(c)}_\omega(x,x')P^{(c)}_{\omega}(x',x)\right]\,
  \\\nonumber
  &\times
  \frac{\partial_{x'}P_d(x',L-\ell_e)}{\ell_e}\,\frac{\partial_{x'}P_d(x',\ell_e)}{\ell_e}
  \:.
\end{align}

\begin{figure}[!ht]
\centering
\diagramw{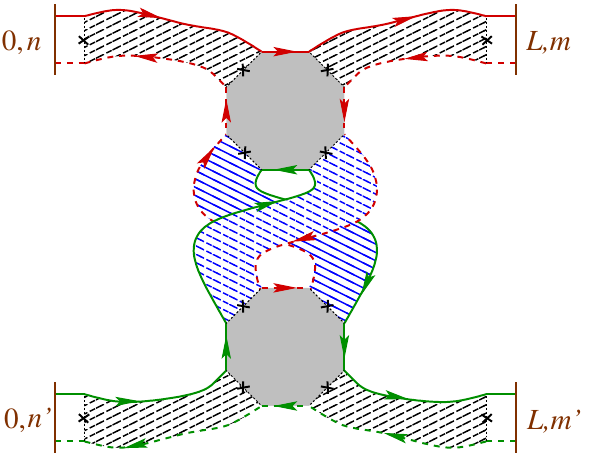}{0.3\textwidth}{-2cm}
$+\left(G^\mathrm{R}\leftrightarrow G^\mathrm{A}\right)$
\hspace{1cm}
\diagramw{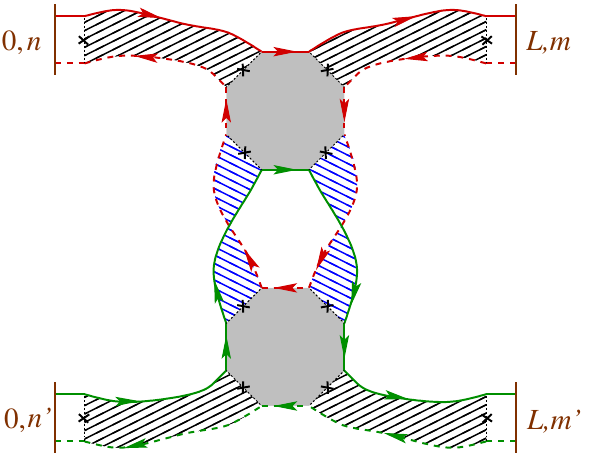}{0.3\textwidth}{-2cm}
$+\left(G^\mathrm{R}\leftrightarrow G^\mathrm{A}\right)$
\caption{(color online) \it Four diagrams contributing to the conductance correlations (``DoS correlation diagrams'').}
\label{fig:ConductanceCorr34}
\end{figure}

\end{widetext} 
The contributions (\ref{eq:ConductanceCorrelator3},\ref{eq:ConductanceCorrelator4}) can be interpreted as DoS correlations~\cite{AltShk86,AkkMon07} (they are similar to the DoS correlations recalled in Subsection~\ref{subsec:DoScorr}).
Compared to the diffusion constant correlation terms, the DoS correlation terms are more symmetric~:
they are in perfect correspondence through the substitution $P^{(d)}\leftrightarrow P^{(c)}$ in the loop, while the ``external'' Diffusons are not affected by this exchange, as it was the case for the diffusion constant correlation terms (Fig.~\ref{fig:ConductanceCorr12}).
This observation will play an important role later on.

We now consider a diffusive wire of length $L$ for which the Diffuson has the form \eqref{eq:PdWire}. 
Neglecting contributions of intervals of width $\ell_e$ at the boundaries, like $\int_0^{\ell_e}\D x$ and $\int_{L-\ell_e}^L\D x$, the four gradients of Diffusons in (\ref{eq:ConductanceCorrelator1},\ref{eq:ConductanceCorrelator2},\ref{eq:ConductanceCorrelator3},\ref{eq:ConductanceCorrelator4}) simply give a factor $1/L^4$. 
The full expression with ``external'' Diffusons will however be useful for future discussions on injectivity correlations in Section~\ref{sec:InjCorr}.

Two helpful remarks~:
\begin{itemize}
\item 
  The analysis can be simplified as follows~:
  in the ``low temperature'' regime, when $L_T=\sqrt{D/T}\gg\min{L}{L_\varphi}$, it is legitimate to perform the substitution $\delta_T(\omega)\to\delta(\omega)$, thus we have 
  $\smean{\delta g^2}^{(3)}=(1/2)\smean{\delta g^2}^{(1)}$ and 
  $\smean{\delta g^2}^{(4)}=(1/2)\smean{\delta g^2}^{(2)}$.
  In the ``high temperature'', $L_T=\sqrt{D/T}\ll\min{L}{L_\varphi}$, the two contributions $\smean{\delta g^2}^{(3)+(4)}$ can be neglected.

\item  
  In the wire, there is a perfect symmetry between Diffuson and Cooperon contributions, what allows us to deduce $\smean{\delta g^2}^{(2)}$ from $\smean{\delta g^2}^{(1)}$ and $\smean{\delta g^2}^{(4)}$ from $\smean{\delta g^2}^{(3)}$ by performing the simple substitution $\gamma_d\to\gamma_c$.
\end{itemize}
Note that these two simplifications only hold for a two-terminal device (see Ref.~\onlinecite{TexMon16} for a discussion of the multiterminal case).
These remarks will allow us to consider only one contribution to the correlator and deduce the others by these symmetry arguments.
In a first step, we will discuss the contributions $^{(1)+(3)}$. Restricting to these two contributions corresponds to consider the high magnetic field limit, when $\gamma_c\gg\gamma_d$ (i.e. $L_c\ll L_d$). 
The other contributions $^{(2)+(4)}$ and the full magnetic field dependence will be discussed in a second step.

\subsection{Action of functional derivatives}

Knowing the correlations of conductance, it is now straightforward to obtain the correlator \eqref{eq:DefChiG}, what requires to determine the action of the two functional derivations $\delta/\delta U(\vec{r})$ and $\delta/\delta U(\vec{r}\,')$ on the diagrams of Fig.~\ref{fig:ConductanceCorr12}.
We must take care that the two functional derivatives act on two \textit{different} conductances.
The functional derivation corresponds to interrupt a Green's function line~:
\begin{equation}
  \derivf{G^\mathrm{R,A}(\vec{r}\,',\vec{r}\,'';\varepsilon)}{U(\vec{r})} 
  = 
  G^\mathrm{R,A}(\vec{r}\,',\vec{r};\varepsilon)\,
  G^\mathrm{R,A}(\vec{r},\vec{r}\,'';\varepsilon)
  \:.
\end{equation} 
Thus the interruption of a ladder involves the box~:
\begin{equation}
\label{eq:H3}
  \diagram{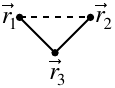}{0.75}{-0.5cm} \simeq -2\I\pi\rho_0\, \tau_e^2\,
  \delta(\vec{r}_1-\vec{r}_2)\,\delta(\vec{r}_1-\vec{r}_3)
  \:,
\end{equation}
or its complex conjugate, where $\rho_0=\DoS_0/2_s$ is the DoS per spin channel.
Using \eqref{eq:Rule1} with this expression, we deduce~:
\begin{align}
\label{eq:FunctionalDerivation}
  P_\omega^{(d,c)}(x',x'')
  &\overset{\delta/\delta U(\vec{r}) }{\longrightarrow}
  P_\omega^{(d,c)}(x',x)
  \left( \mp \frac{\I}{D}\right)
  P_\omega^{(d,c)}(x,x'')
\end{align}
depending whether the derivation acts on a retarded ($-$) or an advanced ($+$) Green's function line.

A last important remark is that the action of the functional derivation on the ``external'' Diffusons gives zero, which follows from the fact that the two Green's functions belong to the \textit{same} conductance~:
\begin{align}
  \label{eq:Property1}
  \diagram{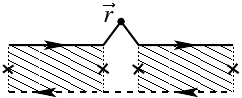}{0.6}{-0.25cm} 
  + 
  \diagram{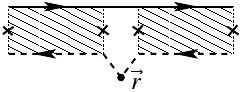}{0.6}{-0.5cm} 
  =0
  \:,
\end{align}
as the addition of the box \eqref{eq:H3} and its conjugate vanishes.
Note that this remark also ensures the equivalence of the Landauer approach followed here and the Kubo approach, which involves local conductivity diagrams without external Diffusons.

In conclusion, the action of the two functional derivatives on the first diagram of Fig.~\ref{fig:ConductanceCorr12} leads to the six diagrams of Fig.~\ref{fig:ChiG}.
The full correlator $\chi_g(\vec{r},\vec{r}\,')$ is thus given by 36 such diagrams. Hopefully, the remaining 30 contributions will be obtained by simple symmetry arguments.

\subsection{Correlator $\chi_g(\vec{r},\vec{r}\,')$ for the wire}

When the external Diffusons are replaced by a constant factor, we can simplify (\ref{eq:ConductanceCorrelator1},\ref{eq:ConductanceCorrelator2}) by noticing that the remaining integrals are the expression of the trace
$
  \tr{ (\gamma_\omega-\partial_x^2)^{-1} (\gamma_\omega^*-\partial_x^2)^{-1} }
$.
As a result, the six diagrams of Fig.~\ref{fig:ChiG} give~:~\cite{footnote7} 
\begin{widetext} 
\begin{align}
  \label{eq:RootEqChiG}
  \mean{\derivf{g}{U(\vec{r})}\,\derivf{g}{U(\vec{r}\,')}}^{(1)}
  = \frac{4}{D^2L^4} \int\D\omega\,\delta_T(\omega)
  \bigg\{
    &-\bra{x}\frac{1}{\gamma_\omega-\partial_x^2}\frac{1}{\gamma_\omega^*-\partial_x^2}\ket{x'}
    \bra{x'}\frac{1}{\gamma_\omega^*-\partial_x^2}\frac{1}{\gamma_\omega-\partial_x^2}\ket{x}
    \nonumber\\
    &+ \bra{x}\frac{1}{\gamma_\omega-\partial_x^2}\ket{x'}
    \bra{x'}\frac{1}{\gamma_\omega-\partial_x^2}\frac{1}{\gamma_\omega^*-\partial_x^2}\frac{1}{\gamma_\omega-\partial_x^2}\ket{x}
    \nonumber\\
    &+\bra{x}\frac{1}{\gamma_\omega-\partial_x^2}\frac{1}{\gamma_\omega^*-\partial_x^2}\frac{1}{\gamma_\omega-\partial_x^2}\ket{x'}\bra{x'}\frac{1}{\gamma_\omega-\partial_x^2}\ket{x}
    \nonumber\\
    &+(\mathrm{3\ similar\ terms\ with\ }\gamma_\omega\leftrightarrow\gamma_\omega^*)
  \bigg\}
\end{align}
The sign difference arises from the fact that the two first diagrams of Fig.~\ref{fig:ChiG} involve twice the box~\eqref{eq:H3}, while the four other diagrams involve the box and its complex conjugate.
Eq.~\eqref{eq:RootEqChiG} is a central result.
\begin{figure}[!ht]
\centering
\includegraphics[width=0.4\textwidth]{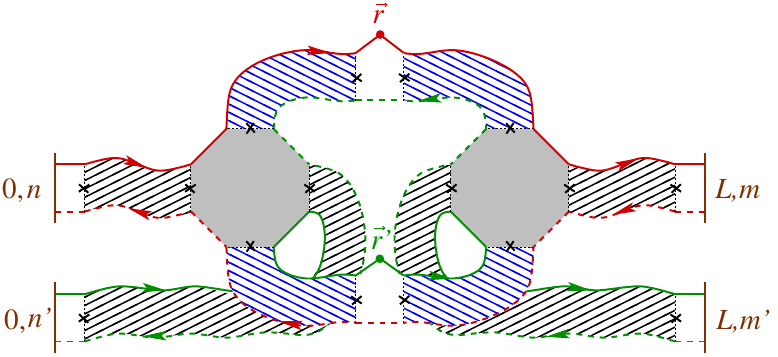}\hspace{1cm}
\includegraphics[width=0.4\textwidth]{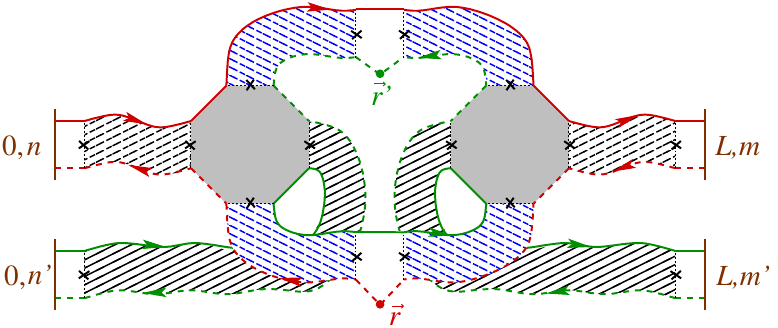}
\\
\includegraphics[width=0.4\textwidth]{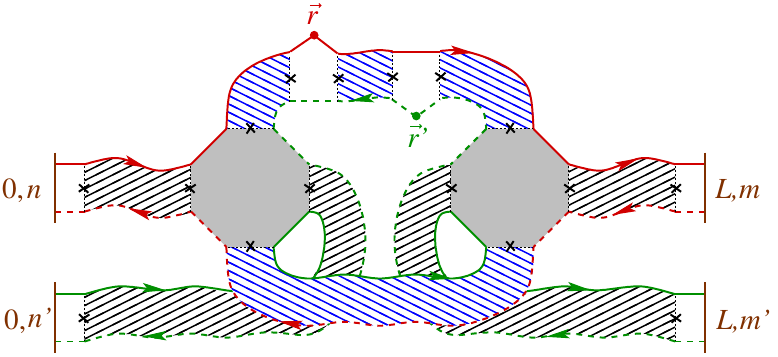}\hspace{1cm}
\includegraphics[width=0.4\textwidth]{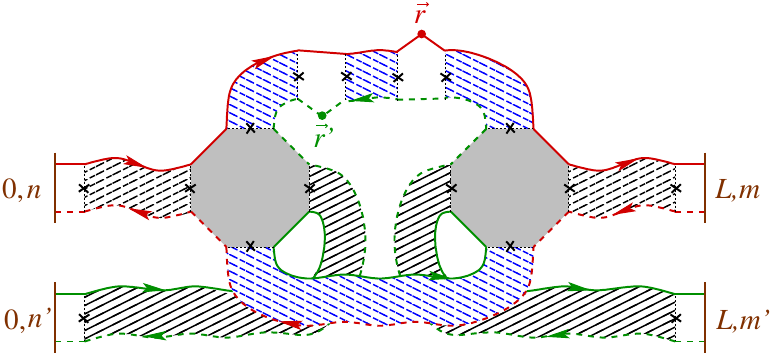}
\\
\includegraphics[width=0.4\textwidth]{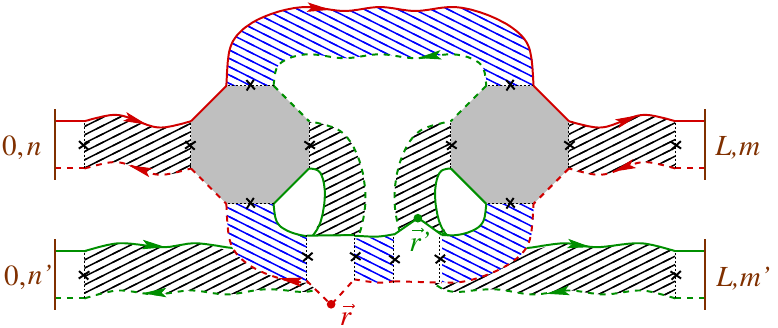}\hspace{1cm}
\includegraphics[width=0.4\textwidth]{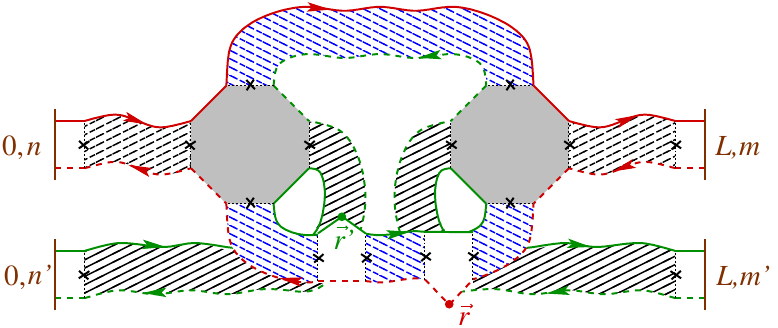}
\caption{(color online) \it Contributions to the correlator $\chi_g(\vec{r},\vec{r}\,')$. The six diagrams obtained by action of the two functional derivatives \eqref{eq:FunctionalDerivation} on the diagram of the top of Fig.~\ref{fig:ConductanceCorr12}.
}
\label{fig:ChiG}
\end{figure}
\end{widetext} 

\subsubsection{``Low'' temperature ($L_\varphi\ll L_T$)}

We first simplify the expression of the correlator by considering the ``low'' temperature limit ($L_T\gg\min{L}{L_\varphi}$) when it is justified to replace the thermal function by a Dirac function.
We deduce
\begin{align}
  \label{eq:ChiGconvenientRepresentation}
  &\mean{ \derivf{g}{U(\vec{r})}\derivf{g}{U(\vec{r}\,')} }^{(1)} 
   = 
   \frac{4}{D^2L^4}
   \\\nonumber
   &\times
   \left(
        \derivp{}{\gamma} 
      - \derivp{}{\gamma'}
    \right)^2
    \bra{x} \frac{1}{\gamma-\partial_x^2} \ket{x'}
    \bra{x'}\frac{1}{\gamma'-\partial_x^2} \ket{x}
    \bigg|_{\gamma'=\gamma}
\end{align}
In order to simplify the calculation, we now assume that the wire is longer than the phase coherence length, $L\gg L_\varphi$. In this case we can use the expression of the propagator in bulk  (for an infinitely long wire) \eqref{eq:DiffusonInfiniteWire}, for which we obtain~:
\begin{align}
  \derivp{}{\gamma} \bra{x}\frac{1}{\gamma-\partial_x^2}\ket{x'}
  \simeq 
  -\frac{1  + \sqrt{\gamma} |x-x'|}{2\gamma}
   \bra{x}\frac{1}{\gamma-\partial_x^2}\ket{x'}
  \:.
\end{align}
We finally end with the expression 
\begin{align}
  \label{eq:Chi1LargeLT}
  & \chi_g(\vec{r},\vec{r}\,')^{(1)} = 
   \mean{ \derivf{g}{U(\vec{r})}\derivf{g}{U(\vec{r}\,')} }^{(1)} 
  \nonumber\\
  &\simeq 2 \frac{L^2}{D^2}\left(\frac{L_\varphi}{L}\right)^6 
  \left( 1 + \frac{|x-x'|}{2L_\varphi} \right) \EXP{-2|x-x'|/L_\varphi}
  \:.
\end{align}
We have therefore obtained that correlations are \textit{short range}.
The correlator in the high magnetic field regime is simply obtained by multiplying this expression by a factor $3/2$ in order to account for the second Diffuson diagram (left diagram of Fig.~\ref{fig:ConductanceCorr34}).
The Cooperon contributions can be obtained by the symmetry arguments mentioned above.

The integral of the correlator will be useful below (see footnote~\cite{footnote7})~:
\begin{align}
  \label{eq:IntegralChiGLargeLT}
  &\int\D x\D x'\,\mean{ \derivf{g}{U(\vec{r})}\derivf{g}{U(\vec{r}\,')} }^{(1)} 
  \nonumber\\  
  &
  =\mean{g'(\varepsilon_F)^2}^{(1)}
  \simeq \frac{5}{2}\left(\frac{L_\varphi}{L}\right)^7
  \tau_D^2
  \:,
\end{align}
where $\tau_D=L^2/D$ is the Thouless time of the diffusive wire.
This result can be compared to the linear conductance fluctuations $\mean{\delta g^2}^{(1)}\simeq(L_\varphi/L)^3$ obtained in the same regime~; hence each derivative $\delta/\delta U(r)$ is responsible for a factor $(L_\varphi/L)^2\tau_D=\tau_\varphi$ where $\tau_\varphi=L_\varphi^2/D$ is the phase coherent time, in agreement with the heuristic argument of Section~\ref{sec:MainResults}.

\subsubsection{``High'' temperature ($L_T\ll L_\varphi$)}

\paragraph{Sum rule.--}

The high temperature is more difficult to analyse. 
A good starting point is to consider the integral of the correlator, which is easy to compute, as the spatial integration of the brackets in Eq.~\eqref{eq:RootEqChiG} admits the simple expression~:
\begin{align}
  \left(
    \derivp{}{\gamma_\omega} - \derivp{}{\gamma_\omega^*}
  \right)^2
  \tr{ \frac{1}{\gamma_\omega-\partial_x^2}\frac{1}{\gamma_\omega^*-\partial_x^2} }
  \:.
\end{align}
Using 
$\partial/\partial\gamma_\omega-\partial/\partial\gamma_\omega^*=-\I D\,\partial/\partial\omega$, we obtain 
\begin{align}
  &\int\D x\D x'\,
  \mean{\derivf{g}{U(\vec{r})}\,\derivf{g}{U(\vec{r}\,')}}^{(1)}
  \\\nonumber
  =& -\frac{4}{L^4} \int\D\omega\,\delta_T''(\omega)\,
  \tr{ \frac{1}{\gamma_\omega-\partial_x^2}\frac{1}{\gamma_\omega^*-\partial_x^2} }
  \:.
\end{align}
This simple structure has a clear interpretation~: the double derivative arises from the fact that the non-interaction non-linear conductance can also be written as an integral of the correlator 
$
\partial_{\varepsilon}\partial_{\varepsilon'}\mean{g(\varepsilon)g(\varepsilon')}
=-\partial_{\varepsilon}^2\mean{g(\varepsilon)g(\varepsilon')}$ 
weighted by Fermi functions.

In the limit $L_T\ll L_\varphi$, we may treat the function $\delta_T(\omega)$ as a ``broad'' function (cf. Appendix~\ref{Appendix:ThermalFctsSS1}).
We use $\delta_T''(0)=-(4/15)/(2T)^3$ and 
\begin{align}
  &\int\frac{\D\omega}{D}\,
  \tr{ \frac{1}{\gamma_\omega-\partial_x^2}\frac{1}{\gamma_\omega^*-\partial_x^2} }
  =\pi\tr{ \frac{1}{\gamma-\partial_x^2} }
  \nonumber\\
  &=\pi\, L\,P_{\omega=0}^{(d)}(x,x)\simeq \frac{\pi\,L}{2\sqrt{\gamma}}
\end{align}
from which we deduce
\begin{align}
  \label{eq:IntegralChiG}
  &\int\D x\D x'\,
  \mean{\derivf{g}{U(\vec{r})}\,\derivf{g}{U(\vec{r}\,')}}^{(1)}
  \nonumber
  \\
  &=\int\D(\varepsilon-\varepsilon')\,\delta_T(\varepsilon-\varepsilon')\,
  \mean{g'(\varepsilon)g'(\varepsilon')}^{(1)}
  \nonumber
    \\
  &\simeq
  \frac{\pi}{15}\left(\frac{L_T}{L}\right)^6\frac{L_\varphi}{L}\tau_D^2
  \:.
\end{align}
This can be compared to the linear conductance fluctuations $\mean{\delta g^2}^{(1)}\simeq(\pi/3)(L_T/L)^2(L_\varphi/L)$~; each derivative $\delta/\delta U(r)$ is now responsible for a factor $(L_T/L)^2\tau_D=1/T$.

\vspace{0.25cm}

\paragraph{Spatial structure.--}

In order to analyse the spatial structure of the correlator we make two remarks~:
\begin{align}
  &\bra{x}\frac{1}{\gamma_\omega-\partial_x^2}\frac{1}{\gamma_\omega^*-\partial_x^2}\frac{1}{\gamma_\omega-\partial_x^2}\ket{x'}
  \nonumber\\
  =&-\derivp{}{\gamma_\omega}
  \bra{x}\frac{1}{\gamma_\omega-\partial_x^2}\frac{1}{\gamma_\omega^*-\partial_x^2}\ket{x'}
\end{align}
and 
\begin{align}
  &\bra{x}\frac{1}{\gamma_\omega-\partial_x^2}\frac{1}{\gamma_\omega^*-\partial_x^2}\ket{x'}
  \\\nonumber
  &=\frac{1}{\gamma_\omega^*-\gamma_\omega}
  \left[
     \bra{x}\frac{1}{\gamma_\omega-\partial_x^2}\ket{x'}
    -\bra{x}\frac{1}{\gamma_\omega^*-\partial_x^2}\ket{x'}
  \right]
  \:,
\end{align}
hence all contributions in Eq.~\eqref{eq:RootEqChiG} may be expressed in terms of the two propagators 
$P_{\omega}^{(d)}(x,x')=\bra{x}(\gamma_\omega-\partial_x^2)^{-1}\ket{x'}$ and $P_{-\omega}^{(d)}(x,x')=\bra{x}(\gamma_\omega^*-\partial_x^2)^{-1}\ket{x'}$.
After some algebra, we obtain that the bracket in Eq.~\eqref{eq:RootEqChiG} is 
\begin{align}
  \label{eq:ChiGbracket}
&\bigg\{\cdots\bigg\}
=
-4\left( \frac{P_{\omega}^{(d)}(x,x')-P_{-\omega}^{(d)}(x,x')}{\gamma_\omega-\gamma_\omega^*} \right)^2
\\\nonumber
&+\frac{1}{\gamma_\omega-\gamma_\omega^*}
\left[
    \derivp{ P_{\omega}^{(d)}(x,x') ^2}{\gamma_\omega}
  - \derivp{ P_{-\omega}^{(d)}(x,x') ^2}{\gamma_\omega^*}
\right]
\:.
\end{align}
An important observation is that the result of integration over frequency leads to a behaviour $\sim1/T$, different from the $1/T^3$ obtained for the spatial integral of the correlator, Eq.~\eqref{eq:IntegralChiG}. Precisely, we obtain that the correlator at coinciding point behaves as 
$\chi_g(\vec{r},\vec{r})\sim(\tau_D/L)^2(L_T/L)^2(L_\varphi/L)^4$.
The origin of this observation can be understood by writing formally the correlator as 
$
  \chi_g(\vec{r},\vec{r}) = \int\D\omega\,\delta_T(\omega)
  \,\Phi(|x-x'|;\omega)
$,
where the function $\Phi(|x-x'|;\omega)$, proportional to the bracket \eqref{eq:ChiGbracket}, has a width $\sim L_\varphi$ in space and a width $\sim1/\tau_\varphi$ in frequency.
The sum rule discussed above has revealed that the spatial integral of this function can be written as the second derivative of a function of the frequency~: 
$\int_0^\infty\D X\,\Phi(X;\omega)=f''(\omega)$, what is responsible for 
$\int\D(x-x')\,\chi_g(\vec{r},\vec{r}\,')=\int\D\omega\,\delta_T(\omega)\,f''(\omega)\sim1/T^3$.
In order to simplify the analysis of the limit $L_T\ll L_\varphi$, we split the correlator into two contributions as 
$\chi_g(\vec{r},\vec{r}\,')=\delta_T(0)\int\D\omega\,\Phi(|x-x'|;\omega)
+\int\D\omega\,\big[\delta_T(\omega)-\delta_T(0)\big]\,\Phi(|x-x'|;\omega)$.
The first term vanishes after spatial integration whereas the second term ensures the sum rule~\eqref{eq:IntegralChiG}.
In order to simplify the calculations, we assume that we can decouple the temperature dependence and the spatial decay over $L_\varphi$ in the second term, leading to the structure~:
\begin{align}
  \label{eq:SplittingChiG}
  &\chi_g(\vec{r},\vec{r}\,') \simeq 
  \frac{2}{3}\frac{\tau_D^2}{L^2}
  \left(\frac{L_T}{L}\right)^2\left(\frac{L_\varphi}{L}\right)^4
  \nonumber\\
  &\times
  \left[
    \phi\left(\frac{|x-x'|}{L_\varphi}\right)
    +\left(\frac{L_T}{L_\varphi}\right)^4\,
    \Upsilon\left(\frac{|x-x'|}{L_\varphi}\right)
  \right]
\end{align}
where $\phi$ and $\Upsilon$ are two dimensionless narrow functions of order unity, with the important property 
\begin{equation}
  \label{eq:PropertyPhi}
  \int_0^\infty\D u\,\phi(u)=0
  \:.
\end{equation}
The sum rule \eqref{eq:IntegralChiG} corresponds to 
\begin{equation}
  \label{eq:IntegralUpsilon}
  \int_0^\infty\D u\,\Upsilon(u)=\frac{\pi}{20}
  \:.
\end{equation}

\begin{figure}[!ht]
\centering
\includegraphics[scale=0.75]{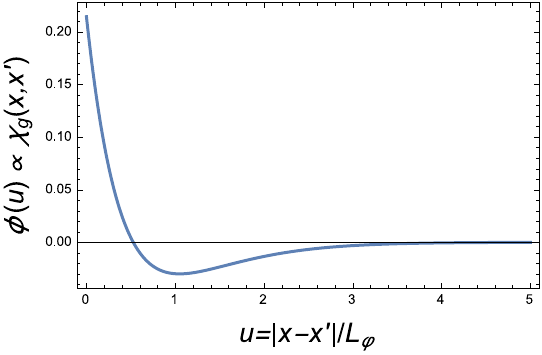}
\caption{\it Function \eqref{eq:DefPhi} controlling the correlator $\chi_g$ in the limit $L_T\ll L_\varphi$.}
  \label{fig:phiDEu}
\end{figure}

Let us now study more precisely the dimensionless function $\phi(u)$, defined by
\begin{align}
  \label{eq:DefPhi}
  \phi(u)
  &=\int_{-\infty}^{+\infty}\,\D\theta\,
  \bigg\{
  -4\left( \frac{Q_\theta(u)-Q_\theta^*(u)}{\Gamma_\theta-\Gamma_\theta^*} \right)^2
  \\\nonumber
  &-\frac{1}{\Gamma_\theta-\Gamma_\theta^*}
  \left[
    \frac{1+\sqrt{\Gamma_\theta}\,u}{\Gamma_\theta}\,Q_\theta(u)^2 
    - 
    \mathrm{c.c.}
  \right]
  \bigg\}
  \:,
\end{align}
where $\Gamma_\theta=1-\I\theta$ and $Q_\theta(u)=\EXP{-\sqrt{\Gamma_\theta}u}/(2\sqrt{\Gamma_\theta})$.
The value at the origin can be computed explicitely
\begin{align}
  \phi(0) = \int_0^\infty\frac{\D\theta}{1+\theta^2}
  \left(  \frac{1}{1+\theta^2} - \frac{\sqrt{1+\theta^2}-1}{\theta^2}
  \right)
  =1-\frac{\pi}{4}
  \:.
\end{align}
The function is represented in Fig.~\ref{fig:phiDEu}.

\section{Correlations of injectivities ($\chi_\DoS$)}
\label{sec:InjCorr}

The analysis of the injectivity correlator \eqref{eq:DefChiU} is quite similar to the analysis of the  conductance correlations, as the injectivity can be expressed in terms of Green's functions as  
\begin{equation}
  \label{eq:Injectivity}
  \DoS_1(\vec{r};\varepsilon) = 2_s\sum_{n=1}^{\Nc} \frac{v_n}{2\pi}
  \bigg|
    \int\D y'\,\chi_n(y')\,G^\mathrm{R}(\vec{r},\vec{r}\,';\varepsilon)
  \bigg|^2
\end{equation}
where $x'=0$ and $v_n$ is the group velocity in channel $n$.
The injectivity thus involves a pair of Green's functions starting from the boundary, as the conductance in  the Fisher and Lee relation~\eqref{eq:GfromFisherLee}.

\subsection{Preliminary~: Averaged injectivity}

As a first simple illustration, we analyse the mean injectivity.
We use the rule \eqref{eq:Rule2} and take into account the additional factor $2_s/(2\pi)$~:
\begin{align}
  &\mean{\DoS_1(\vec{r};\varepsilon)}
  =\frac{2_s}{2\pi}\sum_n\diagram{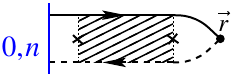}{0.75}{-0.4cm}
  \\
  &= \frac{2_s}{2\pi} \,\frac{1}{\tau_e}\frac{P_d(x,\ell_e)}{\ell_e}\,2\pi\rho_0\tau_e
  =\DoS_0\, \frac{P_d(x,\ell_e)}{\ell_e}
  \:,
\end{align}
where the termination has involved the box
\begin{equation}
  \diagram{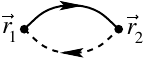}{0.75}{-0.25cm}
  \simeq 2\pi\rho_0\tau_e\,\delta(\vec{r}_1-\vec{r}_2)
  \:.
\end{equation}
In the wire, using \eqref{eq:PdWire}, we recover the expression given in Ref.~\onlinecite{GraBut99}~:
\begin{equation}
  \label{eq:MeanInjectivity}
  \mean{\DoS_1(\vec{r};\varepsilon)} = \DoS_0\,\left(1 - \frac{x}{L}\right)
  \:.
\end{equation}
A similar analysis gives the second injectivity $\mean{\DoS_2(\vec{r};\varepsilon)}=\DoS_0\,{P_d(x,L-\ell_e)}/{\ell_e}=\DoS_0\,x/L$.
We can check the sum rule~\eqref{eq:SumRuleInjectivities}.

\subsection{DoS correlations}
\label{subsec:DoScorr}

Before going to the more complicated matter of injectivity correlations, let us recall the expression of the DoS correlator. Considering only long range diagrams, we have~\cite{AkkMon07}~:
  \begin{align}
    \label{eq:DoSCorrelartions}
      &\frac{\mean{ \DoS(\vec{r};\varepsilon)\,\DoS(\vec{r}\,';\varepsilon') }_c}{\DoS_0^2}
      \\\nonumber
      &=\diagram{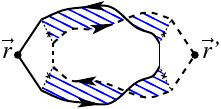}{0.7}{-0.5cm}+\mathrm{c.c.}
      +\Big(P^{(d)}\to P^{(c)}\Big)
      \\\nonumber
    &= 2\left(\frac{2\pi}{\xiloc}\right)^2
   \re\left[
      P_\omega^{(d)}(x,x')^2+P_\omega^{(c)}(x,x')^2
      \right]
   \:,
  \end{align}
  where $\omega=\varepsilon-\varepsilon'$ and $\xiloc$ is the localisation length, Eq.~\eqref{eq:LocalisationLength}.
The box with three corners was given above, Eq.~\eqref{eq:H3}.
In a coherent wire of length $L$, we estimate $P_\omega^{(d)}(x,x')\to P_d(x,x')\sim L$, thus
\begin{equation}
    \label{eq:DoSCorrelartions2}
    \frac{\mean{ \DoS(\vec{r};\varepsilon)\,\DoS(\vec{r}\,';\varepsilon') }_c}{\DoS_0^2}
    \sim \left(\frac{L}{\xiloc}\right)^2 = \frac{1}{g^2}
    \ll1
    \:.
\end{equation}
These correlations are small as the validity of the diagrammatic approach describes the diffusive regime $L\ll\xiloc$.
In the weakly coherent wire, $L_\varphi\ll L$, the correlations decay exponentially over the scale~$L_\varphi$.

\subsection{General expression of the injectivity correlator}

The injectivity correlator 
$\mean{\DoS_1(\vec{r};\varepsilon)\DoS_1(\vec{r}\,';\varepsilon')}_c$ is given by the diagrams shown in Fig.~\ref{fig:InjectivityCorr12} (note that the two injectivities could also be correlated with only one Hikami box, however this leads to a contribution reduced by a factor $\ell_e/L$).~\cite{footnote8} 
Hence, the analysis is quite similar to the conductance correlator, with however two differences~: 
(\textit{i})
the termination of the Green's functions line is a specific coordinate and not a contact (Fig.~\ref{fig:InjectivityCorr12}).
(\textit{ii})
One must take into account the additional factor $2_s/(2\pi)$ per injectivity, see Eq.~\eqref{eq:Injectivity}.
As the question of symmetrisation with respect to magnetic field reversal will be of importance here, we consider the two first contributions arising from Diffuson and Cooperon at the same time (two  diagrams of Fig.~\ref{fig:InjectivityCorr12}).
The application of the simple rules (\ref{eq:Rule1},\ref{eq:Rule2},\ref{eq:Rule3}) with the additionnal factor $[2_s/(2\pi)]^2$ for the pair of injectivities gives~:
\begin{widetext} 
\begin{align}
  \label{eq:InjectivityCorr12}
    \chi_\DoS(\vec{r},\vec{r}\,')^{(1)+(2)} =
    \frac{4}{\xiloc^2}
  \int\D\omega\,\delta_T(\omega)
  \int\D\xi\D\xi'
  \,  
  \bigg[
     \partial_{\xi}P_d(x,\xi)\,
     \partial_{\xi}P_d(x',\xi)\,
     \left( \frac{\partial_{\xi'}P_d(\xi',\ell_e)}{\ell_e}\right)^2
     &\left| P_\omega^{(d)}(\xi,\xi') \right|^2
     \nonumber\\
   +\, \partial_{\xi}P_d(x,\xi)\,
     \partial_{\xi'}P_d(x',\xi')\,
     \frac{\partial_{\xi}P_d(\xi,\ell_e)}{\ell_e}
     \frac{\partial_{\xi'}P_d(\xi',\ell_e)}{\ell_e}
     &\left| P_\omega^{(c)}(\xi,\xi') \right|^2
  \bigg]
  \:,
\end{align}
where the localisation length $\xiloc$ was defined above, Eq.~\eqref{eq:LocalisationLength}.
We use that $\xi,\:\xi'>\ell_e$ and now consider the case of the diffusive wire characterised by Eq.~\eqref{eq:PdWire}. Thus we can simplify the expression in brackets by making use of 
$\partial_{\xi}P_d(\xi,\ell_e)/\ell_e=-1/L$.
 
  Symmetrisation/antisymmetrisation with respect to magnetic field reversal follows from the discussion of section~\ref{subsec:SUMFR}. We recall that  $P_\omega^{(d)}$ is a function of $\mathcal{B}-\mathcal{B}'$ and $P_\omega^{(c)}$ a function of $\mathcal{B}+\mathcal{B}'$.
    Thus the change $\mathcal{B}'\leftrightarrow-\mathcal{B}'$ implies
    $P_\omega^{(d)}\leftrightarrow P_\omega^{(c)}$.
Some algebra eventually gives
  \begin{align}
  \label{eq:ACentralResult}
    \chi_\DoS^{s,a}(\vec{r},\vec{r}\,')^{(1)+(2)} &=
    \frac{4}{\xiloc^2}
    \int\D\omega\,\delta_T(\omega)\,  
    \int\frac{\D\xi\D\xi'}{L^2}
  \\\nonumber
  &\times
    \frac{\partial_{\xi}P_d(x,\xi)\pm\partial_{\xi'}P_d(x,\xi')}{2}
    \frac{\partial_{\xi}P_d(x',\xi)\pm\partial_{\xi'}P_d(x',\xi')}{2}
     \left(
     \left| P_\omega^{(d)}(\xi,\xi') \right|^2
    \pm  
     \left| P_\omega^{(c)}(\xi,\xi') \right|^2
  \right)
  \:.
  \end{align}
We emphasize that the existence of a non-zero antisymmetric part $\chi_\DoS^{a}$ crucially depends on the fact that the ``external'' Diffusons have different configurations in the two diagrams of Fig.~\ref{fig:InjectivityCorr12}~:
in the first case the two Diffusons starting from the boundary $x=0$ reach the same Hikami box whereas in the second case they end at two different Hikami boxes.
The expression allows one to discuss the order of magnitude of the injectivity correlations~: 
setting $T=0$ and considering the coherent limit, we estimate $\chi_\DoS^{s,a}(\vec{r},\vec{r}\,')\sim \xiloc^{-2} (P_d)^2\sim (L/\xiloc)^2\sim1/g^2$, like the DoS fluctuations \eqref{eq:DoSCorrelartions2}, as expected.

\begin{figure}[!ht]
\centering
\includegraphics[width=0.4\textwidth]{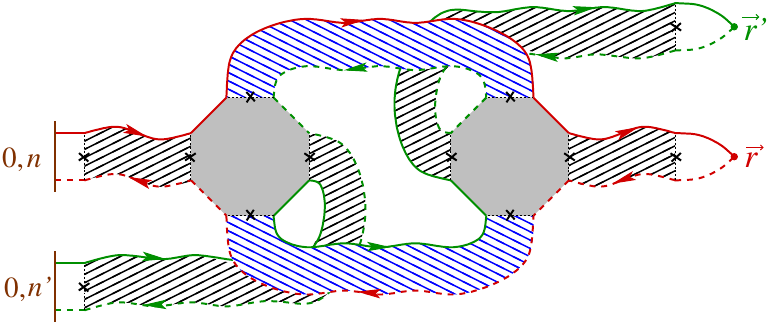}
\hfil
\includegraphics[width=0.4\textwidth]{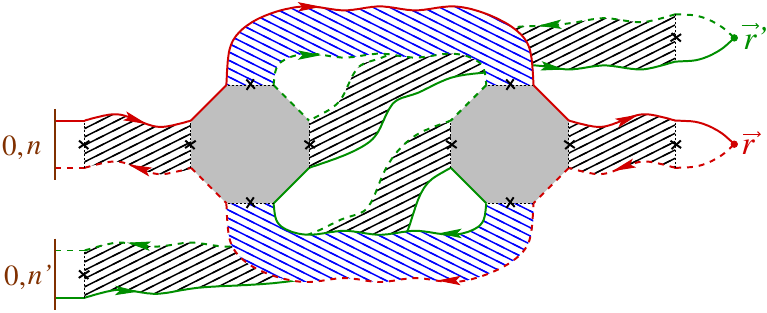}
\caption{(color online) \it The two contributions to the injectivity correlations which produce the antisymmetry in magnetic field.}
\label{fig:InjectivityCorr12}
\end{figure}

  
We now consider the other contributions, which are similar to the diagrams of Fig.~\ref{fig:ConductanceCorr34} (with terminations on the right now corresponding to the two coordinates $\vec{r}$ and $\vec{r}\,'$)~:
\begin{align}
  \label{eq:ChiU3and4Genuine}
  \chi_\DoS(\vec{r},\vec{r}\,')^{(3)+(4)} &=
  \frac{2}{\xiloc^2}
  \int\D\omega\,\delta_T(\omega)\int\D\xi\D\xi'\,  
     \\ \nonumber
     &\times
     \partial_{\xi}P_d(x,\xi)\,
     \partial_{\xi'}P_d(x',\xi')\,
     \frac{\partial_{\xi}P_d(\xi,\ell_e)}{\ell_e}
     \frac{\partial_{\xi'}P_d(\xi',\ell_e)}{\ell_e}
     \left(
         \re\left[ P_\omega^{(d)}(\xi,\xi')^2 \right]
       + \re\left[ P_\omega^{(c)}(\xi,\xi')^2 \right]
     \right)
     \:.
\end{align}
We use again the expression of the Diffuson in the wire, Eq.~\eqref{eq:PdWire}, in order to simplify the expression as 
\begin{align}
  \label{eq:ChiU3and4}
  \chi_\DoS(\vec{r},\vec{r}\,')^{(3)+(4)} =
  \frac{2}{\xiloc^2}
  \int\D\omega\,\delta_T(\omega)\int\frac{\D\xi\D\xi'}{L^2}\,  
     &
     \partial_{\xi}P_d(x,\xi)\,
     \partial_{\xi'}P_d(x',\xi')\,
     \left(
         \re\left[ P_\omega^{(d)}(\xi,\xi')^2 \right]
       + \re\left[ P_\omega^{(c)}(\xi,\xi')^2 \right]
     \right)
     \:.
\end{align}  
\end{widetext}
Eqs.~(\ref{eq:ACentralResult},\ref{eq:ChiU3and4}) are central results.
An important difference with the two first contributions is that the ``external'' Diffusons now factorize.
This has important consequences for the symmetrisation as
\begin{align}
\chi_\DoS^s(\vec{r},\vec{r}\,')^{(3)+(4)} &= \chi_\DoS(\vec{r},\vec{r}\,')^{(3)+(4)} 
 \\
\chi_\DoS^a(\vec{r},\vec{r}\,')^{(3)+(4)} &= 0
\:.
\end{align}

\vspace{0.25cm}

\paragraph*{Remark~:}
 
In order to avoid the spurious divergences produced by using the expression of the box \eqref{eq:Rule3} originally derived by Hikami~\cite{Hik81}, the derivation of conductance correlations (Figs.~\ref{fig:ConductanceCorr12} and~\ref{fig:ConductanceCorr34}) has involved the procedure proposed by Kane, Serota and Lee (KSL) \cite{KanSerLee88}, which relies on current conservation condition for the non-local conductivity tensor (we refer to Ref.~\onlinecite{TexMon04} for a discussion of the weak localisation case).
The origin of these unphysical divergences comes from the fact that the organisation of the perturbation theory with Diffusons/Cooperons and Hikami boxes does not automatically fulfill elementary conservation laws such as current conservation, which should be imposed through Ward identities (see also Ref.~\onlinecite{HasStoBar94} where the set of current conserving diagrams is constructed for the weak localisation). 
Using the similarity between conductance correlation diagrams (Fig.~\ref{fig:ConductanceCorr12}) and injectivity correlation diagrams (Fig.~\ref{fig:InjectivityCorr12}), we have used the same prescription in this latter case.
This point would however deserves a more rigorous justification.

\subsection{Symmetric part}

In order to calculate the spatial integrals in \eqref{eq:ACentralResult} we use the decoupling between the long range Diffuson $P_d(\cdot,\cdot)$ and the short range Diffuson $P_\omega^{(d)}(\xi,\xi')$, which constraints $\xi\approx\xi'$.
Therefore we may rewrite the Diffuson contribution of \eqref{eq:ACentralResult} as 
  \begin{align}
   \chi_\DoS^{s}(\vec{r},\vec{r}\,')^{(1)} & \simeq
    \frac{4}{\xiloc^2}
    \int\D\omega\,\delta_T(\omega)\,  
  \nonumber\\
  &\times
    \int_0^L\frac{\D\xi}{L^2}\,
    \partial_{\xi}P_d(x,\xi)\,\partial_{\xi}P_d(x',\xi)\,
  \nonumber\\
  &\times
    \int\D(\xi-\xi')\,
     \left| P_\omega^{(d)}(\xi,\xi') \right|^2
  \:.
  \end{align}
We now introduce the useful property
\begin{align}
  \label{eq:UsefulIntegralPd}
  \int_0^L\D\xi\,\partial_\xi P_d(x,\xi)\,\partial_\xi P_d(\xi,x') = P_d(x,x')
\end{align}
whose proof simply follows from an integration by parts. 
Using \eqref{eq:UsefulIntegralPd}, we find
\begin{align}
  \label{eq:IntermediateChiU1}
  \chi_\DoS^{s}(\vec{r},\vec{r}\,')^{(1)}
  &\simeq \left(\frac{L}{\xiloc}\right)^2 \frac{P_d(x,x')}{L}
  \\\nonumber
  &\times\frac{1}{L^3}\int\D\omega\,\delta_T(\omega)\,  
  \frac{2}{|\gamma_\omega|\left(\sqrt{\gamma_\omega}+\sqrt{\gamma_\omega^*}\right)}
  \:.
\end{align}
Performing the same approximation in \eqref{eq:ChiU3and4}, we obtain 
\begin{align}
  \label{eq:IntermediateChiU3}
  \chi_\DoS(\vec{r},\vec{r}\,')^{(3)}
  &\simeq  \frac{1}{2}
      \left(\frac{L}{\xiloc}\right)^2
      \frac{P_d(x,x')}{L}
  \\\nonumber
  &\times\frac{1}{L^3}\int\D\omega\,\delta_T(\omega)\,  
  \re\left[ \gamma_\omega^{-3/2} \right]
  \:.
\end{align}

The injectivity correlations are thus \textit{long range}.

\subsubsection{Limit $L_\varphi\ll L_T,\, L$}

When one can neglect the effect of thermal broadening, one simply performs the substitution 
$\delta_T(\omega)\to\delta(\omega)$ in the previous integral. One gets
\begin{equation}
  \label{eq:Chi2sLargeLT}
      \chi_\DoS^{s}(\vec{r},\vec{r}\,')^{(1)}
      \simeq 
      \left(\frac{L}{\xiloc}\right)^2
      \left(\frac{L_\varphi}{L}\right)^3
      \frac{P_d(x,x')}{L}
   \:.
\end{equation}
The second Diffuson contribution \eqref{eq:IntermediateChiU3} obviously leads to the same result, up to a factor $1/2$, therefore the correlator in the high magnetic field regime is 
\begin{equation}
  \label{eq:ChiU1plus3}
  \chi_\DoS^s(\vec{r},\vec{r}\,')^{(1)+(3)} = \frac{3}{2}\,\chi_\DoS^s(\vec{r},\vec{r}\,')^{(1)}
  \:.
\end{equation}

In the coherent limit $L\sim L_\varphi$, we see that the injectivity correlations are of the same order as the DoS correlations, Eq.~\eqref{eq:DoSCorrelartions2}, as expected.

\subsubsection{Limit $L_T \ll L_\varphi\ll  L$}

In this case, we treat $\delta_T(\omega)$ as a ``broad'' function. We perform
the substitution $\delta_T(\omega)\to\delta_T(0)=1/(6T)=L_T^2/(6D)$ in Eq.~\eqref{eq:IntermediateChiU1}.
The remaining integral over frequency is 
\begin{align}
  &\int \frac{\D\omega}{D}
  \frac{1}{|\gamma_\omega|\left(\sqrt{\gamma_\omega}+\sqrt{\gamma_\omega^*}\right)}
  \nonumber\\
  &= L_\varphi \int \frac{\D\omega}{\sqrt{2(\omega^2+1)(1+\sqrt{\omega^2+1}})}
  = L_\varphi \pi 
  \:.
\end{align}
In conclusion we get
\begin{equation}
  \label{eq:Chi2sSmallLT}
      \chi_\DoS^{s}(\vec{r},\vec{r}\,')^{(1)}
      \simeq \frac{\pi}{3} 
      \left(\frac{L}{\xiloc}\right)^2
      \left(\frac{L_T}{L}\right)^2\frac{L_\varphi}{L}\,
      \frac{P_d(x,x')}{L}
 \:.
\end{equation}
The two regimes therefore lead to the same behaviour,  \eqref{eq:Chi2sLargeLT} and \eqref{eq:Chi2sSmallLT}, up to the substitution $(L_\varphi/L)^2\to(L_T/L)^2$.
In this regime, the contribution \eqref{eq:IntermediateChiU3} is much smaller [it vanishes within the same approximation $\delta_T(\omega)\to\delta_T(0)$] and we simply have
\begin{equation}
  \chi_\DoS^s(\vec{r},\vec{r}\,')^{(3)} \simeq 0
  \:.
\end{equation}

\subsection{Antisymmetric part}

The study of the antisymmetric part of the injectivity correlator requires a more precise analysis.
We repeat that the existence of a finite $\chi_\DoS^{a}$ crucially relies on the spatial structure of the ``external'' Diffusons in the correlator of injectance (Fig.~\ref{fig:InjectivityCorr12}).
We can use that the weight in \eqref{eq:ACentralResult} is 
\begin{equation}
  \frac{\partial_{\xi}P_d(x,\xi)-\partial_{\xi'}P_d(x,\xi')}{2}
  =\frac{\heaviside(x-\xi)-\heaviside(x-\xi')}{2}
  \:,
\end{equation}
where $\heaviside(x)$ is the Heaviside function.
For the antisymmmetric part, this constraints the integral \eqref{eq:ACentralResult} as 
\begin{align}
  \label{eq:ChiAconvenientRepresentation}
  &\chi_\DoS^{a}(\vec{r},\vec{r}\,')^{(1)+(2)}  =
    \frac{2}{\xiloc^2}
    \int\D\omega\,\delta_T(\omega)  
    \\\nonumber
    &\times\int_0^{x_<}\frac{\D\xi}{L}\int_{x_>}^L\frac{\D\xi'}{L}\,
     \left(
     \left| P_\omega^{(d)}(\xi,\xi') \right|^2
    -
     \left| P_\omega^{(c)}(\xi,\xi') \right|^2
  \right)
  \:,
\end{align}
where $x_<=\min{x}{x'}$ and $x_>=\max{x}{x'}$. Eq.~\eqref{eq:ChiAconvenientRepresentation} is one of the key result of the paper.
We now restrict ourselves to the contribution of the Diffuson. Using \eqref{eq:DiffusonInfiniteWire} one gets
\begin{align}
  \label{eq:ChiUAinetermediate}
  \chi_\DoS^{a}(\vec{r},\vec{r}\,')^{(1)}
  \simeq 
  2 \left(\frac{L}{\xiloc}\right)^2
  \frac{1}{L^4} 
  \int\D\omega\,\delta_T(\omega) 
  \frac{\left| P_\omega^{(d)}(x,x') \right|^2}
       {\left(\sqrt{\gamma_\omega}+\sqrt{\gamma_\omega^*}\right)^2}
  \:.
\end{align}
We recall that $\chi_\DoS^{a}(\vec{r},\vec{r}\,')^{(3)+(4)}=0$.

\subsubsection{$L_\varphi\ll L,\: L_T$}

Neglecting thermal broadening, i.e. performing $\delta_T(\omega)\to\delta(\omega)$, one gets 
\begin{equation}
  \label{eq:Chi2aLargeLT}
  \chi_\DoS^{a}(\vec{r},\vec{r}\,')^{(1)}  \simeq 
  \frac{1}{8} \left(\frac{L}{\xiloc}\right)^2
  \left(\frac{L_\varphi}{L}\right)^4\,\EXP{-2|x-x'|/L_\varphi}
  \:.
\end{equation}
In contrast with the symmetric part \eqref{eq:Chi2sLargeLT}, which is long range, the antisymmetric part of the injectivity correlator is \textit{short range}.

\subsubsection{$L_T\ll L_\varphi\ll L$}

In this regime, we can perform the substitution $\delta_T(\omega)\to\delta_T(0)=1/(6T)=L_T^2/(6D)$ in \eqref{eq:ChiUAinetermediate}, what leads to the structure~:
\begin{align}
  \label{eq:Chi2aSmallLT}
  \chi_\DoS^{a}(\vec{r},\vec{r}\,')^{(1)} 
  \simeq
  \frac{1}{3}
  &\left(\frac{L}{\xiloc}\right)^2
  \left(\frac{L_T}{L}\right)^2
  \left(\frac{L_\varphi}{L}\right)^2
  \psi\left(\frac{|x-x'|}{L_\varphi}\right)
  \:,  
\end{align}
where the dimensionless function is 
\begin{align}
  \psi(u) = \int_{-\infty}^{+\infty}\,\D\theta\,
  \frac{|Q_\theta(u)|^2}{\left(\sqrt{\Gamma_\theta}+\sqrt{\Gamma_\theta^*}\right)^2}
  \:,
\end{align}
with $\Gamma_\theta=1-\I\theta$ and $Q_\theta(u)=\EXP{-\sqrt{\Gamma_\theta}u}/(2\sqrt{\Gamma_\theta})$.
A change of variable leads to the convenient integral representation
\begin{align}
  \label{eq:DefPsiDEu}
  \psi(u) &= \frac18
  \int_1^\infty\D y\,\frac{\EXP{-2uy}}{y^2\sqrt{y^2-1}}
  \\
   & 
  \begin{cases}
    =\frac18 & \mbox{if } u=0 \\
    \displaystyle
    \simeq\frac{1}{16}
    \sqrt{\frac{\pi}{u}}\,\EXP{-2u}
    & \mbox{if } u\gg 1
  \end{cases}
  \:.
\end{align}

\section{Correlations of the non-linear conductance}
\label{sec:NLC}

We can now combine the results of Sections~\ref{sec:CondCorr} and~\ref{sec:InjCorr} in order to derive the correlations of the non-linear conductance for weakly disordered wires.
We refer to the notation introduced in Subsection~\ref{subsec:SUMFR}.

We first remark that due to \eqref{eq:Property1}, we have $\mean{\delta g/\delta U(\vec{r})}=0$ and therefore
\begin{equation}
   \mean{ \Gnonint }
  =\mean{ \Gint }
  =0
  \:,
\end{equation}
thus $\mean{ \mathcal{G}_{s,a} }=0$.
As a consequence, the non-linear conductance can only be characterised through its correlator and we will be interested below in $\smean{ \mathcal{G}_{s,a} ^2 }$.

\subsection{Preliminary~: coherent QD}
\label{subsec:PreliminaryQD}

We briefly come back to the case of coherent quantum dots, what will be helpful in order to clarify the future calculations in diffusive wires.
In the ergodic regime, a great simplification used in Refs.~\onlinecite{SanBut04,PolBut06,PolBut07a} is to neglect all spatial dependences.
In particular, at $T=0$, Eq.~\eqref{eq:ButtikerChristen} is simplified as~\cite{PolBut06,PolBut07a}
$
  g_{\alpha\beta\gamma}
  = (1/2)\big[ 
    g_{\alpha\beta}'(\varepsilon_F)
    \delta_{\beta\gamma}
    - g_{\alpha\beta}'(\varepsilon_F)\, u_\gamma
    - g_{\alpha\gamma}'(\varepsilon_F)\,u_\beta
  \big]
$.
In the two-terminal configuration, we write $g=-g_{21}$ and $\mathcal{G}=\mathcal{G}_s+\mathcal{G}_a=-g_{211}$ and get
\begin{align}
  \mathcal{G}_s =
  g'(\varepsilon_F)
  \left( u_1^{s} - \frac{1}{2} \right)
  \quad\mbox{and}\quad
  \mathcal{G}_a =
  g'(\varepsilon_F)\,u_1^{a}
\end{align}
where $u_1^{s,a}=\big[u_1(\mathcal{B})\pm u_1(-\mathcal{B})\big]/2$.
As a result, using that $\smean{g'(\varepsilon_F)\,u_1}=0$ and $\smean{u_1^a}=0$, the fluctuations are 
\begin{align}
  \label{eq:127}
  \smean{  \mathcal{G}_s ^2 } 
  &= \mean{ g'(\varepsilon_F)^2 }
  \left[
     \left(\frac12 - \mean{u_1}\right)^2 + \mean{\big(\delta u_1^s\big)^2}
  \right]
  \\
  \label{eq:128}
   \smean{  \mathcal{G}_a ^2 } 
  &= \mean{ g'(\varepsilon_F)^2 } \mean{\big(u_1^a\big)^2}
  \:.
\end{align}
Starting from these formulae, B\"uttiker, Polianski and S\'anchez~\cite{SanBut04,PolBut06,PolBut07a} have obtained the results recalled in the introduction, Eqs.~(\ref{eq:Polianski2007s},\ref{eq:Polianski2007a}).
DoS correlations are small, $\mean{\delta u_1^2}\sim1/g^2$, hence the symmetric part is dominated by the first term 
$\smean{ \left(\mathcal{G}_s\right)^2 } 
\simeq 
\mean{ g'(\varepsilon_F)^2 }\big(1/2-\mean{u_1}\big)^2$.
The presence of the characteristic potential encodes the strong renormalisation of the potential inside the QD due to screening, as $u_1\sim1$ (free electron result is recovered by setting $u_1=0$).
If the QD has contacts characterised by $N_1$ and $N_2$ channels, we have $\mean{u_1}=\gamma_\mathrm{int}\,N_1/N$ (see for example Refs.~\onlinecite{PolBut07a,Tex16})~; this shows that the dominant term vanishes for perfect screening ($\gamma_\mathrm{int}=1$) and symmetric contacts $N_1=N_2$.

\subsection{Metal in the diffusive regime}
\label{subsec:Structure}

\begin{widetext}
In the diffusive devices, the fluctuations present structures analogous to (\ref{eq:127},\ref{eq:128}) with additional spatial integrations~:
\begin{align}
  \label{eq:g211sDW}
   &\smean{  \mathcal{G}_s ^2 } 
  = 
   \int\D\vec{r}\D\vec{r}\,'\, 
  \mean{ \derivf{g}{U(\vec{r})} \derivf{g}{U(\vec{r}\,')} }
  \left[
     \left(\frac12 - \mean{u_1(\vec{r})}\right)
     \left(\frac12 - \mean{u_1(\vec{r}\,')}\right)
     + \mean{\delta u_1^s(\vec{r})\delta u_1^s(\vec{r}\,')}
  \right]
  \\
  \label{eq:g211aDW}
   &\smean{  \mathcal{G}_a ^2 } 
  = 
   \int\D\vec{r}\D\vec{r}\,'\, 
  \mean{ \derivf{g}{U(\vec{r})} \derivf{g}{U(\vec{r}\,')} }
  \,
  \mean{u_1^a(\vec{r})\, u_1^a(\vec{r}\,')}
  \:,
\end{align}
where $u_1^{s,a}(\vec{r})$ denotes the symmetric and antisymmetric parts of the injectivity with respect to magnetic field reversal and $\delta u_1(\vec{r})=u_1(\vec{r})-\smean{u_1(\vec{r})}$ the fluctuating part (sample to sample fluctuations).
\end{widetext}
Our purpose is now to analyse the two correlators (\ref{eq:g211sDW},\ref{eq:g211aDW}) in the different regimes, what will require a detailed analysis of the several contributions.

\subsection{Without interaction}

We first consider the non-interaction part of the non-linear conductance, which is symmetric with respect to magnetic field reversal~:
this corresponds to the term $(1/2)^2$ of the bracket $[\cdots]$ in Eq.~\eqref{eq:g211sDW}.
As pointed out above, the fluctuations can be written in terms of the correlator studied in Section~\ref{sec:CondCorr}~:
\begin{equation}
  \label{eq:CorrNoninteracting}
 \smean{ \Gnonint^2 }
 = \frac{1}{4} \int\D x\D x'\, \chi_g(\vec{r},\vec{r}\,')
 \:.
\end{equation}

\subsubsection{$L_\varphi\ll L,\: L_T$}

Integration of the correlator \eqref{eq:Chi1LargeLT} gives
\begin{equation}
  \smean{ \Gnonint ^2}^{(1)}
  \simeq 
  \frac{5}{8}   \left(\frac{L_\varphi}{L}\right)^7 \tau_D^2
\end{equation}
where $\tau_D=L^2/D$ is the Thouless time.
We obtain the high magnetic field result by multiplying the expression by $3/2$~:
\begin{equation}
  \label{eq:FluctGOlowT}
  \smean{ \Gnonint ^2}^{(1)+(3)}
  \simeq 
 \frac{15}{16}   
 \left(\frac{L_\varphi}{L}\right)^7 \tau_D^2
 \:.
\end{equation}
Note that this result, which also characterizes the behaviour of the differential conductance $\smean{\big[\DifG(V)-\DifG(0)\big]^2}\simeq(2eV)^2\smean{ \mathcal{G}^2}$ for low voltage, $eV\ll1/\tau_\varphi$, presents the same power of $L_\varphi$ than the differential conductance at high voltage,  $eV\gg1/\tau_\varphi$, obtained by LK~\cite{LarKhm86,KhmLar86} $\smean{\delta\DifG(V)^2}\sim(eV/\EThouless)(L_\varphi/L)^7$ (see also Ref.~\onlinecite{LudBlaMir04} and Section~\ref{subsec:KLDiffCond}).

\subsubsection{$L_T\ll L_\varphi\ll L$}

Using \eqref{eq:CorrNoninteracting} and \eqref{eq:IntegralChiG} we can immediatly write~:
\begin{equation}
  \label{eq:FluctGOhighT}
  \smean{ \Gnonint ^2}^{(1)}
    \simeq 
   \frac{\pi}{60} \left(\frac{L_T}{L}\right)^6\frac{L_\varphi}{L}\, \tau_D^2
   \:,
\end{equation}
which is therefore the high field result as $\smean{ \Gnonint ^2}^{(3)}$ is negligible in this case.

\subsubsection{Correlations at different Fermi energies}
\label{subsec:CorrelationG0}

Note that the correlations at different Fermi energies can be simply characterised~:
we simply have $\smean{ \Gnonint(\varepsilon_F)\Gnonint(\varepsilon_F-\omega)}=-(1/4)\,\mathscr{C}''(\omega)$ where the conductance correlator $\mathscr{C}(\omega)$ is given by Eqs.~(\ref{eq:CorrelCondOmega},\ref{eq:CorrelCondOmegaIncoh}).
The correlator thus changes in sign and presents a negative tail
\begin{equation}
  \smean{ \Gnonint(\varepsilon_F)\Gnonint(\varepsilon_F-\omega)}
  \simeq -(45\sqrt{2}/16)\EThouless^{3/2}|\omega|^{-7/2}
\end{equation}
for $\omega\gg\mathrm{max}(1/\tau_\varphi,1/\tau_D)$.

\subsection{With interaction}
\label{subsec:Section7D}

\begin{figure}[!ht]
\centering
\includegraphics[width=0.4\textwidth]{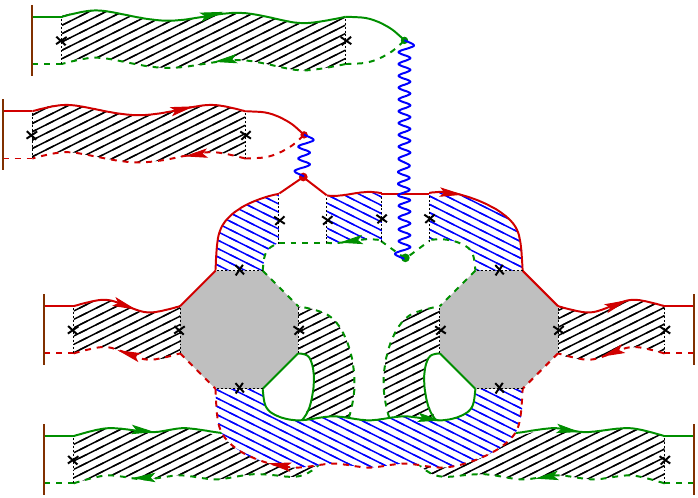}
\caption{(color online) \it One contribution to the non-linear conductance correlations $\smean{\left(\Gint\right)^2}_\mathrm{uncorr}$, Eq.~\eqref{eq:UncorrelInterPart}.}
\label{fig:NonlinearDiagramUncorr}
\end{figure}

\begin{figure}[!ht]
\centering
\includegraphics[width=0.4\textwidth]{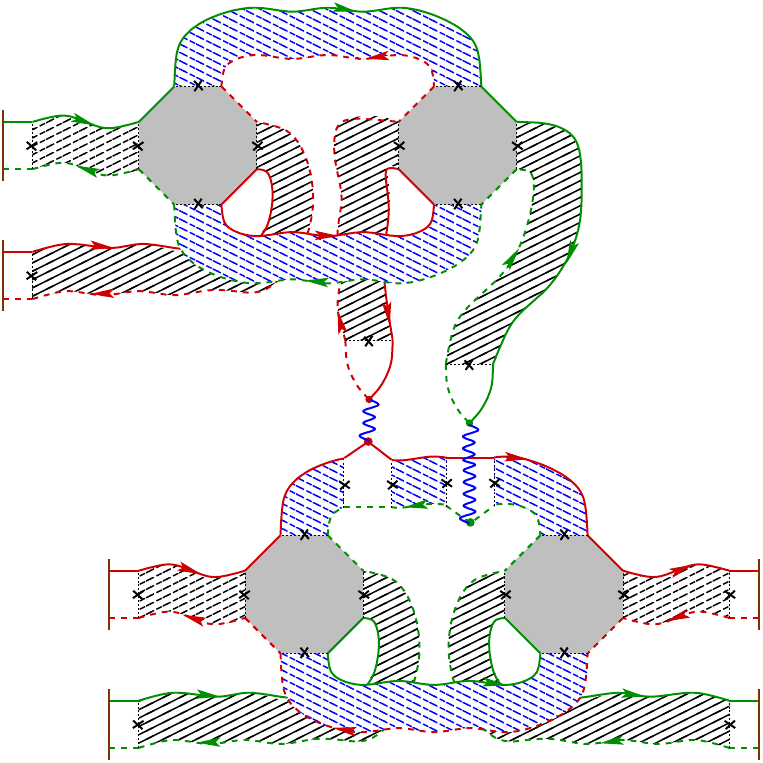}
\caption{(color online) \it One contribution to the non-linear conductance correlations $\smean{\left(\Gint\right)^2}_\mathrm{corr}$, Eq.~\eqref{eq:NLCproductCorr}, among the $6\times36=216$ diagrams.}
\label{fig:NonlinearDiagram}
\end{figure}

The interaction part of the non-linear conductance involves a product of $\delta g/\delta U(\vec{r})$ by an injectivity.
Considering $\smean{(\Gint)^2}$, the two conductance's functional derivatives must necessarily be correlated because $\mean{\delta g/\delta U(\vec{r})}=0$, which follows from \eqref{eq:Property1}.
On the other hand the possibility to correlate or not the two injectivities give rise to two contributions~:
\begin{equation}
  \smean{\left(\Gint\right)^2}
  = \smean{\left(\Gint\right)^2}_\mathrm{uncorr}
  + \smean{\left(\Gint\right)^2}_\mathrm{corr}
  \:,
\end{equation}
corresponding to the case of uncorrelated (Fig.~\ref{fig:NonlinearDiagramUncorr}) and correlated injectivities (Fig.~\ref{fig:NonlinearDiagram}), respectively (in Section~\ref{sec:MainResults}, we have used the notation $\smean{\left(\Gint\right)^2}_\mathrm{corr}\equiv\smean{\big(\mathcal{G}^\mathrm{int,\,fluc}\big)^2}$).
Because the asymmetry in magnetic field arises from the injectivity, we have $\smean{\mathcal{G}_a^2}_\mathrm{uncorr}=0$.
The first contribution reads (Fig.~\ref{fig:NonlinearDiagramUncorr})
\begin{align}
  \label{eq:UncorrelInterPart}
    \smean{ \left(\Gint_s\right)^2 }_\mathrm{uncorr}
    = \int\D\vec{r}\D\vec{r}\,'\,\chi_g(\vec{r},\vec{r}\,')\,
    \frac{\mean{\DoS_1(\vec{r};\varepsilon_F)}\mean{\DoS_1(\vec{r}\,';\varepsilon_F)}}{\DoS_0^2}
\end{align}
(we have disregarded the Fermi functions associated to the injectivities as the average injectivity has a smooth energy dependence).
Eq.~\eqref{eq:UncorrelInterPart} corresponds to diagrams of the type represented in Fig.~\ref{fig:NonlinearDiagramUncorr}.
We clearly identify this contribution in \eqref{eq:g211sDW}.
It is equivalent to the term $\smean{ g'(\varepsilon_F)^2 }\smean{u_1}^2$ of Eq.~\eqref{eq:127} for QDs.

The second contribution involves the correlator of conductance's functional derivatives and the correlator of injectivities~:
\begin{equation}
  \label{eq:NLCproductCorr}
  \smean{ \left(\Gint_{s,a}\right)^2 }_\mathrm{corr}
  = \int\D\vec{r}\D\vec{r}\,'\, 
  \chi_g(\vec{r},\vec{r}\,')\,\chi_\DoS^{s,a}(\vec{r},\vec{r}\,')
  \:,
\end{equation}
where the structure was discussed in Subsection~\ref{subsec:SUMFR}~;
note that \eqref{eq:NLCproductCorr} corresponds with the last term of \eqref{eq:g211sDW} and \eqref{eq:g211aDW}.
This corresponds to diagrams of the type represented in Fig.~\ref{fig:NonlinearDiagram}.
We recall that $\delta g/\delta U(\vec{r})$ and the injectivity $\DoS_1(\vec{r};\varepsilon)$ are uncorrelated (see Appendix~\ref{Appendix:CorrInjFdc}).

Finally we also have to discuss the correlation $\smean{\Gint_s\Gnonint}$, what will be needed in order to analyse the fluctuations of the symmetric part of the conductance $\smean{ \big( \Gint_s \big)^2 }=\smean{(\Gnonint+\Gint_s)^2}$.
This corresponds to the terms $-(1/2)\big[\smean{u_1(\vec{r})}+\smean{u_1(\vec{r}\,')}\big]$ in the bracket $[\cdots]$ of Eq.~\eqref{eq:g211sDW}, and has the same origin as the term $-\smean{ g'(\varepsilon_F)^2 }\smean{u_1}$ of Eq.~\eqref{eq:127} for QDs.
This contribution is similar to the diagram of Fig.~\ref{fig:NonlinearDiagramUncorr}, with one injectivity less~:
\begin{align}
  \label{eq:CorrG0Gs}
    \mean{ \Gint_s\Gnonint }
    = -\frac12\int\D\vec{r}\D\vec{r}\,'\,\chi_g(\vec{r},\vec{r}\,')\,
    \frac{\mean{\DoS_1(\vec{r};\varepsilon_F)}}{\DoS_0}
    \:,
\end{align}
where the $1/2$ arises from~\eqref{eq:NoninteractingNLC}.
Since $\DoS_1(\vec{r};\varepsilon_F)/\DoS_0\sim1$, this contribution is of the same order as $\mean{\Gnonint^2 }$. 
The minus sign, i.e. the fact that $\Gint_s$ and $\Gnonint$ are anticorrelated, expresses that screening strongly renormalises the electrostatic potential inside the wire~; this point was emphasized in the case of QDs in Subsection~\ref{subsec:PreliminaryQD}. 
In the next subsections, we analyse the three contributions \eqref{eq:UncorrelInterPart}, \eqref{eq:NLCproductCorr} and~\eqref{eq:CorrG0Gs} in the different regimes.
In a first step we will derive the high magnetic field expressions of the correlator, when Cooperon contributions, labelled $^{(2)}$ and $^{(4)}$ above, are suppressed. 
The full magnetic field dependence, i.e. the correlators 
$\smean{ \mathcal{G}_{s,a}(\mathcal{B})\mathcal{G}_{s,a}(\mathcal{B}') }$,
will be discussed in a second step.

\subsubsection{Symmetric part $\mathcal{G}_s$}

\paragraph{$L_\varphi\ll L,\: L_T$.---}

We start by considering the contribution \eqref{eq:UncorrelInterPart}, which is computed by using that the average injectivity \eqref{eq:MeanInjectivity} has a smooth spatial dependence while the correlator $\chi_g$ is short range, thus
\begin{align}
    &\smean{ \left(\Gint_s\right)^2 }_\mathrm{uncorr}
    \\\nonumber
    &\simeq 
    \int_0^L\D x\, \left(\frac{\mean{\DoS_1(\vec{r};\varepsilon_F)}}{\DoS_0}\right)^2
    \int\D(x-x')\,\chi_g(\vec{r},\vec{r}\,')
    \:.
\end{align}
Using \eqref{eq:IntegralChiGLargeLT} and adding a factor $3/2$ in order to take into account the contribution $^{(3)}$, we get
\begin{equation}
  \label{eq:GsUncorrLargeLT}
  \smean{ \left(\Gint_s\right)^2 }_\mathrm{uncorr}^{(1)+(3)}
  \simeq \frac{5}{4}\left(\frac{L_\varphi}{L}\right)^7 \tau_D^2
  \:.
\end{equation}

The calculation of the contribution \eqref{eq:CorrG0Gs} is quite similar, we get
\begin{equation}
  \label{eq:CorrG0GsLargeLT}
  \mean{ \Gint_s\Gnonint }^{(1)+(3)}
  \simeq - \frac{15}{16}\left(\frac{L_\varphi}{L}\right)^7 \tau_D^2
  \:.
\end{equation}
This term exactly coincides with \eqref{eq:FluctGOlowT}, up to the sign, which is due to the fact that averaging the injectivity provides the factor $1/2$ which is missing in front of \eqref{eq:CorrG0Gs} compared to \eqref{eq:CorrNoninteracting}.

The contribution \eqref{eq:NLCproductCorr} is given by combining \eqref{eq:Chi1LargeLT} and \eqref{eq:Chi2sLargeLT}.
By using that $\chi_g(x,x')$ is short range and $\chi_\DoS^s(x,x')$ is long range, we can simplify the double integral as 
\begin{align}
  \label{eq:SimplificationGs}
  \smean{ \left(\Gint_s\right)^2 }_\mathrm{corr}^{(1)}
  &\simeq
   \int_0^L\D x\,\chi_\DoS^s(x,x)^{(1)}
   \int 
   \D(x-x')\,\chi_g(x,x')^{(1)}
   \nonumber\\
  &\simeq
  \frac{5}{12}
  \left(\frac{L}{\xiloc}\right)^2
  \left(\frac{L_\varphi}{L}\right)^{10}\, \tau_D^2
  \:.
\end{align}
Finally we add a factor $(3/2)^2$ (one for each correlator) in order to account for the second Diffuson contribution~:
\begin{equation}
    \label{eq:FluctuationsGsLowT}
      \smean{ \left(\Gint_s\right)^2 }_\mathrm{corr}^{(1)+(3)}
      \simeq
      \frac{15}{16}
      \left(\frac{L}{\xiloc}\right)^2
      \left(\frac{L_\varphi}{L}\right)^{10}\, \tau_D^2
      \:.
\end{equation}
Compared to the result without interaction \eqref{eq:FluctGOlowT}, the contribution from interaction is reduced, by a factor $(L/\xiloc)^2(L_\varphi/L)^3\ll1$ originating from the correlations of the characteristic potential.

Gathering all results we finally get
\begin{align}
&\smean{ \mathcal{G}_s^2 }
  =  \smean{ \left( \Gnonint+\Gint_s\right)^2 }
  \\\nonumber
 &=
 \mean{ \Gnonint^2 } 
 +2\mean{ \Gint_s\Gnonint } 
 + \smean{ \left(\Gint_s\right)^2 }_\mathrm{uncorr} 
 + \smean{ \left(\Gint_s\right)^2 }_\mathrm{corr} 
 \nonumber
 \\
 &\simeq
 \frac{5}{16}
 \left[ 1 + 3\left(\frac{L}{\xiloc}\right)^2\left(\frac{L_\varphi}{L}\right)^{3} \right]
 \left(\frac{L_\varphi}{L}\right)^{7}\, \tau_D^2
\end{align}
in the regime of high magnetic field, much larger than $\Bcorr\sim \phi_0/(L_\varphi w)$ (cf.~subsection~\ref{subsubsec:NLCa} below).
The second subdominant term corresponds to $\smean{ \left(\Gint_s\right)^2 }_\mathrm{corr} $.

\vspace{0.25cm}

\paragraph{$L_T\ll L_\varphi\ll L$.---}

We now have to use \eqref{eq:SplittingChiG}, leading to 
\begin{align}
  &\smean{ \left(\Gint_s\right)^2 }_\mathrm{uncorr}^{(1)}
    \simeq 
    \frac{2}{3}\tau_D^2
  \left(\frac{L_T}{L}\right)^2\left(\frac{L_\varphi}{L}\right)^4
  \\\nonumber
  &\times
  \int_0^L\frac{\D x}{L}\int_0^L\frac{\D x'}{L}
  \left(1-\frac{x}{L}\right)
  \left(1-\frac{x'}{L}\right)
  \\\nonumber
  &\times
  \left[
    \phi\left(\frac{|x-x'|}{L_\varphi}\right)
    +\left(\frac{L_T}{L_\varphi}\right)^4\,
    \Upsilon\left(\frac{|x-x'|}{L_\varphi}\right)
  \right]
  \:.
\end{align}
The spatial integrals can be calculated by changing the variables as $x=R+\rho/2$ and $x'=R-\rho/2$, leading to 
\begin{align}
  &-\frac{1}{2L^3} \int_{0}^{\infty}\D\rho\,\rho^2\,\phi\left(\frac{\rho}{L_\varphi}\right)
  \\\nonumber
  &+ \frac{2}{L}\left(\frac{L_T}{L_\varphi}\right)^4
  \int_0^L\frac{\D R}{L}\left(1-\frac{R}{L}\right)^2\int_{0}^{\infty}\D\rho\,
  \Upsilon\left(\frac{\rho}{L_\varphi}\right)
  \:,
\end{align}
where we made use of \eqref{eq:PropertyPhi}. 
We obtain numerically $\int_0^\infty\D u\,u^2\,\phi(u)\simeq-0.122718$.
Using \eqref{eq:IntegralUpsilon} we get
\begin{align}
  \label{eq:GsUncorrSmallLT}
  &\smean{ \left(\Gint_s\right)^2 }_\mathrm{uncorr}^{(1)}
  \\\nonumber
  &\simeq 
   C_0
   \left(\frac{L_T}{L}\right)^2 \left(\frac{L_\varphi}{L}\right)^7 \tau_D^2 
   +
   \frac{\pi}{45}
   \left(\frac{L_T}{L}\right)^6 \frac{L_\varphi}{L} \, \tau_D^2 
\end{align}
where $C_0\simeq0.0409$
(with $\smean{ \left(\Gint_s\right)^2 }_\mathrm{uncorr}^{(3)}$ negligible).
Although we expect that the first term dominates as $L_T\ll L_\varphi\ll L$, the second term would be important if $L_T$ and $L_\varphi$ would get closer, and ensures the crossover towards \eqref{eq:GsUncorrLargeLT} when we simply set~$L_T\sim L_\varphi$ in Eq.~\eqref{eq:GsUncorrSmallLT}.
When $L_T\ll L_\varphi$, the term $\smean{ \left(\Gint_s\right)^2 }_\mathrm{uncorr}$ gives therefore the dominant contribution to the non-linear conductance~; the calculation has shown that this observation crucially relies on the non-trivial spatial structure of the correlator $\chi_g(\vec{r},\vec{r}\,')$.

The correlation is determined by using similar arguments.
Introducing the decomposition \eqref{eq:SplittingChiG} in Eq.~\eqref{eq:CorrG0Gs}, we obtain 
\begin{align}
  \mean{ \Gint_s\Gnonint }^{(1)}
  \simeq -\frac13 \left(\frac{L_T}{L}\right)^6
  &\int_0^L\frac{\D R}{L}\left(1-\frac{ R}{L}\right)
  \nonumber\\
  \times&
  \int_{-\infty}^{+\infty}\frac{\D\rho}{L}\,\Upsilon\left(\frac{|\rho|}{L_\varphi}\right)
  \:.
\end{align}
As a result
\begin{equation}
  \mean{ \Gint_s\Gnonint }^{(1)}
  \simeq -\frac{\pi}{60} \left(\frac{L_T}{L}\right)^6 
  \frac{L_\varphi}{L} \, \tau_D^2
  \:,
\end{equation}
which now coincides with \eqref{eq:FluctGOhighT}, up to the sign, for a similar reason that \eqref{eq:CorrG0GsLargeLT} coincides with \eqref{eq:FluctGOlowT}.

We analyse the last contribution \eqref{eq:NLCproductCorr}.
The starting point of the analysis are the expressions of the two correlators \eqref{eq:SplittingChiG} and \eqref{eq:Chi2sSmallLT}.
The calculation of the double integral \eqref{eq:NLCproductCorr} is conveniently performed by setting $x=R+\rho/2$ and $x'=R-\rho/2$.
The Diffuson, which controls the injectivity correlator, takes the form
\begin{equation}
  \label{eq:PdRelativeCoordinates}
  P_d(x,x')=R \left(1-\frac{R}{L}\right) - \frac{|\rho|}{2}\left(1-\frac{|\rho|}{2L}\right)
  \:.
\end{equation}
This leads to 
\begin{align}
  \smean{  \left(\Gint_s\right)^2  }_\mathrm{corr}^{(1)}
  &\simeq   \frac{2\pi}{9}\frac{\tau_D^2}{L^2}
   \left(\frac{L}{\xiloc}\right)^2
  \left(\frac{L_T}{L}\right)^4\left(\frac{L_\varphi}{L}\right)^5
  \nonumber\\
  &\times
  \int_0^L\frac{\D R}{L}\int_{-\infty}^{+\infty}\D\rho
  \left[  R \left(1-\frac{R}{L}\right) - \frac{|\rho|}{2} \right]
  \nonumber\\
  &\times
  \left[
    \phi\left(\frac{|\rho|}{L_\varphi}\right)
    +\left(\frac{L_T}{L_\varphi}\right)^4\,
    \Upsilon\left(\frac{|\rho|}{L_\varphi}\right)
  \right]  
\end{align}
where we have dropped the last term $\sim\rho^2/L$ of \eqref{eq:PdRelativeCoordinates} which brings a negligible contribution as $\rho\lesssim L_\varphi$.
Due to the properties \eqref{eq:PropertyPhi}  
the double spatial integral simplifies as 
\begin{align}
   &- \int_0^\infty\D\rho\,\rho\,\phi\left(\frac{\rho}{L_\varphi}\right)
   \\\nonumber
   &+2\left(\frac{L_T}{L_\varphi}\right)^4\,
   \int_0^L\frac{\D R}{L}\,R \left(1-\frac{R}{L}\right)
   \int_0^\infty\D\rho\,\Upsilon\left(\frac{\rho}{L_\varphi}\right)
   \:.
\end{align}
Finally, we deduce 
\begin{align}
    \label{eq:FullExpressionGsHighT}
      \smean{ \left(\Gint_s\right)^2 }_\mathrm{corr}^{(1)}
      &\simeq 
      \tau_D^2\left(\frac{L}{\xiloc}\right)^2
      \left(\frac{L_T}{L}\right)^{4}\left(\frac{L_\varphi}{L}\right)^{2}
      \\\nonumber
      &\times   
      \left[
         C_s\left(\frac{L_\varphi}{L}\right)^{5}
         +
         \frac{\pi^2}{270}
         \left(\frac{L_T}{L}\right)^{4}
      \right]
      \:,
\end{align}
where $C_s=-(2\pi/9)\int_0^\infty\D u\,u\,\phi(u)\simeq0.0414$.
Although the first term
\begin{equation}
  \label{eq:SimplifiedExpressionGsHighT}
    \smean{ \left(\Gint_s\right)^2 }_\mathrm{corr}
    \simeq 
    C_s\left(\frac{L}{\xiloc}\right)^2
   \left(\frac{L_T}{L}\right)^{4}\left(\frac{L_\varphi}{L}\right)^{7}
   \tau_D^2
\end{equation}
dominates, the full expression \eqref{eq:FullExpressionGsHighT} shows that the fluctuations crosses over towards the result \eqref{eq:FluctuationsGsLowT} when~$L_T\sim L_\varphi$, thanks to the second term.

Interestingly we have obtained that the $L_T$ dependence of the non-interaction and interaction parts differ~: $L_T^6$ versus $L_T^4$.
We recall that the origin of the difference lies in the property~\eqref{eq:PropertyPhi}.

Gathering once again all the contributions, we deduce the non-linear conductance in the high field regime
(for $\mathcal{B}\gg\Bcorr$)~:
\begin{align}
  &\smean{ \mathcal{G}_s^2 }
  \simeq
  \bigg[ 
    \frac{\pi}{180}\left(\frac{L_T}{L}\right)^{6}\frac{L_\varphi}{L}
    \\\nonumber
  &  + C_0\left(\frac{L_T}{L}\right)^{2}\left(\frac{L_\varphi}{L}\right)^{7}
    + C_s\left(\frac{L}{\xiloc}\right)^2
    \left(\frac{L_T}{L}\right)^{4}\left(\frac{L_\varphi}{L}\right)^{7}
  \bigg]
  \tau_D^2
  \:.
\end{align}
The last subdominant term corresponds to $\smean{ \left(\Gint_s\right)^2 }_\mathrm{corr} $.

\subsubsection{Antisymmetric part $\mathcal{G}_a$}

Calculations follow the same lines.

\vspace{0.25cm}

\paragraph{$L_\varphi\ll L,\: L_T$.---}

We now combine the two correlators \eqref{eq:Chi1LargeLT} and \eqref{eq:Chi2aLargeLT}, which are both short range. We deduce
\begin{equation}
      \mean{ \mathcal{G}_{a}^2 }^{(1)}
      \simeq \frac{9}{64} 
      \left(\frac{L}{\xiloc}\right)^2
      \left(\frac{L_\varphi}{L}\right)^{11}\, \tau_D^2
  \:.
\end{equation}
We multiply this result by a factor $(3/2)$ to account for the contribution $^{(3)}$ of the correlator $\chi_g$ (we recall that ${\chi_\DoS^a}^{(3)}=0$). We obtain
\begin{equation}
      \mean{ \mathcal{G}_{a}^2 }^{(1)+(3)}
      \simeq \frac{27}{128} 
      \left(\frac{L}{\xiloc}\right)^2
      \left(\frac{L_\varphi}{L}\right)^{11}\, \tau_D^2
\end{equation}
which is the high field result.
The antisymmetric part is therefore reduced compared to the equivalent contribution to the symmetric part~:
using the notation $\smean{\left(\Gint\right)^2}_\mathrm{corr}\equiv\smean{\big(\mathcal{G}_s^\mathrm{int,\,fluc}\big)^2}$ of Section~\ref{sec:MainResults} for Eq.~\eqref{eq:FluctuationsGsLowT}, we can write
$$
   \mathcal{G}_a \sim \mathcal{G}_s^\mathrm{int,\,fluc}\,\sqrt{\frac{L_\varphi}{L}} \ll \mathcal{G}_s^\mathrm{int,\,fluc}
$$
(we recall that $\mathcal{G}_a\equiv\Gint_a$).

\vspace{0.25cm}

\paragraph{$L_T\ll L_\varphi\ll L$.---}

The determination of the fluctuations requires to combine the two short range correlators \eqref{eq:SplittingChiG} and \eqref{eq:Chi2aSmallLT}, leading to 
\begin{align}
  \mean{ \mathcal{G}_{a}^2 }^{(1)}
  \simeq \frac49
  \tau_D^2
  \left(\frac{L}{\xiloc}\right)^2
  \left(\frac{L_T}{L}\right)^{4}
  \left(\frac{L_\varphi}{L}\right)^{7}
  \int_0^\infty\hspace{-0.25cm}\D u\,\phi(u)\,\psi(u)
  \:,
\end{align}
where the two dimensionless functions were defined above, see Eqs.~\eqref{eq:DefPhi} and \eqref{eq:DefPsiDEu}.
The remaining integrals are more conveniently computed by integrating first over $u$, as the integrants behave exponentially, and then over the dimensionless frequencies $\theta$ in the two remaining integrals.
We obtain numerically 
$\int_0^\infty\D u\,\phi(u)\,\psi(u)\simeq0.006733$.
As a result 
\begin{align}
  \mean{ \mathcal{G}_{a}^2 }
  \simeq C_a
  \left(\frac{L}{\xiloc}\right)^2
  \left(\frac{L_T}{L}\right)^{4}
  \left(\frac{L_\varphi}{L}\right)^{7}
  \tau_D^2
  \:,
\end{align}
where $C_a\simeq0.00299$.
Quite remarkably, we have obtained that the term \eqref{eq:SimplifiedExpressionGsHighT} in the symmetric part and the antisymmetric part are of the same order in this regime, 
$$
  \mathcal{G}_a \sim \mathcal{G}_s^\mathrm{int,\,fluc}
  \:,
$$ 
contrary to what was observed when $L_\varphi\ll L_T$.
The ratio between the two contributions is however quite small
$\sqrt{\mean{ \mathcal{G}_{a}^2 }/\smean{ (\mathcal{G}_s^\mathrm{int,\,fluc})^2 }}\simeq\sqrt{C_a/C_s}\simeq0.27$.

\subsection{Magnetic field dependence}

The field dependence can be easily obtained by using the symmetry between the Diffuson and Cooperon contributions.
The two correlators are related to the high field correlators as 
\begin{align}
  \label{eq:SubstitutionBdependence1}
  &\chi_g(\vec{r},\vec{r}\,')
  \\\nonumber
  &=
  \chi_g(\vec{r},\vec{r}\,')^{(1)+(3)}\big|_{L_\varphi\to L_d}
  +
  \chi_g(\vec{r},\vec{r}\,')^{(1)+(3)}\big|_{L_\varphi\to L_c}
  \\
  \label{eq:SubstitutionBdependence2}
  &\chi_\DoS^{s,a}(\vec{r},\vec{r}\,')
  \\\nonumber
  &=
  \chi_\DoS^{s,a}(\vec{r},\vec{r}\,')^{(1)+(3)}\big|_{L_\varphi\to L_d}
  \pm
  \chi_\DoS^{s,a}(\vec{r},\vec{r}\,')^{(1)+(3)}\big|_{L_\varphi\to L_c}
\end{align}
where the two lengths were defined above, Eq.~\eqref{eq:LdLc}.

\subsubsection{Symmetric part}

\paragraph{$L_\varphi\ll L,\: L_T$.---}

As the calculation of \eqref{eq:CorrNoninteracting} has involved a single correlation function, we may simply perform in Eq.~\eqref{eq:FluctGOlowT} the substitution $L_\varphi\to L_d$ in $\smean{ \Gnonint ^2}^{(1)+(3)}$, then, performing $L_\varphi\to L_c$, we deduce $\smean{ \Gnonint ^2}^{(2)+(4)}$~:
\begin{equation}
  \smean{ \Gnonint(\mathcal{B})\Gnonint(\mathcal{B}') }
    \simeq 
 \frac{15}{16}
   \left[
     \left(\frac{L_d}{L}\right)^7 + \left(\frac{L_c}{L}\right)^7
   \right] 
   \tau_D^2
\end{equation}
where the two lengths were defined above, Eq.~\eqref{eq:LdLc}.
The correlation term also coincides with this result
$\smean{ \Gint_s(\mathcal{B})\Gnonint(\mathcal{B}') }
=-\smean{ \Gnonint(\mathcal{B})\Gnonint(\mathcal{B}') }$.

The contribution \eqref{eq:UncorrelInterPart} can be obtained by a similar argument.
The simple substitutions in \eqref{eq:GsUncorrLargeLT} lead to 
\begin{equation}
  \mean{ \Gint_s(\mathcal{B})\Gint_s(\mathcal{B}') }_\mathrm{uncorr}
  \simeq \frac{5}{4}    
    \left[
      \left(\frac{L_d}{L}\right)^{7} +  \left(\frac{L_c}{L}\right)^{7} 
    \right]
 \tau_D^2
 \:.
\end{equation}

The contribution 
$  \smean{ \Gint_s(\mathcal{B})\Gint_s(\mathcal{B}') }_\mathrm{corr} $ 
can also be straightforwardly obtained thanks to the decoupling between the short range correlations of conductance's functional derivatives and the long range correlations of the injectivities.
Using \eqref{eq:SimplificationGs}, we obtain
\begin{align}
  &\smean{ \Gint_s(\mathcal{B})\Gint_s(\mathcal{B}') }_\mathrm{corr}
  \simeq \frac{15}{16}
    \left(\frac{L}{\xiloc}\right)^2
    \\\nonumber
  &\times  
    \left[
      \left(\frac{L_d}{L}\right)^{7} +  \left(\frac{L_c}{L}\right)^{7} 
    \right]
    \left[
      \left(\frac{L_d}{L}\right)^{3} +  \left(\frac{L_c}{L}\right)^{3} 
    \right]
    \tau_D^2
  \:.
\end{align}

\vspace{0.25cm}

\paragraph{$L_T\ll L_\varphi\ll L$.---}

The study of the other regime follows the same lines. We deduce from \eqref{eq:FluctGOhighT}
\begin{equation}
   \smean{ \Gnonint(\mathcal{B})\Gnonint(\mathcal{B}')  }
    \simeq 
   \frac{\pi}{60}
   \left(\frac{L_T}{L}\right)^6
     \frac{L_d+L_c}{L}
   \, \tau_D^2 
  \:.
\end{equation}

\begin{widetext}
Similarly, the first interaction contribution is deduced from \eqref{eq:GsUncorrSmallLT}
\begin{align}
  \mean{ \Gint_s(\mathcal{B})\Gint_s(\mathcal{B}') }_\mathrm{uncorr}
  \simeq
  \: C_0
   \left(\frac{L_T}{L}\right)^2
    \left[
      \left(\frac{L_d}{L}\right)^7 +  \left(\frac{L_c}{L}\right)^7
    \right]
   \, \tau_D^2
   + \frac{\pi}{45}
   \left(\frac{L_T}{L}\right)^6
   \left[
      \left(\frac{L_d}{L}\right) +  \left(\frac{L_c}{L}\right)
    \right]
   \, \tau_D^2 
   \:.
\end{align}

Finally, we only consider the dominant term in \eqref{eq:FullExpressionGsHighT}~:
\begin{align}
  \smean{ \Gint_s(\mathcal{B})\Gint_s(\mathcal{B}') }_\mathrm{corr} 
  \simeq 
      C_s
      \left(\frac{L}{\xiloc}\right)^2
      \left(\frac{L_T}{L}\right)^4
    \left[ 
         \left(\frac{L_d}{L}\right)^6 +  \left(\frac{L_c}{L}\right)^6
      \right]\, 
      \left[ 
         \frac{L_d}{L} +  \frac{L_c}{L}
      \right]\, 
      \tau_D^2
      \:.
\end{align}
\end{widetext}

\subsubsection{Antisymmetric part}
\label{subsubsec:NLCa}

\paragraph{Preliminary~: antisymmetric part of the injectance correlations.---}

As the asymmetry of the non-linear conductance under magnetic field reversal is due to the asymmetry of the injectivity, a good preliminary exercice is to characterise this latter.
In order to simplify the discussion, we analyse the integral of the inject\textit{ivity}, denoted the inject\textit{ance}, 
$\overline{\DoS}_1(\varepsilon)=\int\D\vec{r}\,\DoS_1(\vec{r};\varepsilon)$, 
whose correlator is given by integration of the correlator \eqref{eq:Chi2aLargeLT}~:
$\smean{\delta\overline{\DoS}_1^2}_a^{(1)}/(L^2\DoS_0^2)\simeq(1/8)(L/\xiloc)^2(L_\varphi/L)^5$.
Using the substitution \eqref{eq:SubstitutionBdependence2} we obtain that the antisymmetric part of the injectance fluctuations is 
\begin{align}
  \frac{\smean{\delta\overline{\DoS}_1^2}_a}{L^2\DoS_0^2}
  &\simeq \frac{1}{8}\left(\frac{L}{\xiloc}\right)^2
  \left[
    \left(\frac{L_d}{L}\right)^5
    -
    \left(\frac{L_c}{L}\right)^5
  \right]
  \\
  &\underset{\mathcal{B}\to0}{\simeq}
  \frac{5}{16}\left(\frac{L}{\xiloc}\right)^2\left(\frac{L_\varphi}{L}\right)^5
  \left(\frac{\mathcal{B}}{\Bcorr}\right)^2
  \:,
\end{align}
where $\Bcorr=\big[\sqrt{3}/(2\pi)\big]\,\phi_0/(L_\varphi\sw)$.
Therefore $\sqrt{\smean{\delta\overline{\DoS}_1^2}_a}\propto\mathcal{B}$ and we expect a similar behaviour for the conductance $\mathcal{G}_a\propto\mathcal{B}$.
However we will see that, surprisingly, the linear behaviour is not always obtained.

\vspace{0.25cm}

\paragraph{Regime $L_\varphi\ll L_T,\ L$.-- } 

As the correlations $\smean{ \mathcal{G}_a(\mathcal{B})\mathcal{G}_a(\mathcal{B}') } $ involves the integration of two short range correlators, one has this time to calculate the integrals by using the substitutions (\ref{eq:SubstitutionBdependence1},\ref{eq:SubstitutionBdependence2}).
Some algebra gives 
\begin{widetext} 
\begin{align}
  \smean{ \mathcal{G}_a(\mathcal{B})\mathcal{G}_a(\mathcal{B}') } 
  &\simeq \frac{3}{8}\left(\frac{L}{\xiloc}\right)^2\tau_D^2
 \\
  &\times  \left\{
    \frac{9}{16}
    \left[
       \left(\frac{L_d}{L}\right)^{11} + \left(\frac{L_c}{L}\right)^{11}
    \right]
- \left(\frac{L_dL_c}{L^2}\right)^{4}
  \frac{L_{d\parallel c}}{L}
  \left[
    \left(\frac{L_d}{L}\right)^{2}
    \left(1+\frac{L_{d\parallel c}}{4L_d}\right) 
    +
    \left(\frac{L_c}{L}\right)^{2}
    \left(1+\frac{L_{d\parallel c}}{4L_c}\right) 
  \right]
  \right\}    
  \:,
  \nonumber
\end{align}
\end{widetext}
where $1/L_{d\parallel c}=1/L_d+1/L_c$.

The low field expansion shows that the quadratic term $\mathcal{B}^2$ vanishes and we obtain 
\begin{equation}
  \smean{ \mathcal{G}_a(\mathcal{B})^2 } 
  \underset{\mathcal{B}\to0}{ \simeq }
  \frac{1599}{1024} \left(\frac{L}{\xiloc}\right)^2
  \left(\frac{L_\varphi}{L}\right)^{11}\tau_D^2\,
  \left(\frac{\mathcal{B}}{\Bcorr}\right)^4
  \:.
\end{equation}
This characterises a quadratic behaviour of the non-linear conductance
\begin{equation}
  \mathcal{G}_a(\mathcal{B})  \underset{\mathcal{B}\to0}{\sim}
  \mathcal{G}_a(\infty)\,\mathrm{sign}(\mathcal{B})\,(\mathcal{B}/\Bcorr)^2
\end{equation}
in the low field regime.
As it is shown below, the vanishing of the linear term seems rather accidental as it will not be obtained for $L_T\ll L_\varphi$, neither in the coherent limit when $L\ll L_\varphi$. 
In these two other situations the linear behaviour is obtained~$\mathcal{G}_a(\mathcal{B})\sim\mathcal{B}$.

\vspace{0.25cm}

\paragraph{Regime $L_T\ll L_\varphi\ll L$.-- } 

The calculation is more complicated in this regime, hence we will only analyse the $\mathcal{B}\to0$ limit.
The starting point combines the two correlators \eqref{eq:SplittingChiG} and \eqref{eq:Chi2aSmallLT} leading to
\begin{align}
    \smean{ \mathcal{G}_a(\mathcal{B})\mathcal{G}_a(\mathcal{B}') } 
  &\simeq\frac{4}{9}\tau_D^2\left(\frac{L}{\xiloc}\right)^2
  \left(\frac{L_T}{L}\right)^4
  \\\nonumber
  \times\int_0^\infty\frac{\D\rho}{L}
  &\left[
    \left(\frac{L_d}{L}\right)^4 \phi\left(\frac{\rho}{L_d}\right) 
    + (L_d\to L_c)
  \right]
  \\\nonumber
  \times&\left[
    \left(\frac{L_d}{L}\right)^2 \psi\left(\frac{\rho}{L_d}\right) 
    - (L_d\to L_c)
  \right]
  \:.
\end{align}
The calculation of the integral is a little bit complicated, however the expansion as $\mathcal{B}=\mathcal{B}'\to0$ leads to tracktable calculations.
We write $L_d=L_\varphi$ and $L_c\simeq L_\varphi\,\big[1-(\mathcal{B}/\Bcorr)^2/2\big]$, leading to 
\begin{align}
  \label{eq:GaHighTLowField}
  \smean{ \mathcal{G}_a(\mathcal{B})^2 } 
  \simeq
  K_a\left(\frac{L}{\xiloc}\right)^2
  \left(\frac{L_T}{L}\right)^4
  \left(\frac{L_\varphi}{L}\right)^{7}
  \left(\frac{\mathcal{B}}{\Bcorr}\right)^2
  \tau_D^2
\end{align}
for $\mathcal{B}\ll\Bcorr$,
where
\begin{align}
  K_a = \frac{8}{9}
  \int_0^\infty\D u\,\phi(u)\,\left[ \psi(u) - \frac12u\,\psi'(u) \right]
  \simeq 0.00642
  \:.
\end{align}
Thus we have obtained the expected linear behaviour 
\begin{equation}
  \mathcal{G}_a(\mathcal{B})  \underset{\mathcal{B}\to0}{\sim}
  \mathcal{G}_a(\infty)\,(\mathcal{B}/\Bcorr)
  \:.
\end{equation}


\section{Coherent regime}
\label{sec:Coherent}

In the coherent limit, $L\lesssim L_\varphi$, we cannot anymore use the translation invariant property inside the wire and the effect of the boundaries must be treated properly.
We have thus to reconsider the analysis of the two main correlators in this regime.

\subsection{Correlators}

A convenient starting point for the determination of the correlators in this case is to expand the propagator \eqref{eq:PropagatorFiniteWire} for $\omega=0$ as 
\begin{align}
  P^{(d)}_0(x,x')\underset{\gamma\to0}{=}
  P_d(x,x')
  \Big[1&+ \gamma\,A_1(x,x') 
  \\\nonumber
  &+ \gamma^2\,A_2(x,x') +\mathcal{O}(\gamma^3)
  \Big]
\end{align}
with
\begin{align}
  A_1(x,x') = \frac{x_<^2+(L-x_>)^2-L^2}{6}
\end{align}
and
\begin{align}
  A_2(x,x') &= \frac{1}{360}
  \Big\{
    7\, L^4 -10\,L^2\,\left[ x_<^2 + (L-x_>)^2 \right]
    \nonumber\\
    + & 3\,x_<^4 + 10\, x_<^2\, (L-x_>)^2 + 3\, (L-x_>)^4 
  \Big\}
  \:.
\end{align}

\subsubsection{Correlator $\chi_g$}

\begin{figure}[!ht]
\centering
\includegraphics[width=0.4\textwidth]{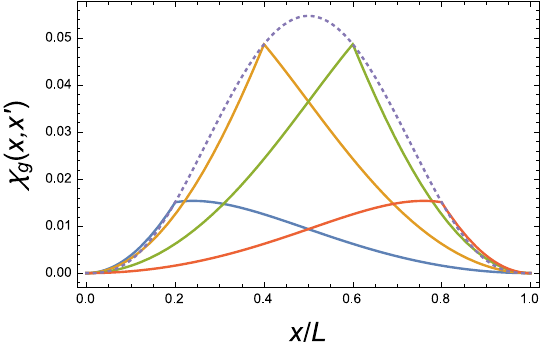}
\caption{(color online) \it Correlator $\chi_g(\vec{r},\vec{r}\,')$ for $x'/L=0.2$, $0.4$, $0.6$ and $0.8$.
  The dotted line is the envelope $\chi_g(\vec{r},\vec{r})$.}
\label{fig:chiG}
\end{figure}

We now deduce from \eqref{eq:ChiGconvenientRepresentation}
$\chi_g(\vec{r},\vec{r}\,')^{(1)}=[4/(D^2L^4)]\,2\,P_d(x,x')^2\,\big[2A_2(x,x')-A_1(x,x')^2\big]$, i.e.
\begin{align}
  \label{eq:ChiGcoherent}
  \chi_g(\vec{r},\vec{r}\,')^{(1)} 
  =  \frac{4\tau_D^2}{45L^2}
  \left(\frac{P_d(x,x')}{L}\right)^2\left( 1 - \frac{x_<^4+(L-x_>)^4}{L^4} \right) 
  \:.
\end{align}
The correlator is plotted in Fig.~\ref{fig:chiG}.

\subsubsection{Correlator $\chi_\DoS^s$}

\begin{figure}[!ht]
\centering
\includegraphics[width=0.4\textwidth]{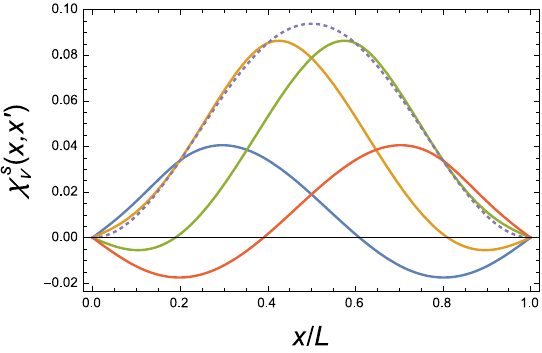}
\caption{(color online) \it Correlator $\chi_\DoS^{s}(\vec{r},\vec{r}\,')$ for  $x'/L=0.2$, $0.4$, $0.6$ and $0.8$.
  The dotted line is the envelope $\chi_\DoS^{s}(\vec{r},\vec{r})$.}
\label{fig:chiNuS}
\end{figure}

The symmetric part of the injectivity correlator may be calculated from \eqref{eq:ACentralResult}. Some algebra gives~:
\begin{widetext} 
\begin{align}
  \label{eq:ChiScoherent}
  \chi_\DoS^s(\vec{r},\vec{r}\,')^{(1)}  &= \frac{2}{45}
  \left(\frac{L}{\xiloc}\right)^2\frac{P_d(x,x')}{L}
  \\  \nonumber
  &\times\left\{
    \frac{6\,[x_<^4+(L-x_>)^4]-10\,x_<^2(L-x_>)^2}{L^4}
    -15\,\frac{x_<^3+(L-x_>)^3}{L^3}
    +10\,\frac{x_<^2+(L-x_>)^2}{L^2}
    -1
  \right\}
  \:,
\end{align}
which can change in sign as shown by Fig.~\ref{fig:chiNuS}.
\end{widetext}

\subsubsection{Correlator $\chi_\DoS^a$}

The antisymmetric part is more easy to determine thanks to \eqref{eq:ChiAconvenientRepresentation}.

\vspace{0.25cm}

\paragraph{High magnetic field.--}

In the high field regime, the Cooperon contributions are suppressed, hence, replacing $P_\omega^{(d)}$ by $P_d$ in \eqref{eq:ChiAconvenientRepresentation} we obtain straightforwardly the simple expression
\begin{equation}
  \label{eq:ChiAcoherentHighField}
  \chi_\DoS^{a}(\vec{r},\vec{r}\,')^{(1)}
  = \frac{2}{9} 
  \left(\frac{L}{\xiloc}\right)^2\left[\frac{P_d(x,x')}{L}\right]^3
  \:.
\end{equation}

\vspace{0.25cm}

\paragraph{Low magnetic field.--}

\begin{figure}[!ht]
\centering
\includegraphics[width=0.4\textwidth]{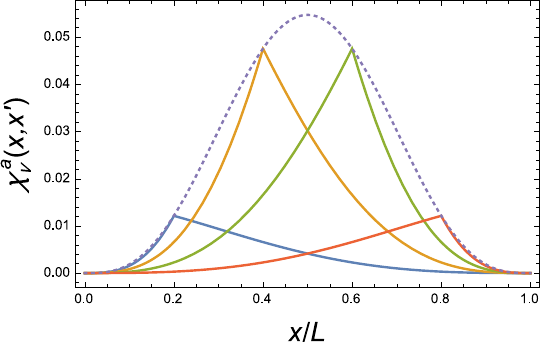}
\caption{(color online) \it Correlator $\chi_\DoS^{a}(\vec{r},\vec{r}\,')$ in the low magnetic field regime for $x'/L=0.2$, $0.4$, $0.6$ and $0.8$.
The dotted line is the envelope $\chi_\DoS^{a}(\vec{r},\vec{r})$.}
\label{fig:chiNuA}
\end{figure}

A non-trivial result is obtained by keeping the first order term in $\gamma$ in the expansion of the propagator, hence
\begin{align}
  &\chi_\DoS^{a}(\vec{r},\vec{r}\,')^{(1)+(2)}  \simeq
    \frac{4}{\xiloc^2}
    \frac{(\gamma_d-\gamma_c)}{L^2}
    \nonumber\\
    &\times\int_0^{x_<}\D\xi\int_{x_>}^L\D\xi'\,
    P_d(\xi,\xi')^2\,A_1(\xi,\xi')
    \:.
\end{align}
As a result we obtain
\begin{align}
  \label{eq:ChiAcoherent}
  &\chi_\DoS^{a}(\vec{r},\vec{r}\,')^{(1)+(2)}  
  \simeq
  \frac{2}{135}
   \left( \frac{L}{\xiloc}\right)^2
   (\gamma_c-\gamma_d)L^2\,
  \nonumber\\
  &\times \left(\frac{P_d(x,x')}{L}\right)^3\left( 5 - 3\,\frac{x_<^2+(L-x_>)^2}{L^2} \right) 
  \:.
\end{align}
The spatial dependence is plotted in Fig.~\ref{fig:chiNuA}.

\subsection{Non-linear conductance}

\subsubsection{Weak magnetic field}

The symmetric part of the non-linear conductance depends weakly on the magnetic field. 
We determine the zero field value.
Eq.~\eqref{eq:CorrNoninteracting} with \eqref{eq:ChiGcoherent} gives 
$\mean{ \Gnonint^2 }^{(1)}=\tau_D^2/4725$.
As all terms contribute the same in this limit, the result must be multiplied by~$2\times(3/2)$~:
\begin{equation}
  \label{eq:GOcoherent}
  \mean{ \Gnonint^2 }
  = \frac{\tau_D^2}{1575}
  \:.
\end{equation}

Now we analyse the symmetric part of the interaction contribution. 
The first term \eqref{eq:UncorrelInterPart} is 
\begin{equation}
  \label{eq:Gs2CoherentUncorr}
   \smean{ \left(\Gint_s\right)^2 }_\mathrm{uncorr}
  = \frac{139}{207\,900}\tau_D^2
  \:.
\end{equation}
The correlation is 
\begin{equation}
  \label{eq:G0GsCoherent}
  \mean{ \Gnonint\Gint_s }
  = -\frac{\tau_D^2}{1575}
  \:.
\end{equation}
The contribution from interaction combines \eqref{eq:ChiGcoherent} and \eqref{eq:ChiScoherent}. We find 
$\smean{ \left(\Gint_s\right)^2 }^{(1)}_\mathrm{corr}=(8/45^2)(19/38\,610)\,(L/\xiloc)^2\tau_D^2$.
As each correlator $\chi_g$ and $\chi_\DoS^s$ receive a factor $3$ in order to account for other contributions, the result must be multiplied by a factor~$9$~: 
\begin{equation}
  \label{eq:Gs2CoherentCorr}
  \smean{ \left(\Gint_s\right)^2 }_\mathrm{corr}
  = \frac{76}{4\,343\,625}\left( \frac{L}{\xiloc}\right)^2\tau_D^2
  \:.
\end{equation}
Gathering all contributions, we obtain 
\begin{align}
  \label{eq:NLCfullyCoherentZeroField}
  \smean{ \mathcal{G}_s^2 }
  &= \mean{ \Gnonint^2 }  + 2 \mean{ \Gnonint\Gint_s }
  + \smean{ \left(\Gint_s\right)^2 }_\mathrm{uncorr} + \smean{ \left(\Gint_s\right)^2 }_\mathrm{corr}
  \nonumber\\
  &= \left[
    \frac{1}{29\,700} + \frac{76}{4\,343\,625}\,\left( \frac{L}{\xiloc}\right)^2
  \right]
  \tau_D^2
\end{align}
in the zero field limit. 
It is worth stressing the origin of the small dimensionless prefactor 
$\smean{ \mathcal{G}_s^2 }\simeq3.3\times10^{-5}\,\tau_D^2$ has its origin in the compensation between the free electron result, with much larger prefactor $1/1575\simeq6.3\times10^{-4}$, and negative contributions from screening which renormalises the disordered potential.

Finally we analyse the weak magnetic field behaviour of the antisymmetric part by combining \eqref{eq:ChiGcoherent} and \eqref{eq:ChiAcoherent}. 
We now include the factor $3$ in the correlator $\chi_g$ and get
\begin{align}
  \mean{ \mathcal{G}_a(\mathcal{B})\mathcal{G}_a(\mathcal{B}') }
  &\simeq \frac{38}{51\,121\,125}
  \\\nonumber
  &\times
  \left( \frac{L}{\xiloc}\right)^2
  \left[\left(\frac{L}{L_c}\right)^2-\left(\frac{L}{L_d}\right)^2\right]\tau_D^2
  \:.
\end{align}
In particular, for $L_\varphi=\infty$ we have $L_d=\infty$ and $L_c=L_\mathcal{B}$, hence
\begin{equation}
  \mean{ \mathcal{G}_a(\mathcal{B})^2 }
  \underset{\mathcal{B}\to0}{\simeq} \frac{38}{51\,121\,125}
  \left( \frac{L}{\xiloc}\right)^2
  \left(\frac{\mathcal{B}}{\mathcal{B}_{c0}}\right)^2
  \tau_D^2
  \:,  
\end{equation} 
where the crossover field is now $\mathcal{B}_{c0}=\big[\sqrt{3}/(2\pi)\big]\,\big[\phi_0/(L\sw)\big]$.

\subsubsection{High magnetic field}

We now discuss the high field result, $\mathcal{B}\gg\mathcal{B}_{c0}$.
The dominant contributions to the symmetric part of the non-linear conductance all involve a single correlator $\chi_g$, hence Eqs.~(\ref{eq:GOcoherent},\ref{eq:Gs2CoherentUncorr},\ref{eq:G0GsCoherent}) should all be divided by a factor of $2$ in the high field regime in order to account for the suppression of the Cooperon contribution (while the negligible contribution \eqref{eq:Gs2CoherentCorr} should be divided by a factor $4$ as it involves the product of two correlators), so that 
\begin{equation}
  \smean{ \mathcal{G}_s^2 }
  \simeq\frac{\tau_D^2}{59\,400}
  \:.
\end{equation}

Combining \eqref{eq:ChiGcoherent} and \eqref{eq:ChiAcoherentHighField} we deduce
$\smean{\mathcal{G}_a^2}^{(1)}=(8/7\,818\,525)(L/\xiloc)^2\tau_D^2$. 
The result must be multiplied by a factor $(3/2)$ in order to account for the second Diffuson contribution ${\chi_g}^{(3)}$ (recall that ${\chi_\DoS^{a}}^{(3)}=0$)~:
\begin{equation}
  \mean{ \mathcal{G}_a^2 }^{(1)+(3)}
  = \frac{4}{2\,606\,175}\left( \frac{L}{\xiloc}\right)^2\tau_D^2  
  \:.
\end{equation}


\section{Conclusion}
\label{sec:Conclusion}

We have studied the non-linear conductance of weakly disordered wires and have analysed the effect of Coulomb interaction by combining the scattering formalism introduced by B\"uttiker and diagrammatic techniques.
We have derived general formulae for injectivity correlators $\chi_\DoS^{s,a}(\vec{r},\vec{r}\,')$, Eqs.~(\ref{eq:ACentralResult},\ref{eq:ChiU3and4}), and the correlators $\chi_g(\vec{r},\vec{r}\,')$ of conductance's functional derivatives, Eq.~\eqref{eq:RootEqChiG} (although some of the external Diffusons were simplified in these general expressions, assuming the wire geometry, the external Diffusons can be reintroduced straightforwardly).
We recall that the existence of a non-zero antisymmetric part, $\chi_\DoS^{a}\neq0$, crucially relies on the \textit{external} Diffusons in the injectivity correlator~\eqref{eq:InjectivityCorr12}, i.e. on the asymmetry between the two diagrams of Fig.~\ref{fig:InjectivityCorr12}.

Although we have applied our formalism to the simple geometry of a wire connected at two-terminals (Fig.~\ref{fig:PotentialInWire}), it can in principle be applied to more complex geometries, like it was done for other physical quantities in Refs.~\onlinecite{TexMon04,TexDelMon09,TexMon16}. The calculation in the general case would however become quite heavy as the main formulae are given under the form of multiple integrals (six spatial integrals and two energy integrals, as we have seen).

The correlator $\smean{ \mathcal{G}^2 }$ for a wire has been splitted into different contributions which have been analysed separately.
From the experimental point of view, it is only possible to distinguish contributions from their symmetry under magnetic field reversal, i.e. 
$\mathcal{G}_{s,a}(\mathcal{B})=\big[\mathcal{G}(\mathcal{B})\pm\mathcal{G}(-\mathcal{B})\big]/2$.

We now summarize the new results obtained in the article.

\subsection{Coherent regime}

In the coherent regime $L\lesssim L_\varphi$ and in absence of magnetic field we have obtained in Section~\ref{sec:Coherent} that
\begin{align}
  \smean{ \mathcal{G}_s(0)^2 }
  = \left[
    \frac{1}{29\,700} + \frac{76}{4\,343\,625}\,g^{-2}
  \right]
  \EThouless^{-2}
  \:,
\end{align}
where $g=\xiloc/L$ is the dimensionless (Drude) conductance.
The first term thus mixes the non-interaction term $\Gnonint$ and some of the  contributions to $\Gint_s$.
The value obtained for free electrons, $\smean{\mathcal{G}_\nonint^2}=\EThouless^{-2}/1575$, is compensated by terms of the same order, leading to a result smaller by a factor $\sim1/20$~; this is due to screening, which strongly renormalises the electrostatic potential inside the wire.  
This reminds us that Coulomb interaction has a strong effect and brings contributions of the same order as the one given by the free electron (Landauer-B\"uttiker) theory.
The last contribution 
$\smean{ \left(\Gint_s\right)^2 }_\mathrm{corr}\equiv\smean{\big(\mathcal{G}_s^\mathrm{int,\,fluc}\big)^2}\sim(g\EThouless)^{-2}$ 
originating from the mesoscopic fluctuations of the electrostatic potential, is negligible due to the factor $g^{-2}\ll1$ (note that $76/4\,343\,625\simeq1.7\times10^{-5}$ is approximatively one half of $1/29\,700\simeq3.4\times10^{-5}$).
In the high field regime ($\mathcal{B}\gg\mathcal{B}_{c0}$), the fluctuations are twice smaller~:
\begin{equation}
  \label{eq:G211Scoherent}
  \smean{ \mathcal{G}_s(\infty)^2 }
  \simeq\frac{1}{59\,400}
  \EThouless^{-2}
\end{equation}
whereas the antisymmetric part is~:
\begin{equation}
  \label{eq:G211Acoherent}
  \smean{ \mathcal{G}_a(\infty)^2 }
  \simeq\frac{4}{2\,606\,175}
  (g\EThouless)^{-2}
  \:.
\end{equation}
Varying the magnetic field, the magnetic field dependence is linear at low field, 
$\mathcal{G}_a(\mathcal{B})\sim\mathcal{G}_a(\infty)\, (\mathcal{B}/\mathcal{B}_{c0})$, 
where the crossover field $\mathcal{B}_{c0}\sim\phi_0/(L\sw)$ corresponds to one quantum flux in the wire.

\subsection{Weakly coherent regime}

In the weakly coherent regime $L_\varphi\ll L_T,\:L$ we have obtained in Section~\ref{sec:NLC} that
\begin{align}
  \label{eq:G211Sincoherent}
  \smean{ \mathcal{G}_s(\infty)^2 }
  \simeq \frac{5}{16}
  &\left[ 
    1
    + \frac{3}{g^2}
    \left(\frac{L_\varphi}{L}\right)^{3}
  \right]
  \left(\frac{L_\varphi}{L}\right)^{7}
  \EThouless^{-2}
\end{align}
in the high field regime $\mathcal{B}\gg\Bcorr\sim\phi_0/(L_\varphi\sw)$.
In practice, we can expect that the crossover between the coherent \eqref{eq:G211Scoherent} and incoherent \eqref{eq:G211Sincoherent} results occurs by equating the two expressions, i.e. for $L_\varphi/L\simeq5.7$.
The antisymmetric part is given in this regime by 
\begin{align}
  \label{eq:G211Aincoherent}
  \smean{ \mathcal{G}_a(\infty)^2 }
  \simeq \frac{27}{128}
  \left(\frac{L_\varphi}{L}\right)^{11}
  (g\EThouless)^{-2}
  \:.
\end{align}
Crossover with \eqref{eq:G211Acoherent} occurs for $L_\varphi/L\simeq0.341$.

We have also shown that, surprisingly, the $\mathcal{B}\to0$ behaviour is quadratic $\mathcal{G}_a(\mathcal{B})\sim \mathcal{G}_a(\infty)\,\mathrm{sign}(\mathcal{B})\,(\mathcal{B}/\Bcorr)^2$, where $\Bcorr\sim\phi_0/(L_\varphi\sw)$.

\subsection{Thermal fluctuations}

In the regime $L_T\ll L_\varphi\ll L$ we derived (for $\mathcal{B}\gg\Bcorr$)~:
\begin{align}
  \label{eq:g221sConclu0}
  &\smean{ \mathcal{G}_s(\infty)^2 }
  \simeq
  \bigg[ 
    \frac{\pi}{180}\left(\frac{L_T}{L}\right)^{6}\frac{L_\varphi}{L}
   \\ \nonumber
  &  + C_0\left(\frac{L_T}{L}\right)^{2}\left(\frac{L_\varphi}{L}\right)^{7}
    + \frac{C_s}{g^2}
    \left(\frac{L_T}{L}\right)^{4}\left(\frac{L_\varphi}{L}\right)^{7}
  \bigg]
  \EThouless^{-2}
\end{align}
where $C_0\simeq0.0409$ and $C_s\simeq0.0414$.
The dominant term is produced by the contribution $\smean{ \left(\Gint_s\right)^2 }_\mathrm{uncorr}$ and the result of a calculation for free electrons is negligible, underlying once more the importance of screening.
This observation, which has led in particular to the behaviour $\mathcal{G}\propto T^{-1/2}$, compared to $\Gnonint\propto T^{-3/2}$, crucially relies on the non-trivial spatial structure of the correlator $\chi_g(\vec{r},\vec{r}\,')$.
The first term in \eqref{eq:g221sConclu0} allows one to understand how the matching with the dominant term of \eqref{eq:G211Sincoherent} is realised at $L_T\sim L_\varphi$.
However, for $L_T/L_\varphi$ sufficiently small, we can neglect it, leading to
\begin{align}
  \label{eq:g221sConclu}
  \smean{ \mathcal{G}_s(\infty)^2 }
  \simeq
     C_0
  \left(\frac{L_T}{L}\right)^{2}\left(\frac{L_\varphi}{L}\right)^{7}
    \EThouless^{-2}
  \:.
\end{align}
The antisymmetric part is 
\begin{align}
  \label{eq:g221aConclu}
  \smean{ \mathcal{G}_a(\infty)^2 }
  \simeq
  \frac{C_a}{g^2}
  \left(\frac{L_T}{L}\right)^{4}
 \left(\frac{L_\varphi}{L}\right)^{7}\,
  \EThouless^{-2}
  \:,
\end{align}
where $C_a\simeq0.00299$, i.e. of the same order as the last term of \eqref{eq:g221sConclu0}.

The behaviour for $\mathcal{B}\to0$ was also obtained~:
\begin{align}
  \label{eq:GaHighTLowFieldConclu}
  \smean{ \mathcal{G}_a(\mathcal{B})^2 } 
  \simeq
  \frac{K_a}{g^2}
  \left(\frac{L_T}{L}\right)^4
  \left(\frac{L_\varphi}{L}\right)^{7}
  \left(\frac{\mathcal{B}}{\Bcorr}\right)^2
  \EThouless^{-2}
\end{align}
where
$  K_a \simeq 0.00642$.

In all regimes, the contribution $\smean{ \left(\Gint_s\right)^2 }_\mathrm{corr}\equiv\smean{ \left(\mathcal{G}_s^\mathrm{int,\,fluc}\right)^2 }$  is negligible.
However the contribution $\smean{ \mathcal{G}_a^2 }$, with same physical origin, can in principle be identified from its magnetic field dependence.

\subsection{Decoherence by electronic interaction}

At low temperature ($T\lesssim1\:$K), decoherence is dominated by electronic interactions, which leads to the following temperature dependence of the phase coherence length
\begin{equation}
  \label{eq:PCLee}
  L_\varphi = \sqrt{2}\,(\xiloc L_T^2/\pi)^{1/3}
  \propto T^{-1/3}
  \:,
\end{equation}
valid for $L_T\ll L_\varphi$ (thus $L_\varphi\ll\xiloc$)~;
the main behaviour was first derived in the seminal paper \cite{AltAroKhm82}, although the prefactor given in this reference is incorrect~; See Refs.~\onlinecite{PieGouAntPotEstBir03,TexMon05b,AkkMon07,TexDelMon09} and references therein.~\cite{footnote9} 
As a result, if \eqref{eq:PCLee} is substituted in (\ref{eq:g221sConclu},\ref{eq:g221aConclu}) we obtain the behaviours (at high field $\mathcal{B}\gg\Bcorr$)
\begin{align}
  \smean{ \mathcal{G}_s(\infty)^2 }
  &\simeq \frac{8\sqrt{2}C_0}{\pi^{7/3}}
  g^{7/3}
  \left(\frac{L_T}{L}\right)^{20/3}
  \EThouless^{-2}
  \\
  \smean{ \mathcal{G}_a(\infty)^2 }
  &\simeq \frac{8\sqrt{2}C_a}{\pi^{7/3}}
  g^{1/3}
  \left(\frac{L_T}{L}\right)^{26/3}
  \EThouless^{-2}
  \:.
\end{align}
It is also interesting to rewrite the low field behaviour \eqref{eq:GaHighTLowField}, or \eqref{eq:GaHighTLowFieldConclu}. 
In order to identify the $L_\varphi$-dependence we write
$\mathcal{B}/\Bcorr=(L_\varphi/L)(\mathcal{B}/\mathcal{B}_{c0})$, where $\mathcal{B}_{c0}=\big[\sqrt{3}/(2\pi)\big]\,\big[\phi_0/(L\sw)\big]$ is the correlation field for the coherent wire. As a result, for $\mathcal{B}\ll\Bcorr$ we find 
\begin{align}
  \smean{ \mathcal{G}_a(\mathcal{B})^2 }
  \simeq
  \frac{16\sqrt{2}K_a}{\pi^{3}}\,
  g\,
  \left(\frac{L_T}{L}\right)^{10}
  \left(\frac{\mathcal{B}}{\mathcal{B}_{c0}}\right)^2
  \EThouless^{-2}
  \:.
\end{align}

All these results show that,  as the temperature is increased, the non-linear conductance decays quite fast since it involves high powers of the phase coherence length. Measurement should be favored by considering small coherent devices, with not to large conductance (i.e. wires etched in a two-dimensional electron gas).

\subsection{Open questions}

In the coherent limit, the simple description of contacts which we have adopted (absorbing boundary conditions for the diffusion propagators) might not be fully appropriate as the contacts have usually a two-dimensional character. This problem was considered for the weak localisation in Ref.~\onlinecite{ChaSanPro91} (see also Ref.~\onlinecite{TexMon05}). A study of this effect would therefore be useful in order to provide more accurate predictions to be confronted with experiments in the coherent regime.

As we discussed, the sign of the non-linear conductance is fluctuating, which implies $\smean{\mathcal{G}}=0$ and is reflected by the change in sign of the zero temperature correlator (see Subsection~\ref{subsec:CorrelationG0}).
Deyo, Spivak and Zuyzin~\cite{DeySpiZyu06} argued that this is related to the non monotonous dependence with the temperature.
We have not considered this problem in the article, which would therefore deserve further investigation in the diffusive regime.

We come back on our assumption of perfect screening (Subsection~\ref{subsec:CharacPot}).
In this regime, it is expected that Hartree and Fock contributions to low temperature properties of weaky disordered metals are equally important. For several well studied quantities, this can be accounted for through interaction constants, without changing the functional dependence in the characteristic scales (like the thermal length, etc)~: 
this is the case for the DoS anomaly and the Altshuler-Aronov correction to the conductivity~\cite{AltAro85,AkkMon07}.
For the non-linear conductance in zero-dimension (quantum dots), as studied in Refs.~\onlinecite{SanBut04,SpiZyu04,PolBut06,PolBut07a,DeySpiZyu06}, the nature of screening was accounted for through a global dimensionless constant, denoted $\gamma_\mathrm{int}=C_\mu/C$ in Refs.~\onlinecite{SanBut04,PolBut06,PolBut07a,AngZakDebGueBouCavGenPol07} and $\beta$ in Refs.~\onlinecite{SpiZyu04,DeySpiZyu06}.
In these last references, the possibility of introducing an interaction constant for the non-linear transport can be related to the local nature of the response of the electrostatic density to the density of injected charge carriers, in other terms, following Ref.~\onlinecite{DeySpiZyu06}, one would have to add a dimensionless interaction constant in our Eq.~\eqref{eq:CharacPotPerfectScreeningLimit}, what would have very simple consequences on the calculations of the present article.
The precise justification of such a simple prescription seems however not obvious, in particular beyond the zero-dimensional limit where spatial structures or correlation functions are important, as we have seen. 
This point should therefore be further studied.

In the experiment \cite{AngZakDebGueBouCavGenPol07,Ang07}, measurements were performed in a different geometry (ring). It would thus be interesting to extend our calculations for the wire to more complex geometries, starting from the general expression (\ref{eq:CorrelatorGintGint},\ref{eq:CorrelatorGnonintGnonint}) [the correlator $\chi_g$ in the general case can be obtained by using \eqref{eq:FunctionalDerivation} on 
(\ref{eq:ConductanceCorrelator1},\ref{eq:ConductanceCorrelator2},\ref{eq:ConductanceCorrelator3},\ref{eq:ConductanceCorrelator4}) while the general expression of $\chi_\nu$ is given by (\ref{eq:ACentralResult},\ref{eq:ChiU3and4Genuine})] and with the help of the formalism of Refs.~\onlinecite{TexMon04,TexDelMon09,TexMon16} (note that the non-linear response of a ring made of strictly 1D wire was analysed in Ref.~\onlinecite{HerLew09}).

\section*{Acknowledgements}

JM thanks LPTMS for hospitality.
CT acknowledges many stimulating discussions and remarks from Lionel Angers, H\'el\`ene Bouchiat, Sophie Gu\'eron, David S\'anchez and Denis Ullmo.

\begin{appendix}


\section{Absence of correlation between injectivity and functional derivative of the conductance}
\label{Appendix:CorrInjFdc}

\begin{figure}[!ht]
\centering
\diagram{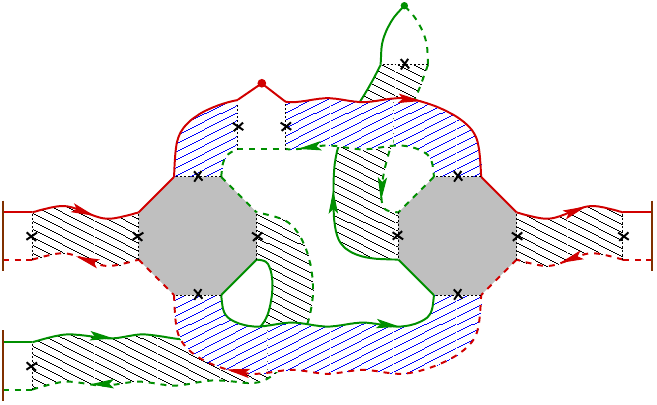}{0.5}{-1.25cm}
$+\ \big(\mathrm{R}\leftrightarrow\mathrm{A}\big)=0$ 
\caption{(color online) \it Correlation $\smean{\DoS_1(\vec{r};\varepsilon)\,\delta g(\varepsilon')/\delta U(\vec{r}\,')}$ vanishes.}
\label{fig:CorrelationInjecFDcond}
\end{figure}

We briefly consider the correlation between the injectivity $\DoS_1(\vec{r};\varepsilon)$ and the functional derivative $\delta g(\varepsilon')/\delta U(\vec{r}\,')$, which was not considered in the paper.
The key observation is that this requires to correlate a single injectivity diagram
\begin{equation}
\diagram{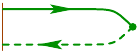}{0.75}{0cm}
\end{equation}
with a pair of conductance lines
\begin{equation}
\diagram{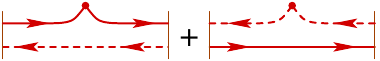}{0.75}{0cm}
\end{equation}
(see Fig.~\ref{fig:DiagramNonLinear}).
As a result, a diagram contributing to the correlator $\smean{\DoS_1(\vec{r};\varepsilon)\,\delta g(\varepsilon')/\delta U(\vec{r}\,')}$ is always paired with a similar diagram in which retarded and advanced lines are exchanged, like in Fig.~\ref{fig:CorrelationInjecFDcond}, which vanishes by virtue of Eq.~\eqref{eq:Property1}.

Therefore, in the weak disorder limit and within the diffusion approximation, the injectivity and the conductance's functional derivative are uncorrelated~:
\begin{equation}
  \mean{\DoS_1(\vec{r};\varepsilon)\,\derivf{g(\varepsilon')}{U(\vec{r}\,')}}
  =0
  \:.
\end{equation}

\section{Thermal functions}
\label{Appendix:ThermalFcts}

Finite temperature calculations involve the thermal function
\begin{equation}
  \Pi_V(\varepsilon) = f(\varepsilon-eV/2) - f(\varepsilon+eV/2)  
  \:,
\end{equation}
for $eV>0$, where $f(\varepsilon)$ is the Fermi-Dirac distribution.
It is convenient to introduce the Fourier transform
\begin{equation}
  \widehat{\Pi}_V(t) = \int\frac{\D\varepsilon}{2\pi} \, \Pi_V(\varepsilon)\,\EXP{-\I\varepsilon t}
\end{equation}
in order to express the function
\begin{align}
  \label{eq:DefThermalFctF}
  F(\omega;V,T) &= \int\D\varepsilon\,\Pi_V(\varepsilon) \,\Pi_V(\varepsilon-\omega) 
  \\
  &= 2\pi \int\D t\, \widehat{\Pi}_V(t)\, \widehat{\Pi}_V(-t)\, \,\EXP{\I\omega t}
\end{align}
which has a width $\mathrm{max}(T,eV)$.
Using the property
\begin{equation}
  \int\D\omega\, F(\omega;V,T) = \left[\int\D\varepsilon\, \Pi_V(\varepsilon)\right]^2 = (eV)^2
  \:
\end{equation}
we deduce that the function has height $eV\,\mathrm{min}(1,eV/T)$.
This function is useful to express integrals of the form
\begin{equation}
  \label{eq:IntegralWithThermalFcts}
   \int\D\varepsilon\D\varepsilon'\, \Pi_V(\varepsilon)\, \Pi_V(\varepsilon')\, \Phi(\varepsilon-\varepsilon')
   =\int\D\omega\, F(\omega;V,T) \, \Phi(\omega)
\end{equation}
where $\Phi(\omega)$ is a known function.

The function 
\begin{align}
  \label{eq:DefThermalFctG}
  G(\omega;V,T)
  =2\pi \int\D t\, \derivp{\widehat{\Pi}_V(t)}{(eV)}\, \derivp{\widehat{\Pi}_V(-t)}{(eV)}\, \,\EXP{\I\omega t}
  \:. 
\end{align}
will also be useful.

\subsection{Linear regime $V\to0$}
\label{Appendix:ThermalFctsSS1}

In the small voltage regime, we have 
\begin{equation}
  \lim_{V\to0} \frac{1}{eV} \widehat{\Pi}_V(t)
  =\frac{1}{2\pi} \left(\frac{\pi Tt}{\sinh\pi Tt }\right)
\end{equation}
from which we deduce that
\begin{equation}
  \lim_{V\to0} \frac{1}{(eV)^2} F(\omega;V,T) = \delta_T(\omega)
\end{equation}
is a normalised function of width $T$. 
Explicitly~:
\begin{equation}
  \label{eq:DefDeltaOmega}
   \delta_T(\omega) = \frac{1}{2T}\, h(\omega/(2T))
   \hspace{0.25cm}\mbox{with}\hspace{0.25cm}
   h(x) = \frac{x\coth x-1}{\sinh^2 x}
   \:.
\end{equation}
In other term, we have the useful property 
\begin{align}
\label{eq:ThermFctProperty1}
  \int \D\varepsilon\D\varepsilon'\,
    \derivp{f(\varepsilon)}{\varepsilon}\,
    \derivp{f(\varepsilon')}{\varepsilon'}\,
  \Phi(\varepsilon-\varepsilon') 
  = \int\D\omega\,
  \delta_T(\omega)\,\Phi(\omega)
  \:.
\end{align}

\subsection{Non-linear regime at $T=0$}
\label{Appendix:ThermalFctsSS2}

The zero temperature limit is also easy to discuss as 
$\Pi_V(\varepsilon)=\heaviside(eV/2-|\varepsilon|)$ (where the Fermi energy is at zero for simplicity).
As a result
\begin{equation}
  \label{eq:ThermalFctFZeroT}
  F(\omega;V,0)=
  \begin{cases}
      eV-|\omega| & \mbox{for } \omega\in[-eV,+eV]
      \\
      0  & \mbox{otherwise}
  \end{cases}
\end{equation}
We also deduce easily the function \eqref{eq:DefThermalFctG}~:
\begin{equation}
  \label{eq:ThermalFctGZeroT}
  G(\omega;V,0)
  =\frac12\, \delta(\omega) + \frac14\, \delta(\omega-eV) + \frac14\, \delta(\omega+eV)
  \:.
\end{equation}


\section{Relation with the results of Khmelnitskii and Larkin}
\label{Appendix:KnownResultsNLC}

Although the scattering approach was criticized in Ref.~\onlinecite{DeySpiZyu06} when applied to problems with electronic interactions, the equivalence of the scattering formalism of B\"uttiker \cite{But93,ChrBut96a} with the result of a similar calculation (Hartree within Thomas-Fermi approximation) within the non equilibrium Green's function (Keldysh) method was established by Hern\'andez and Lewenkopf~\cite{HerLew13}.
In this appendix we would like to discuss this equivalence specifically for the case of disordered metals. In the present article, we have used the scattering formalism of B\"uttiker \cite{But93,ChrBut96a} describing non-linear transport and including the effect of interaction in a Thomas-Fermi approximation. 
The general formulae of this formalism were used as a starting point to apply the standard diagrammatic techniques for weakly disordered metals.
We propose here another equivalent approach, which starts from the main result of Khmelnitskii and Larkin \cite{LarKhm86,KhmLar86} $I$-$V$ characteristic in disordered metals, and we show how we can include the effect of electronic interactions on the top of this theory in order to establish the correspondence with the results obtained from the scattering formalism applied to weakly disordered metals.

Before establishing this correspondence, we first recall the main formula of Refs.~\onlinecite{LarKhm86,KhmLar86} and briefly discuss few outcomes which are useful for the paper.

\subsection{Mesoscopic fluctuations of the $I$-$V$ characteristic}
\label{subsec:KLmainRes}

Khmelnistskii and Larkin (KL) used the non-equilibrium Green's function method, which allows to deal directly with the observable (the current) in the non-equilibrium situation, which is simpler than correlation functions (conductivity) which appear in the linear response theory.
Denoting by $I(V)$ the current-voltage relation, one obtains the formula for the current correlations in a disordered wire~:~\cite{footnote10}
\begin{widetext}  
\begin{align}
  \label{eq:LarkinKhmelnitskii1986}
  \mean{ \delta I(V_1)\,\delta I(V_2) }
  =&\left( \frac{2_se}{h} \right)^2
   \int\D\varepsilon\D\varepsilon'\,
  \big[ f(\varepsilon-eV_{1L})  - f(\varepsilon-eV_{1R})  \big]
  \big[ f(\varepsilon'-eV_{2L}) - f(\varepsilon'-eV_{2R}) \big]
  \nonumber \\
  &\times 
  \int \frac{\D x\D x'}{L^4}
  \left\{
    4 \left|P_{\varepsilon-\varepsilon'}^{(d)}(x,x')\right|^2
  + 2 \re{ \left[P_{\varepsilon-\varepsilon'}^{(d)}(x,x')^2\right] }
  + \big( P_{\omega}^{(d)}\ \to\ P_{\omega}^{(c)} \big)
  \right\}
\end{align}
where $V_{1L}$ and $V_{1R}$ are the potentials at the two contacts in the configuration $1$, etc.
We emphasize that \eqref{eq:LarkinKhmelnitskii1986} describes the \textit{mesoscopic} (sample to sample) fluctuations of the \textit{averaged} current, where averaging is made over quantum and thermal fluctuations (this should not to be confused with the current noise).
The Diffuson and Cooperon solve the equation
\begin{equation}
  \label{eq:DiffusonLarkinKhmelnitskii}
  \left[
    \frac{1}{L_\varphi^2} -\I\frac{\omega-U_1(\vec{r})+U_2(\vec{r})}{D} 
    -\left( \vec{\nabla} -2\I e\vec{A}_\mp \right)^2
  \right]
  P^{(d,c)}_\omega(\vec r,\vec{r}\,')=\delta(\vec r-\vec{r}\,')
  \:,
\end{equation}
\end{widetext} 
where $U_{1,2}(\vec{r})=eV_{1,2}\,(1-x/L)$ is the potential inside the wire for a voltage $V_{1,2}$.
The vector potentials $\vec{A}_\pm$ are defined in subsection~\ref{subsec:Ladders}.

\subsubsection{Case $V_1=V_2$ : current fluctuations and linear conductance}

The fluctuations $\mean{\delta I(V)^2}$ are easy to compute as the solution of \eqref{eq:DiffusonLarkinKhmelnitskii} is simply obtained in this case.
Let us briefly recall this analysis.
We introduce the correlator of the zero temperature dimensionless conductance
\begin{equation}
  \mathscr{C}(\omega)
  =\mean{ \delta g(\varepsilon_F) \delta g(\varepsilon_F-\omega) }
  \:.
\end{equation}
The first constribution is 
\begin{align}
  \mathscr{C}(\omega)^{(1)}
 = 4 \int \frac{\D x\D x'}{L^4} \left|P_{\omega}^{(d)}(x,x')\right|^2
  \:,
\end{align}
to which one should add the other Diffuson contribution $\mathscr{C}(\omega)^{(3)}$ and the two Cooperon contributions, $\mathscr{C}(\omega)^{(2)}$ and $\mathscr{C}(\omega)^{(4)}$, cf. Eq.~\eqref{eq:LarkinKhmelnitskii1986}.
As a result
$\mathscr{C}(\omega)=\mathscr{C}(\omega)^{(1)}+\mathscr{C}(\omega)^{(2)}+\mathscr{C}(\omega)^{(3)}+\mathscr{C}(\omega)^{(4)}$
and
\begin{align}
  \label{eq:CurrentFluct}
  \mean{\delta I(V)^2} = \left( \frac{2_se}{h} \right)^2
  \int\D\omega\,F(\omega;V,T)\,\mathscr{C}(\omega)
  \:,
\end{align}
where the thermal function $F(\omega;V,T)$  
was defined in Appendix~\ref{Appendix:ThermalFcts}.

The most efficient method furnishing the correlator $\mathscr{C}(\omega)$ is to introduce the spectral determinant~:
\begin{align}
  \mathscr{C}(\omega)^{(1)} 
  &= \frac{4}{L^4}\,\tr{ \frac{1}{\gamma_\omega-\partial_x^2}\frac{1}{\gamma_\omega^*-\partial_x^2} }
  \\
  \nonumber
  &=
  \frac{4}{L^4}\,\frac{1}{\gamma_\omega^*-\gamma_\omega}
  \left(
    \derivp{}{\gamma_\omega}\ln S(\gamma_\omega)
    -\mathrm{c.c.}
  \right)
\end{align}
where $\gamma_\omega=1/L_\varphi^2-\I\omega/D$. The functional determinant $S(\gamma_\omega)=\det(\gamma_\omega-\partial_x^2)$ can be efficiently computed for arbitrary network geometries~\cite{PasMon99,AkkComDesMonTex00,ComDesTex05,Tex10,HarKirTex12}.
For a wire with Dirichlet boundary conditions, we have
$S(\gamma)=\sinh(\sqrt{\gamma}L)/\sqrt{\gamma}$.
In the coherent limit ($L_\varphi=\infty$) and for $\mathcal{B}=0$, we deduce
\begin{align}
  \label{eq:CorrelCondOmega}
  \mathscr{C}(\omega)
  =\frac{3}{2x^3}
  \left(
    \frac{\sinh2x+\sin2x}{\cosh2x-\cos2x}
    - \frac{1}{x}
  \right)
  \:,
\end{align}
where $x=\sqrt{\omega\tau_D/2}$ and $\tau_D=L^2/D$ is the Thouless time.
The low frequency expansion reads 
$\mathscr{C}(\omega)=(2/15)\big[1-(\omega\tau_D)^2/105+\mathcal{O}(\omega^4)\big]$.
The first term corresponds to universal conductance fluctuations of the coherent wire, $\mathscr{C}(0)=\smean{ \delta g^2 }=2/15$.

Below, we simplify the analysis by considering the limit $L\gg L_\varphi$, when boundary conditions can be neglected. We get (at $\mathcal{B}=0$)
\begin{align}
  \label{eq:CorrelCondOmegaIncoh}
  \mathscr{C}(\omega)
  \simeq
  \frac{3\sqrt{2}\,(L_\varphi/L)^3}
       { \sqrt{ \left(1+(\omega\tau_\varphi)^2\right)
                \left(\sqrt{1+(\omega\tau_\varphi)^2}+1\right)  } }
  \:.
\end{align}
The conductance fluctuations are now $\smean{ \delta g^2 }\simeq3(L_\varphi/L)^3$.

We check that the expressions \eqref{eq:CorrelCondOmega} for $\omega\tau_D\gg1$ and \eqref{eq:CorrelCondOmegaIncoh} for $\omega\tau_\varphi\gg1$, present the same large frequency behaviour
\begin{equation}
  \label{eq:CdeOmegaDecay}
  \mathscr{C}(\omega)\simeq3\sqrt2\,|\omega\tau_D|^{-3/2}
  \:,
\end{equation}
 as it should.

\begin{figure}[!ht]
\centering
\includegraphics[scale=0.9]{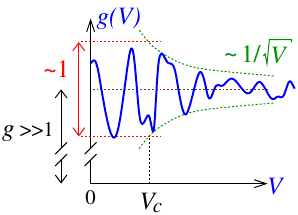}
\caption{(color online) \it Typical structure expected for the linear conductance $g(V)=\tilde{I}(V)/(eV)$ of a coherent wire $L\lesssim L_\varphi$ at $T\ll \EThouless,\: eV$.
  Correlations occur on scale $V_c=\EThouless/e$, which is also the scale for the crossover between linear and non-linear regimes.}
\label{fig:GdeV}
\end{figure}

We consider the case $\mathrm{max}(T,eV)\gg1/\tau_\varphi$, when $\mathscr{C}(\omega)$ is a narrow function compared to $F(\omega;V,T)$ which can be replaced by $F(0;V,T)$ in Eq.~\eqref{eq:CurrentFluct}.
We define the length scale $L_V=\sqrt{D/(eV)}$ similar to the thermal length $L_T=\sqrt{D/T}$.
For convenience we introduce the rescaled current $\tilde{I}=[h/(2_se)]\,I$ (with dimension $[\tilde{I}]=[\mathrm{Energy}]$).
We now apply \eqref{eq:CurrentFluct}~: the function $F(\omega;T,V)$ can be considered as a narrow function of width $\sim T$ in the linear regime $eV\ll T$ (i.e. $L_T\ll L_V,\:L_\varphi$) or of width $\sim eV$ in the non-linear regime $eV\gg T$  (i.e. $L_V\ll L_T,\:L_\varphi$). As a result~:
\begin{align}
  \label{eq:EqC10}
  \smean{\delta\tilde{I}(V)^2}\simeq 
  \frac{2\pi}{\beta}
  \begin{cases}
     \displaystyle
     (eV)^2\,\frac{L_T^2L_\varphi}{3L^3} \propto V^2
       &\mbox{for } eV\ll T\\[0.25cm]
     \displaystyle
     (eV)^2\,\frac{2L_V^2L_\varphi}{L^3} \propto V
       &\mbox{for } eV\gg T
  \end{cases}
\end{align}
where we have simplified the discussion of the $\mathcal{B}$ dependence by introducing the Dyson index $\beta$, which describes the two limiting cases~: zero field ($\beta=1$) or strong field ($\beta=2$).
If we introduce the dimensionless conductance  
\begin{equation}
  g(V) = \frac{\tilde{I}(V)}{eV}
  \:,
\end{equation}
we can rewrite the second line of \eqref{eq:EqC10} as 
\begin{equation}
  \label{eq:FluctConductanceNonlinear}
  \mean{\delta g(V)^2}
  \sim \frac{\EThouless}{eV}\sqrt{\EThouless\tau_\varphi}
  \:.
\end{equation}
The fluctuations of the conductance decay with $V$ (Fig.~\ref{fig:GdeV}).
Correspondingly, the fluctuations of the current grow as 
$\mean{\delta I(V)^2}\sim V$ (Fig.~\ref{fig:IdeV}).

\subsection{Fluctuations of the differential conductance}
\label{subsec:KLDiffCond}

KL also analysed in Refs.~\onlinecite{LarKhm86,KhmLar86} the correlations of the differential conductance 
\begin{equation}
  \DifG(V) = \frac{1}{e}\deriv{\tilde{I}(V)}{V}  = g(V) + V\,\deriv{g(V)}{V}
    \:,
\end{equation}
which requires the knowledge of the correlations \eqref{eq:LarkinKhmelnitskii1986} for $V_1\neq V_2$ as 
$\smean{\delta\DifG(V_1)\delta\DifG(V_2)}=(\partial^2/\partial V_1\partial V_2)\smean{ \delta\tilde I(V_1)\,\delta \tilde I(V_2) }$.
The analysis is more complicate as we must also account for the dependence of the correlator $\mathscr{C}(\omega)$ in the voltage difference $\Delta V=V_1-V_2$, cf. Eq.~\eqref{eq:DiffusonLarkinKhmelnitskii}.
From \eqref{eq:CurrentFluct}, we get
\begin{align}
  \label{eq:FluctuationsDiffConductance}
  \mean{\delta\DifG(V)^2}
  =\int\D\omega\,
  \bigg[
  &
    G(\omega;V,T)\,\mathscr{C}(\omega)
\\\nonumber
&   
    -
    F(\omega;V,T)\,
    \left.\derivp{^2\mathscr{C}(\omega)}{(e\Delta V)^2}\right|_{\Delta V=0}
  \bigg]  
  \:,
\end{align}
where the two thermal functions were defined above, Eqs.~(\ref{eq:DefThermalFctF},\ref{eq:DefThermalFctG}).
For $T=0$, using \eqref{eq:ThermalFctGZeroT}, we get
\begin{align}
  \label{eq:FluctuationsDiffConductanceZero}
  \mean{\delta\DifG(V)^2}
  & 
  = \mathscr{C}(0)
  + \frac{\mathscr{C}(eV) - 2\mathscr{C}(0) + \mathscr{C}(-eV)}{4}
  \nonumber\\
&   
    -
    \int\D\omega\,F(\omega;V,0)\,
    \left.\derivp{^2\mathscr{C}(\omega)}{(e\Delta V)^2}\right|_{\Delta V=0}
  \:,
\end{align}
where the first term is the conductance fluctuations $\mean{\delta g^2}=\mathscr{C}(0)$ at zero voltage.

\subsubsection{Coherent regime}

We consider first the coherent regime $L\lesssim L_\varphi$, when the correlator \eqref{eq:CorrelCondOmega} has height $\mathscr{C}(0)\sim1$ and width $\sim\EThouless$.

For low voltage $eV\ll\EThouless$, we can treat the thermal function $F(\omega;V,0)$ as a narrow function, which leads to rewrite \eqref{eq:FluctuationsDiffConductanceZero} as 
\begin{align}
  \label{eq:DiffCondApproxLowV}
  \mean{\delta\DifG(V)^2} 
  \simeq \mean{\delta g^2}
  +(eV)^2 
  \left[  
     \frac{\mathscr{C}''(0)}{4}
     -\left.\derivp{^2\mathscr{C}(0)}{(e\Delta V)^2}\right|_{\Delta V=0}
  \right]
\end{align}
Simple dimensional analysis shows that the correction to $\mean{\delta\DifG(0)^2}=\mean{\delta g^2}$ is a small correction~:
\begin{equation} 
  \label{eq:DiffCondApproxLowV2}
  \mean{\delta\DifG(V)^2}  - \mean{\delta g^2}
  \sim \left(\frac{eV}{\EThouless}\right)^2
  \:.
\end{equation}
It is important to stress one point~: 
while \eqref{eq:DiffCondApproxLowV2} provides information about the low voltage behaviour of
$\smean{\delta\DifG(V)^2}  - \smean{\delta\DifG(0)^2}$, we have studied in our article the fluctuations of the non-linear conductance 
$\smean{\big[\DifG(V)-\DifG(0)\big]^2}=(2eV)^2\mean{\Gnonint^2}+\mathcal{O}(V^4)$,
which carries the genuine information about the magnetic field asymmetry.
Although the two quantities present similar behaviours, cf. Section~\ref{sec:Coherent} where $\mean{\Gnonint^2}$ for $eV\ll\EThouless$ was derived, they do not coincide exactly.

In the high voltage regime $eV\gg\EThouless$, the thermal function in Eq.~\eqref{eq:FluctuationsDiffConductanceZero} can be treated as a broad function, and thus replaced by $F(0;V,0)=eV$.
This shows that the last term of \eqref{eq:FluctuationsDiffConductanceZero} dominates as it grows with the voltage, while $\mathscr{C}(\pm eV)$ decays with $V$, cf. Eq.~\eqref{eq:CdeOmegaDecay}, thus~:
\begin{align}
  \label{eq:DiffCondApproxHighV}
  \mean{\delta\DifG(V)^2} 
  \simeq - eV 
    \int\D\omega\, 
    \left.\derivp{^2\mathscr{C}(\omega)}{(e\Delta V)^2}\right|_{\Delta V=0}
    \:.
\end{align}
By dimensional analysis, we recover the LK prediction~\cite{LarKhm86}
\begin{equation}
  \label{eq:coherentLK}
  \mean{\delta\DifG(V)^2} \sim \frac{eV}{\EThouless}
  \hspace{0.25cm}\text{for }
  eV\gg\EThouless
  \:;
\end{equation}
see also Ref.~\onlinecite{LudBlaMir04}, where it was argued that this result is not relevant from the experimental point of view, as it only occurs for very large ratio $eV/\EThouless$ in practice.

\begin{figure}[!ht]
\centering
\includegraphics[scale=1]{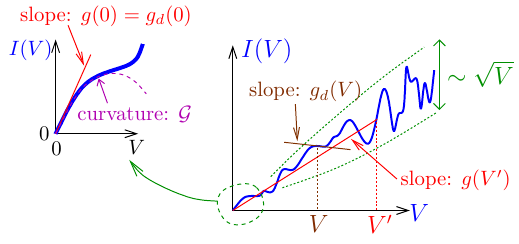}
\caption{(color online) \it Difference between the conductance $g=I/V$ and the differential conductance $\DifG=\D I/\D V$ of a coherent wire. For $V\gtrsim V_c=\EThouless/e$, the current fluctuations grow as $\sim\sqrt{V}$ and correlations of the $I(V)$ curve decay over the scale $V_c$.
The non-linear conductance $\mathcal{G}$ studied in the article corresponds to the curvature of the $I(V)$-curve at origin.}
\label{fig:IdeV}
\end{figure}

The linear growth $\smean{\delta\DifG(V)^2} \sim {eV}/{\EThouless}$, which is related to the existence of regions of \textit{negative} differential conductance (Fig.~\ref{fig:IdeV}), can be confronted with the \textit{decay} of the linear conductance's fluctuations $\mean{\delta g(V)^2}\sim \EThouless/(eV)$.
Let us show that the two behaviours are simply related~:
given that the correlation function $\mean{\delta g(V_1)\delta g(V_2)}$ is a function of height $\mean{\delta g(V)^2}\sim\EThouless/(eV)$ and width (correlation scale) $V_c=\EThouless/e$, we can deduce the correlations of the differential conductance $\DifG=g+V\,\D g/\D V$.
Fluctuations are dominated by the second term
$\smean{\delta\DifG^2}\sim V^2\smean{(\D g/\D V)^2}$.
The derivative $\smean{(\D g/\D V)^2}\sim\smean{\delta g^2}/V_c^2$ therefore
$\smean{\delta\DifG^2}\sim (V/V_c)^2\smean{\delta g^2}\sim eV/\EThouless$. {\sc Qed}.


\subsubsection{Weakly Coherent regime}

In the regime $L_\varphi\ll L$, the analysis is similar than in the previous paragraph, with the difference that the correlator \eqref{eq:CorrelCondOmegaIncoh} is now a function of small height $\mathscr{C}(0)\sim(L_\varphi/L)^3$ and width $1/\tau_\varphi$.
The weakly coherent regime now involves three energy scales ($\EThouless$, $eV$ and $1/\tau_\varphi$), thus one must be careful with scaling arguments.
Diffusion propagators are translation invariant when $L_\varphi\ll L$, hence all quantities can be rescaled by $L_\varphi$.
Inspection of the diffuson
\begin{equation*}
  P_{\omega}^{(d)}(x,x')
  =\bra{x} 
  \frac{1}{ 1/L_\varphi^2 - \I\left( \omega - e\mathcal{E}\,x\right)/D -\partial_x^2 }
  \ket{x'}
  \:,
\end{equation*}
where $\mathcal{E}=\Delta V/L$, makes clear that, when all physical quantities are rescaled by $L_\varphi$, the electric field should be rescaled as $\mathcal{E}\sim1/L_\varphi^3$, and thus the potential scales as $\Delta V\sim L/L_\varphi^3$. 

In the low voltage regime $eV\ll1/\tau_\varphi$, we can start again from \eqref{eq:DiffCondApproxLowV}.
Following the scaling argument, we can replace the derivative by  
$\partial/\partial(e\Delta V)\longrightarrow L_\varphi^3/(DL)$.
This shows that the last term of \eqref{eq:DiffCondApproxLowV} is negligible compare to $\mathscr{C}''(0)\sim\tau_\varphi^2\mathscr{C}(0)$ and leads to 
\begin{equation} 
  \mean{\delta\DifG(V)^2}  - \mean{\delta g^2}
  \sim \left(\frac{eV}{\EThouless}\right)^2  \left(\frac{L_\varphi}{L}\right)^7
  eV\tau_\varphi\ll1
  \:.
\end{equation}
Note that $\mathscr{C}''(0)<0$ (see above), thus $\mean{\delta\DifG(V)^2}$ decays with $V$ at low voltage.

In the high voltage regime $eV\gg1/\tau_\varphi$, we start instead from \eqref{eq:DiffCondApproxHighV}.
The scaling argument gives 
$
\mean{\delta\DifG^2} \sim eV \, (1/\tau_\varphi)\,(L_\varphi^3/DL)^2 \, \mathscr{C}(0) 
$, 
hence
\begin{equation}
  \mean{\delta\DifG(V)^2} 
  \sim \frac{eV}{\EThouless} \left( \frac{L_\varphi}{L} \right)^7
  \hspace{0.25cm}\text{for }
  eV\tau_\varphi\gg1
  \:.
\end{equation}
We have recovered the result of LK~\cite{LarKhm86,KhmLar86} (cf. also the introduction of Ref.~\onlinecite{LudBlaMir04}).
Interestingly, we have obtained the same power $L_\varphi^7$ for $\smean{\mathcal{G}_0^2}$ (regime $eV\tau_\varphi\ll1$) and $\mean{\delta\DifG^2}$ (regime $eV\tau_\varphi\gg1$), however we have seen that exponent $7$ arises from different combinations in the two cases.
Finally we recall LK's result for the correlations (eq.~27 of Refs.~\onlinecite{LarKhm86,KhmLar86})~:
%
\begin{align}
  \label{eq:Eq27LarkinKhmelnitskii}
  \mean{\delta\DifG(V_1)\delta\DifG(V_2)} \sim
  \frac{eV}{\EThouless}\,
  \left(\frac{L_\varphi}{L}\right)^7\,
  Q\!\left(\frac{e\Delta V}{DL}L_\varphi^3\right)
\end{align}
with $V=(V_1+V_2)/2$ and $\Delta V=V_1-V_2$.
The dimensionless function is $Q(0)\sim1$ and decays as $Q(x)\sim x^{-7/3}$.
It involves the scaling variable identified above.
%
%

\subsection{Relation with the formalism of the paper}

We now add the same physical ingredients as the one exposed in Section~\ref{sec:ScatteringFormalism} in order to show how the main expressions obtained within the scattering formalism can be recovered.
As explained in Section~\ref{sec:MainResults}, the contribution of electronic interactions to the non-linear conductance is due to the fact that the electrostatic potential presents mesoscopic fluctuations.
Hence, if we consider the correlator \eqref{eq:LarkinKhmelnitskii1986} for $V_1=V_2$, the mesoscopic potential fluctuations could be accounted for by considering the two potentials in the diffusion equation \eqref{eq:DiffusonLarkinKhmelnitskii} as two mesoscopic fluctuations $\delta U_1(\vec{r})$ and $\delta U_2(\vec{r})$ characterising the two configurations.
I.e. the potentials are $U_{1,2}(\vec{r})=eV\,(1-x/L)+\delta U_{1,2}(\vec{r})$ with the same average but with different mesoscopic fluctuations $\delta U_{1,2}\sim V/g$.

We treat these fluctuations in perturbation theory, thus we write the perturbative correction to the propagator as~:
\begin{align}
&\delta P_\omega^{(d,c)}(\vec{r},\vec{r}\,')
= -\frac{\I}{D}
\int\D\vec{r}\,''\,
P_\omega^{(d,c)}(\vec{r},\vec{r}\,'')
\nonumber\\
&\times
\left[ \delta U_1(\vec{r}\,'')-\delta U_2(\vec{r}\,'') \right]\,
P_\omega^{(d,c)}(\vec{r}\,'',\vec{r}\,')
+\cdots
\:.
\end{align}
Starting from Eq.~\eqref{eq:ConductanceCorrelator1}, either we consider the first order correction for each Diffuson or a second order correction in one of the Diffuson.
Up the potentials, this produces the six contributions represented by the diagrams of Fig.~\ref{fig:ChiG}.
We now average all pairs of potentials $\mean{ \delta U_1(\vec{r})\,\delta U_2(\vec{r}\,')  }$.
One must be careful of the fact that we have considered first order correction in the interaction for each non-linear conductance, therefore each conductance must be interrupted by an interaction line once only, like on Fig.~\ref{fig:NonlinearDiagram}, which corresponds to retain the terms $\mean{ \delta U_1(\vec{r})\,\delta U_2(\vec{r}\,')  }$ (a term with $\mean{ \delta U_1(\vec{r})\,\delta U_1(\vec{r}\,')  }$ would correspond to two interactions for one of the conductance).
For example one contribution obtained from  \eqref{eq:ConductanceCorrelator1} is 
\begin{align}
  \frac{4}{L^4}
  \int\D x\D x'
  &P_{\omega}(x,\xi)\,P_{\omega}(\xi,x')\,
  P_{-\omega}(x',\xi')\,P_{-\omega}(\xi',x)\,
  \nonumber\\
  &\times\left(\frac{\I}{D}\right)^2\,\mean{\delta U_1(\xi)\,\delta U_2(\xi') }
\end{align}
We now use that the potential in the wire is related to the characteristic potential by $U(x)\simeq u_1(x)\,eV$ in order to recover the structure \eqref{eq:NLCproductCorr} where the previous equation exactly corresponds to the first term of \eqref{eq:RootEqChiG} and 
$\mean{\delta U_1(\xi)\,\delta U_2(\xi') }=(eV)^2\chi_\DoS(\xi,\xi')$.
This establishes precisely the correspondence with the scattering formalism of B\"uttiker when applied to disordered metals.

\end{appendix}



\begin{thebibliography}{10}
\bibitem{footnote1}
  Another effect of interaction in the non-linear regime was the question of decoherence which was examined in Ref.~\onlinecite{LudBlaMir04} motivated by several experiments (see \cite{TerBabStrNusSch02} in particular).
\bibitem{footnote2}
  Obviously, we exclude situations where the $I$-$V$ characteristic is non-analytic, like in strickly 1D wires in the presence of interaction (Tomonaga-Luttinger theory) \cite{Gia04} or other non-linear effects such as dynamical Coulomb blockade~\cite{IngNaz92}.
\bibitem{footnote3}
In a good metal like silver, $\ell_\mathrm{TF}$ is of the atomic scale, like the Fermi wavelength $\lambda_F$.
\bibitem{footnote4}
In the smaller samples of Ref.~\onlinecite{ZumMarHanGos06}, ballistic QDs of size $L=1\:\mu$m, the interaction parameter was smaller~:  $\gamma_\mathrm{int}\approx0.7$.
\bibitem{footnote5}
This is not true in more complex networks, as was discussed in Ref.~\onlinecite{TexMon16}.
\bibitem{footnote6}
 Wires etched in 2D electron gas are usually characterised by a different regime where the elastic scattering is dominated by boundary scattering, hence $\ell_e>\sw$.
  The expression of the magnetic length which describes this regime was obtained in Refs.~\onlinecite{DugKhm84,BeeHou88} (see also Refs.~\onlinecite{HouBeeWeeMoo88,ReuBouMai95}).
\bibitem{footnote7}
In principle the functional derivative, with dimension $[\delta/\delta U(\vec{r})]=[$Energy$]^{-1}\times[$Volume$]^{-1}$, introduces a factor $1/\sw^{d-1}$ related to the cross-section of the wire, which is always compensated by the spatial integrals. 
  We will systematically omit this factor.
\bibitem{footnote8}
Note that a related question was studied by Gramespacher and B\"uttiker~\cite{GraBut99} who analysed the correlations of $\DoS_1(\vec{r},\vec{r}\,';\varepsilon)$, the contribution of the scattering states from contact $1$ to the off-diagonal matrix elements $\bra{\vec{r}}\delta(\varepsilon-H)\ket{\vec{r}\,'}$.
\bibitem{footnote9}
  This expression of the phase coherence length is obtained from the analysis of the \textit{weak localisation} correction, i.e. from the averaged current.
  It is a priori not obvious that the analysis of the current fluctuations involve \textit{exactly} the same length scale, however it was shown to be the case in Refs.~\onlinecite{AleBla02,TexMon05b} as the correlators involve the same decoherence functional as the average (see also Ref.~\onlinecite{TreTexYevDelLer11}).
  This has been recently demonstrated experimentally in Ref.~\onlinecite{CapTexMonMaiWieSam13}.
\bibitem{footnote10}
This formula is given in Ref.~\onlinecite{LarKhm86}, from which we have corrected the prefactor as this reference contains several typos.
\end{thebibliography}

\end{document}